\documentclass[cmp,final]{svjour}
\def\makeheadbox{}
\usepackage{longtable,booktabs}
\usepackage{mathptmx}
\usepackage[T1]{fontenc}
\usepackage[utf8]{inputenc}
\usepackage{textcomp}

\usepackage{amsmath, amssymb, csquotes, dsfont, floatflt, textcase, tikz, tikz-cd}
\usepackage[hyperindex,breaklinks,hidelinks]{hyperref}
\usepackage[backend=bibtex8]{biblatex}
\usepackage{graphicx}

\makeatletter
\def\fps@figure{htbp}
\makeatother
\usetikzlibrary{matrix,arrows}
\tikzset{
  commutative diagrams/.cd,
  arrow style=tikz,
  diagrams={>=stealth},
  shift up/.style={
    to path={([yshift=#1]\tikztostart.east) -- ([yshift=#1]\tikztotarget.west) \tikztonodes}},
  shift up left/.style={
    to path={([yshift=#1]\tikztostart.west) -- ([yshift=#1]\tikztotarget.east) \tikztonodes}},
  mathdouble/.style={-,double equal sign distance}
}

\def\sgn{\mathop{\mathrm{sgn}}}
\def\ger{\mathfrak}
\def\sh{\mathcal}
\def\DMO{\DeclareMathOperator}
\newcommand\ev{{\bar 0}}
\newcommand\odd{{\bar 1}}

\mathchardef\ordinarycolon\mathcode`\:
\mathcode`\:=\string"8000
\begingroup \catcode`\:=\active
  \gdef:{\mathrel{\mathop\ordinarycolon}}
\endgroup
\newcommand{\defi}{:=}

\newcommand\vphi{\varphi}
\newcommand\vrho{\varrho}
\newcommand\eps{\varepsilon}
\newcommand\nats{\mathbb{N}}
\newcommand\ints{\mathbb{Z}}
\newcommand\reals{\mathbb{R}}
\newcommand\cplxs{\mathbb{C}}
\newcommand\knums{\mathbb K}

\newcommand\vvoid{\varnothing}
\newcommand\sle{\leqslant}
\newcommand\sge{\geqslant}

\DMO\Ad{\mathrm{Ad}}
\DMO\id{\mathrm{id}}
\DMO\coker{\mathrm{coker}}
\DMO\im{\mathrm{im}}
\DMO\supp{\mathrm{supp}}
\DMO\sign{\mathrm{sign}}
\DMO\blop{\sh L}
\DMO\FF{\sh{F\!\!F}}

\def\size#1#2#3#4#5{#1#2#3#4#5}
\def\Size#1#2#3#4#5{\bigl{#1}#2\bigm{#3}#4\bigr{#5}}
\def\Sizze#1#2#3#4#5{\bigl{#1}#2#3#4\bigr{#5}}

\def\SSSizze#1#2#3#4#5{\Bigl{#1}#2#3#4\Bigr{#5}}
\def\sdual#1#2{\size{\langle}{#1}{|}{#2}{\rangle}}
\def\Sdual#1#2{\Size{\langle}{#1}{|}{#2}{\rangle}}

\def\rscp#1#2{\size{(}{#1}{,}{#2}{)}}

\def\parens#1{\size{(}{#1}{}{}{)}}
\def\Parens#1{\Sizze{(}{#1}{}{}{)}}

\def\Bracks#1{\Sizze{[}{#1}{}{}{]}}
\def\braces#1{\size{\{}{#1}{}{}{\}}}
\def\Braces#1{\Sizze{\{}{#1}{}{}{\}}}

\def\BBBraces#1{\SSSizze{\{}{#1}{}{}{\}}}

\def\Set#1#2{\Size{\{}{#1}{|}{#2}{\}}}
\def\norm#1{\size{\lVert}{#1}{}{}{\rVert}}
\def\Norm#1{\Sizze{\lVert}{#1}{}{}{\rVert}}
\def\ket#1{\size{\lvert}{#1}{}{}{\rangle}}
\def\bra#1{\size{\langle}{#1}{}{}{\rvert}}
\def\abs#1{\size{\lvert}{#1}{}{}{\rvert}}
\def\Abs#1{\Sizze{\lvert}{#1}{}{}{\rvert}}

\def\endo#1{\mathrm{End}\parens{#1}}

\def\auto#1{\mathrm{Aut}\parens{#1}}

\def\Hom#1#2{\mathrm{Hom}\Parens{#1,#2}}

\def\diii{D\mathrm{I\!I\!I}}
\def\aii{A\mathrm{I\!I}}
\def\cii{C\mathrm{I\!I}}
\def\aiii{A\mathrm{I\!I\!I}}
\def\trs{\ints_4}
\def\phs{\ints_2}
\def\Uc{\mathrm U(1)_c}
\def\Usp{\mathrm{SU}(2)_s}

\title{Bulk-boundary correspondence for disordered free-fermion topological
phases}
\author{Alexander Alldridge\inst{1} \and Christopher Max\inst{1,2} \and Martin R. Zirnbauer\inst{2}}
\providecommand{\institute}[1]{}
\institute{Mathematisches Institut, \email{\href{mailto:alldridg@math.uni-koeln.de}{\nolinkurl{alldridg@math.uni-koeln.de}}} \and Institut für theoretische Physik, \email{\href{mailto:cmax@thp.uni-koeln.de}{\nolinkurl{cmax@thp.uni-koeln.de}},
\href{mailto:zirn@thp.uni-koeln.de}{\nolinkurl{zirn@thp.uni-koeln.de}}}}
\date{Received: 2019-07-12}

\def\newtheorem#1{}

\let\BeginKnitrBlock\begin \let\EndKnitrBlock\end
\begin{document}
\maketitle
\begin{abstract}
Guided by the many-particle quantum theory of interacting systems, we
develop a uniform classification scheme for topological phases of
disordered gapped free fermions, encompassing all symmetry classes of
the Tenfold Way. We apply this scheme to give a mathematically rigorous
proof of bulk-boundary correspondence. To that end, we construct real
C\(^\ast\)-algebras harbouring the bulk and boundary data of disordered
free-fermion ground states. These we connect by a natural
bulk-to-boundary short exact sequence, realising the bulk system as a
quotient of the half-space theory modulo boundary contributions. To
every ground state, we attach two classes in different pictures of real
operator \(K\)-theory (or \(KR\)-theory): a bulk class, using Van
Daele's picture, along with a boundary class, using Kasparov's Fredholm
picture. We then show that the connecting map for the bulk-to-boundary
sequence maps these \(KR\)-theory classes to each other. This implies
bulk-boundary correspondence, in the presence of disorder, for both the
``strong'' and the ``weak'' invariants.
\end{abstract}

\hypertarget{sec:intro}{%
\section{Introduction}\label{sec:intro}}

The research field of topological quantum matter stands out by its
considerable scope, pairing strong impact on experimental physics with
mathematical structure and depth. On the experimental side, physicists
are searching for novel materials with technological potential,
\emph{e.g.}, for future quantum computers; on the theoretical side,
mathematicians have been inspired to revisit and advance areas such as
topological field theory and various types of generalised cohomology and
homotopy theory.

Early research on topological quantum matter was concerned primarily
with the integer and fractional quantum Hall effects (QHE) and close
variants thereof. Many of the key principles were already recognised and
developed in that context; on the mathematical side, these included the
interpretation of the Hall conductance as a topological \(K\)-theory
class \autocite{ass,tknn} and also as a Fredholm index
\autocite{b86,bellissard-etal}, and the principle of bulk-boundary
correspondence \autocite{skr00}.

A major boost came around 2005 \autocite{km05}, when the research field
was brought to full scale by the realisation that phenomena similar to
QHE were possible even for systems with symmetries, in particular the
anti-unitary symmetry of time reversal. This led to the notion of
symmetry-protected topological phases as equivalence classes of gapped
Hamiltonians that are connected by homotopies subject to a fixed
symmetry group; see \autocite{ctsr16} for a recent review.

Rapid progress was made in the free-fermion limit with translation
symmetry, where one can use the Fourier--Bloch technique to develop the
theory over the Brillouin torus of conserved momentum. The challenge
nowadays is to advance the theory in the direction of disordered and
interacting systems. In the present paper, we focus on the disorder
aspect, assuming that particle interactions can be handled by mean-field
theory in its most general form, namely the so-called
Hartree--Fock--Bogoliubov (HFB) mean-field approximation, or
free-fermion approximation for short.

The traditional physics approach to disordered free fermions starts from
the Wegner--Efetov supersymmetry formalism, which famously led to the
non-linear \(\sigma\)-model describing disordered metals. For systems
with a small (if not vanishing) density of states, the non-linear
\(\sigma\)-model has also been used, but it remains less well
established there. In the present paper, we build on the mathematically
rigorous approach developed initially for aperiodic systems and the
integer quantum Hall effect by Bellissard and co-workers
\autocite{bellissard-etal}. In that approach one assumes that
translation invariance holds on average over the disorder and in that
sense represents a symmetry of the macroscopic system.

A foundational aspect that has to be emphasised is the significant role
played by superconductors, for theory as well as experiment. In fact,
much of the excitement about topological quantum matter stems from the
technological promise of robust zero-energy modes (so-called Majorana
zero modes) bound to the ends of superconducting wires of symmetry type
\(D\). To mention a second example of well documented interest, planar
substructures in strontium ruthenate
(\(\mathrm{Sr}_2\,\mathrm{Ru}\, \mathrm{O}_4\)) behave as a
two-dimensional topological superconductor with broken time-reversal
symmetry \autocite{sr2ruo4}.

Superconductivity is a true many-body phenomenon that requires the
spontaneous breaking of the fundamental \(\mathrm{U}(1)\) symmetry
underlying charge conservation. As many-body systems with interactions
and collective dynamics, superconductors defy any treatment based on
straight single-particle quantum mechanics. Nonetheless, much of the
phenomenology of superconductors and their quasi-particle excitations
can be captured by the HFB mean-field approximation. In that scheme, one
doubles the single-particle Hilbert space \(\mathcal{V}\) to the Nambu
space of fields, \(\mathcal{W} = \mathcal{V} \oplus \mathcal{V}^\ast\),
and then works with the adjoint action of the free-fermion Hamiltonians
on the fields. The field space \(\mathcal{W}\) is equipped with two
canonical invariant structures: (i) a symmetric bilinear form due to the
canonical anti-commutation relations for fermion fields; (ii) a real
structure \(\mathcal{W}_\mathbb{R} \subseteq \mathcal{W}\) known in
physics as the space of Majorana fields. The resulting formalism, also
known by the name of Bogoliubov--deGennes or Gorkov equations, is
sometimes advertised as single-particle quantum mechanics with a
\enquote{particle-hole symmetry}. In the present work we do better: with
a view towards future extensions to incorporate electron-electron
interactions, we sketch how the formalism emerges from its proper
foundations in many-body quantum theory, and we develop the theory
accordingly.

Our main theme here will be \emph{bulk-boundary correspondence}. Roughly
speaking, this principle says that the topological characteristics of
the bulk system must be reflected in corresponding properties of the
boundary theory. In the general case of interacting systems, one has a
trichotomy of possibilities: the boundary theory (i) may be gapless, or
(ii) may undergo symmetry breaking, or (iii) may be a gapped topological
phase of an exotic nature. For free fermions, however, only the first
alternative is realised.

Historically, early indications of the principle of bulk-boundary
correspondence came from anomalies in quantum field theory, especially
the Adler--Bell--Jackiw anomaly for chiral fermions. Simply put, the
boundary theory in isolation may be anomalous in that a conservation law
is illegally violated and, if so, there must be an exactly opposite
effect from the bulk system to cancel the total anomaly for the boundary
joined with the bulk. For example, the low-energy effective action
functional for Quantum Hall bulk systems contains a Chern--Simons term,
which fails to be gauge-invariant when a boundary is present, and that
gauge anomaly is cancelled by chiral boundary fermions that anomalously
appear or disappear when magnetic flux crosses the boundary.

For the case of the integer quantum Hall effect, the heuristic principle
of bulk-boundary correspondence was established in mathematically
rigorous terms by Kellendonk, Richter, and Schulz-Baldes
\autocite{krs02,skr00}. The starting point of their approach is a short
exact sequence of C\(^\ast\)-algebras, realising the bulk system as a
quotient of the half-space theory taken modulo the boundary theory. The
short exact sequence induces a six-term exact sequence connecting the
\(K\)-theory groups \(K_0\) and \(K_1\) of these C\(^\ast\)-algebras.
Recently, Prodan and Schulz-Baldes \autocite{ps16} have adapted and
applied that early approach to the \enquote{chiral} analogue of IQHE
systems in odd space dimension. As a result, there now exists a fairly
complete understanding of the so-called \enquote{complex symmetry
classes}, known as type \(A\) and \(\aiii\).

Several attempts have been made to give a mathematical proof of
bulk-boundary correspondence for all Tenfold-Way symmetry classes. A
notable result for \(2D\) systems in symmetry class \(\aii\) was proved
by Graf--Porta \autocite{gp}. The first approach encompassing all
symmetry classes in \(2D\) is due to Loring \autocite{loring}, and is
based on the formalism of almost commuting Hermitian matrices. Results
covering all symmetry classes in any dimension were obtained by
Hannabuss, Mathai, and Thiang. In a series of papers
\autocite{hmt,mt1,mt2}, they apply the concept of \(T\)-duality to the
bulk-boundary correspondence. This duality, with its origins in string
theory, relates the \(K\)-theory of two torus bundles over the same base
manifold. In the presence of lattice symmetries, Kubota
\autocite{Kubota} establishes bulk-boundary correspondence at the level
of strong invariants, by using the tools of coarse geometry. Following
ideas of Freed--Moore \autocite{fm}, he combines the \enquote{quantum
symmetries} of the Tenfold Way and the lattice symmetries into a single
group. An alternative route, based on Kasparov \(KK\)-theory, is pursued
by Bourne--Kellendonk--Rennie \autocite{bourne-kellendonk-rennie}.
Building on the previous work of Bourne--Carey--Rennie
\autocite{bourne-carey-rennie}, they show that certain bulk and boundary
classes are related \emph{via} the bulk-to-boundary map of
Kellendonk--Richter--Schulz-Baldes.

Our approach to the bulk and boundary classification of free-fermion
topological phases stands out by its foundation in the physically
well-motivated Nambu space formalism. As explained above, this is
essential for a proper treatment of topological superconductors. Since
the objects we consider have a clear physical meaning, we are moreover
able to draw conclusions with a bearing on the expected physics
phenomenology; for example, our theory immediately yields the emergence
of gapless edge modes for topologically non-trivial bulk phases.

Building on the pioneering work of Bellissard \emph{et al.}
\autocite{b86,b89,bn90,bellissard-etal}, Schulz-Baldes \emph{et al.}
\autocite{krs02,ps16,skr00}, we give a complete and uniform proof of
bulk-boundary correspondence for symmetry-protected topological (SPT)
phases of free fermions with disorder. Our basic physics setting is that
of the Tenfold Way of disordered free fermions \autocite{az97},
encompassing the two complex symmetry classes \(A\) and \(\aiii\), as
well as the eight \enquote{real symmetry classes} \(D\), \(\diii\),
\(\aii\), \(\cii\), \(C\), \(C\mathrm I\), \(A\mathrm I\), and
\(BD\mathrm I\). To avoid lengthy case-by-case considerations, we
introduce the C\(^\ast\)-algebras of Bellissard and
Kellendonk--Richter--Schulz-Baldes in a form suitable for the
introduction of \enquote{real} and \enquote{complex} symmetries, and we
treat the symmetries systematically using Clifford algebras.

In Section \ref{sec:bulk}, we begin by deriving the C\(^\ast\)-algebra,
\(\mathbb A\), for the bulk in the Nambu space setting that emerges from
the many-body theory. The construction puts the real structure of the
Nambu space \(\sh W\) into evidence, as is essential for our treatment
of the symmetries appearing in the Tenfold Way. From our physics-driven
model building, ground states of free-fermion Hamiltonians appear
naturally as real skew-Hermitian unitaries \(J\) on \(\sh W\). These are
called \emph{quasi-particle vacua} (QPV) and figure as the principal
objects of our further investigations. Ultimately, we work in the
tight-binding approximation; we justify its use from physical
principles. Adopting an idea due to Kubota \autocite{Kubota}, our
construction of the bulk algebra \(\mathbb A\) appeals to the uniform
Roe C\(^\ast\)-algebra of the lattice \(\Lambda\) of translations that
leave the crystalline set of atomic sites invariant. That choice of
C\(^\ast\)-algebra reflects the (local) finiteness of the number of
tight-binding degrees of freedom.

We incorporate disorder at this stage by following Bellissard
\autocite{bn90} in encoding translational invariance on average through
a covariance relation for the lattice \(\Lambda\). It turns out (Theorem
\ref{thm:covcrossed}) that this covariance relation yields a particular
structure for the C\(^\ast\)-algebra \(\mathbb A\), namely that of a
crossed-product C\(^\ast\)-algebra. Since the work of Bellissard, the
bulk algebra has conventionally been \emph{defined} as a crossed
product; for us, this structure emerges naturally. To sum up, our real
skew-Hermitian unitaries \(J\in\mathbb A\) are translation-invariant on
average over the disorder. We therefore call them \emph{disordered IQPV}
(where the I stands for \textbf{i}nvariant).

The next step is to include symmetries. Past treatments of the Tenfold
Way have relied on case-by-case considerations, exacerbated further by
the appearance of various case-dependent real and quaternionic
structures. A uniform description avoiding these ramifications was first
put forth by Kitaev \autocite{kitaev} by the judicious use of Clifford
algebras. It was later worked out in detail by Kennedy--Zirnbauer
\autocite{rk-mrz} who introduced the concept of \emph{pseudo-symmetries}
as an organising principle for symmetry classes of translation-invariant
free fermions in the clean limit. In Section \ref{sec:bulkinv}, we
transcribe this concept from the Bloch-bundle setting to that of
C\(^\ast\)-algebras and use it to define symmetry classes of disordered
IQPV \(J\in\mathbb A\).

Our formulation turns out to be well suited for the introduction of
\(K\)-theoretical invariants, and we show how a disordered IQPV
\(J\in\mathbb A\) naturally defines a class in the real \(K\)-theory of
the bulk algebra \(\mathbb A\) (Definition \ref{def:bulk-class}). The
entirely algebraic construction utilises the picture of real
\(K\)-theory introduced by Van Daele \autocite{van_Daele1,van_Daele2}.
The use of this picture was inspired by recent work of Kellendonk
\autocite{kellendonk-VD}. Our construction indeed gives an alternative
definition of real \(K\)-theory. That is, any \(KR\)-theory class of
\(\mathbb A\) is the \enquote{bulk class} of a disordered IQPV in a
certain symmetry class (Theorem \ref{thm:iqpv-kthy}). This fact does not
depend on the specifics of the bulk algebra \(\mathbb A\) but is valid
for any real C\(^\ast\)-algebra.

In Section \ref{sec:bdy}, we introduce a physical system boundary by
defining a half-space algebra \(\smash{\widehat{\mathbb A}}\), following
the ideas of Kellendonk--Richter--Schulz-Baldes \autocite{krs02,skr00}.
We break the translational covariance in one spatial direction and show
that the resulting algebra \(\smash{\widehat{\mathbb A}}\) has the
structure of a crossed product with the half-space semilattice
\(\smash{\hat\Lambda}\subseteq\Lambda\) (Proposition
\ref{prp:halfsp-covcrossed}). This leads to a Toeplitz extension
connecting bulk and boundary (Theorem \ref{thm:bb-ses}).

The Szegő projection defining the Toeplitz extension allows us to
assign, to every disordered IQPV \(J\) in a given symmetry class, a real
skew-Hermitian operator
\(\smash{\widehat J}\in\smash{\widehat{\mathbb A}}\) which satisfies the
same pseudo-symmetries as \(J\) but squares to \(-1\) only up to
boundary contributions. Such operators define real \(K\)-theory classes
for the boundary C\(^\ast\)-algebra \(\mathbb A_\partial\), as we show
by using Kasparov's Fredholm picture for \(KR\)-theory (Definition
\ref{def:bound-IQPV}). Since for free fermions, the appearance of
gapless edge modes signals a topologically non-trivial bulk, it is
essential that the picture of \(K\)-theory used for the boundary classes
allows the spectral gap to close at the boundary. This is an important
aspect not present in previous approaches to the bulk-boundary
correspondence using \(K\)-theory.

Thus, we attach to any disordered IQPV of a given symmetry class two
\(K\)-theory classes: one in the bulk and another one at the boundary.
The main result of our article, in Section \ref{sec:bb}, is to
inter-relate these two classes: We prove in Theorem
\ref{thm:ComplexBound} that the relation is given by the connecting map
in \(KR\)-theory. For the case of the integer QHE, this idea is already
present in the work of Kellendonk--Richter--Schulz-Baldes
\autocite{krs02,skr00}. Our theorem covers all symmetry classes of the
Tenfold Way, and it holds in the presence of disorder, for the strong as
well as the weak invariants. We also show, in Proposition
\ref{prp:CleanLimit}, that the \(K\)-theory of the algebras we have
constructed is invariant w.r.t. certain types of disorder.

In the final Section \ref{sec:discussion}, we complement our results by
a discussion, (i), of the anomalous nature of the boundary phase, (ii),
of the role of the \enquote{trivial} phase, (iii), of the internal space
of our tight-binding model, and (iv), of the construction of numerical
topological invariants. We examine the so-called \enquote{strong}
invariant and show the emergence, for topologically non-trivial systems,
of gapless boundary phases.

\emph{Note on authorship}. This article is part of the PhD thesis work
of CM supervised by AA. The contributions of MZ are primarily in
Sections \ref{sec:intro}--\ref{sec:bulkinv}, where the mathematical
model is developed from physical principles.

\hypertarget{sec:bulk}{%
\section{Bulk systems with disorder}\label{sec:bulk}}

\hypertarget{subs:nambu-bulk}{%
\subsection{\texorpdfstring{From Nambu space to real
C\(^\ast\)-algebras}{From Nambu space to real C\^{}\textbackslash ast-algebras}}\label{subs:nambu-bulk}}

In this subsection, we shall construct the algebra of bulk observables
for a macroscopically translation invariant solid-state system in the
free-fermion approximation and with on-site symmetries present.

In the literature on the subject, the C\(^\ast\)-algebra of a gapped
bulk system is modelled by a crossed product C\(^\ast\)-algebra. This
goes back to Bellissard \autocite{b86,b89}, and it is fundamental for
his mathematical proof of the (integer) quantisation of the Hall
conductance using non-commutative geometry \autocite{bellissard-etal}.
Later work by Schulz-Baldes--Kellendonk--Richter \autocite{skr00} and
Kellendonk--Richter--Schulz-Baldes \autocite{krs02} employed a Toeplitz
extension of the crossed product C\(^\ast\)-algebra of the bulk system
to prove bulk-boundary correspondence for integer quantum Hall systems
in a half-plane geometry. More recently, Prodan--Schulz-Baldes extended
the formalism to incorporate half-space systems in the chiral symmetry
class \(A\mathrm{I\!I\!I}\) (see \autocite[Chapter 5]{ps16} for an
exposition).

Our first goal here is to adapt the mathematical framework in order to
accommodate free-fermion topological phases from the other symmetry
classes of the Tenfold Way. For that purpose, we deem it fit to rebuild
the C\(^\ast\)-algebra of our model from its foundations in many-body
physics. This we shall do step-by-step in the present subsection, with
the final outcome being that our algebra still has a crossed product
structure. The latter will be of crucial importance in Section
\ref{sec:bdy}.

\hypertarget{setting-in-many-body-theory}{%
\subsubsection{Setting in many-body
theory}\label{setting-in-many-body-theory}}

Topological insulators and superconductors are many-body systems
consisting of \(10^{10}\)-\(10^{25}\) interacting particles. To treat
them in any satisfactory way (including particle-particle interactions
as well as disorder), one must apply the formalism of many-body quantum
theory. In the present paper, we treat \emph{disordered} systems,
neglecting residual interactions beyond those captured by mean-field
theory in the most general scheme of Hartree--Fock--Bogoliubov. While
our focus on disorder makes for substantial technical simplifications,
allowing us to do computations in the Nambu space of field operators
(akin to the Hilbert space of single-particle quantum mechanics), we
still consider it appropriate if not necessary to develop (or, at least,
motivate) the mathematical model we use from its very foundations in
many-body theory. A particular case in point is the proper description
of topological superconductors. Indeed, superconductors are
many-electron systems that undergo spontaneous breaking of
\(\mathrm U(1)\) charge symmetry, a many-body phenomenon that cannot be
understood or modelled in a straight single-particle picture.

A basic characteristic of solid bodies and other condensed matter
systems is their \emph{chemical potential} \(\mu\), also known as the
\emph{Fermi energy} in the case of particles obeying Fermi statistics.
By definition, \(\mu\) is the energy level at which the process of
adding or removing a charged particle to or from a macroscopic body
costs no work. Single-electron states with energy above \(\mu\) are
called \emph{conduction} states, those with energy below \(\mu\)
\emph{valence} states. The single-electron conduction (resp. valence)
states span a Hilbert space \(\sh V_+\) (resp. \(\sh V_-\)).

Let us emphasise that it is crucial for us to construct our mathematical
model around the chemical potential \(\mu\). The reason is that we
intend to truncate the locally infinite continuum degrees of freedom to
a so-called tight-binding model supporting only a \emph{finite} number
of energy bands, and that approximation by truncation is physically
justified (at very low temperatures) if and only if the tight-binding
model is designed to capture the band structure close to the chemical
potential.

The basic structure underlying many-body quantum theory is that of a
\emph{Fock space}. In the case of indistinguishable fermions such as
electrons, this is an exterior algebra \[
    \textstyle\bigwedge=\textstyle\bigwedge(\sh V_+\oplus\sh V_-^\ast),
\] completed to a Hilbert space with inner product
\(\sdual\cdot\cdot_\wedge\) which is induced by the Hilbert space
structures of \(\sh V_\pm\). Note that \(\sh V_-^\ast\) denotes the
vector space dual to \(\sh V_-\). Physically speaking, the appearance of
\(\sh V_-^\ast\) (instead of \(\sh V_-\)) means that the Fock vacuum,
\emph{i.e.} the complex line \[
    \textstyle\bigwedge^0=\textstyle\bigwedge^0(\sh V_+\oplus\sh V_-^\ast)\cong\cplxs,
\] is the \enquote{Fermi sea} of occupied valence states. The Fock space
\(\textstyle\bigwedge\) is bi-graded by \[
    \textstyle\bigwedge=\bigoplus_{p,q\sge0}\textstyle\bigwedge^{p,q},\quad\textstyle\bigwedge^{p,q}\defi\textstyle\bigwedge^p(\sh V_+)\otimes\bigwedge^q(\sh V_-^\ast).
\] The integer \(p\) counts the number of single-particle excitations,
\emph{i.e.} the number of electrons populating the conduction states,
while \(q\) is the number of single-hole excitations, \emph{i.e.} the
number of electrons missing from the Fermi sea of occupied valence
states. Abbreviating the language, one calls \(p\) and \(q\) the number
of particles resp. of holes. The difference \(p-q\) is referred to as
the \emph{charge} (relative to \(\mu\)).

Given the Fock--Hilbert space \(\textstyle\bigwedge\) built on the Fock
vacuum \(\textstyle\bigwedge^0\) around the chemical potential \(\mu\),
a \emph{field operator}
\(\psi:\textstyle\bigwedge\longrightarrow\textstyle\bigwedge\) is a
linear combination \[
    \psi=\eps(v_+)+\iota(\vphi_+)+\iota(v_-)+\eps(\vphi_-)\quad(v_\pm\in\sh V_\pm,\ \vphi_\pm\in\sh V_\pm^\ast).
\] Here, the operator of exterior multiplication
\(\eps(v_+):\smash{\textstyle\bigwedge^{p,q}}\longrightarrow\smash{\textstyle\bigwedge^{p+1,q}}\)
increases the number of particles by one, whereas
\(\eps(\vphi_-):\smash{\bigwedge^{p,q}}\longrightarrow\smash{\bigwedge^{p,q+1}}\)
removes an electron from a valence state, thereby increasing the number
of holes by one. Conversely, the operators
\(\iota(\vphi_+):\smash{\bigwedge^{p,q}}\longrightarrow\smash{\bigwedge^{p-1, q}}\)
and
\(\iota(v_-):\textstyle\bigwedge^{p,q}\longrightarrow\textstyle\bigwedge^{p,q-1}\)
decrease the number of particles resp. holes by alternating contraction.

The field operators \(\psi\) span a complex vector space, the so-called
\emph{Nambu space} \(\sh W\) of fields. Note that \(\sh W\) is not
fundamentally a Hilbert space, but will be given a derived Hilbert space
structure in due course. The space \(\sh W\) is equipped with two basic
structures. The first one is a symmetric complex bilinear form,
\(\braces{\cdot,\cdot}:\sh W\otimes\sh W\longrightarrow\cplxs\), which
is canonically defined by \[
    \psi\psi'+\psi'\psi=\Braces{\psi,\psi'}{\id}_\wedge.
\] It is referred to as the \emph{CAR form} (for \textbf{c}anonical
\textbf{a}nticommutation \textbf{r}elations). In explicit terms, using
the notation above, \[
    \Braces{\psi,\psi'}=\vphi_+(v_+')+\vphi_-(v_-')+\vphi_+'(v_+)+\vphi'_-(v_-).
\] The most general many-body operator acting on the Fock space resides
in the complex Clifford algebra \(C\ell(\sh W)\) generated by \(\sh W\)
with the relations given by the CAR form.

Secondly, the Nambu space of fields carries a real structure
\(\gamma:\sh W\longrightarrow\sh W\). Denoting the Fréchet--Riesz
isomorphism for the conduction-state and valence-state Hilbert spaces by
\(h:\sh V_\pm\longrightarrow\sh V_\pm^\ast\), one defines \(\gamma\) as
the anti-linear involution \[
    \eps(v_+)+\iota(\vphi_+)+\iota(v_-)+\eps(\vphi_-)\longmapsto\iota(hv_+)+\eps(h^{-1}\vphi_+)+\eps(hv_-)+\iota(h^{-1}\vphi_-). 
\] Physicists frequently refer to the real subspace \(\sh W_\reals\)
consisting of the \(\gamma\)-fixed vectors as the space of
\emph{Majorana} fields. Notice that the CAR form of \(\sh W\) restricts
to a Euclidean structure of the real vector space \(\sh W_\reals\).

The real structure \(\gamma\) and the CAR form of \(\sh W\) each induce
a corresponding operation on linear transformations \(L\in\endo{\sh W}\)
(and similarly on the anti-linear transformations). In the case of
\(\gamma\), this is the anti-linear involution \[
    L\longmapsto\overline L\defi\gamma\circ L\circ\gamma,
\] and we say that \(L\) is \emph{real} if \(\overline L=L\) (resp.
\emph{imaginary} if \(\overline L=-L\)). In the case of the CAR form,
the induced operation is a linear anti-involution
\(L\longmapsto L^\intercal\) called \emph{transposition}, defined by \[
    \Braces{\psi,L^\intercal\psi'}=\Braces{L\psi,\psi'}\quad(\psi,\psi'\in\sh W).
\] An operator \(L\in\endo{\sh W}\) is called \emph{skew} if
\(L^\intercal=-L\) (resp. \emph{symmetric} if \(L^\intercal=L\)).

Finally, let us remark here that the CAR form \(\braces{\cdot,\cdot}\)
and the real structure \(\gamma\) are fundamental in that they are not
tied to the free-fermion approximation but retain their key roles even
in the presence of electron-electron interactions.

\hypertarget{free-fermion-setting}{%
\subsubsection{Free-fermion setting}\label{free-fermion-setting}}

We proceed to introduce the objects and structures that emerge upon
assuming the free-fermion approximation (or free-fermion limit). For
\(\psi,\psi'\in\sh W_\reals\), consider the expression
\(\psi\psi'-\psi'\psi\) in the Clifford algebra \(C\ell(\sh W_\reals)\).
Such operators, quadratic in the fields and skew-symmetrised, close
under the commutator in \(\endo{\textstyle\bigwedge}\) and span a Lie
algebra isomorphic to \(\ger{so}(\sh W_\reals)\). By exponentiating that
Lie algebra inside \(C\ell(\sh W_\reals)\), one obtains the Lie group
\(\mathrm{Spin}(\sh W_\reals)\subseteq\mathrm U(\textstyle\bigwedge)\).
The latter acts on \(\sh W_\reals\subseteq\endo{\textstyle\bigwedge}\)
by real orthogonal transformations
\(\vrho(g):\sh W_\reals\longrightarrow\sh W_\reals\), defined as \[
    \vrho(g)\psi\defi g\psi g^{-1}\quad(g\in\mathrm{Spin}(\sh W_\reals)). 
\] The significance of \(\mathrm{Spin}(\sh W_\reals)\) is that it
contains all unitary operators of physical interest (in the free-fermion
limit), including the operators for time evolutions, space translations,
space rotations, \emph{etc.}, as well as the unitary Bogoliubov
transformations.

A \emph{free-fermion Hamiltonian} \(\sh H\in C\ell(\sh W)\) is, by
definition, a Clifford algebra element quadratic in the fields and
Hermitian with respect to the Fock space inner product. Because the time
evolutions \(e^{-it\sh H/\hbar}\in\mathrm{Spin}(\sh W_\reals)\)
generated by it act on \(\sh W_\reals\) as real orthogonal
transformations
\(\smash{\vrho(e^{-it\sh H/\hbar})}\in\mathrm{SO}(\sh W_\reals)\), it
follows that \(iH=d\vrho(i\sh H)\) is real and skew:
\(\overline{iH}=+iH=-iH^\intercal\). Put differently, the induced
Hamiltonian \(H\in\endo{\sh W}\) is imaginary and skew: \[
    \overline H=-H=H^\intercal. 
\]

Let us provide a little more detail about the free-fermion Hamiltonians
\(\sh H\). If \(\sh H\) stems from a charge-conserving Hamiltonian
\(\mathsf H\) which is defined on the single-particle Hilbert space
\(\sh V=\sh V_+\oplus \sh V_-\) and decomposed as \[
    \mathsf H=
    \begin{pmatrix}
    A&B\\ C&D
    \end{pmatrix}
    \in
    \begin{pmatrix}
    \endo{\sh V_+}&\Hom{\sh V_-}{\sh V_+}\\ \Hom{\sh V_+}{\sh V_-}&\endo{\sh V_-}
    \end{pmatrix},
\] then for any bases \((e_j^\tau)\) of \(\sh V_\tau\) (\(\tau=\pm\))
with dual bases \((f^j_\tau)\), one has the \emph{second-quantised}
Hamiltonian \[
    \sh H\defi 
    \begin{aligned}[t]
        &\sum_j\eps((A-\mu)e_j^+)\iota(f^j_+)+\sum_\ell\eps(Be^-_\ell)\eps(f^\ell_-)\\
        &-\sum_j\iota(f^j_+)\iota(Ce^+_j)-\sum_\ell\eps(f^\ell_-)\iota((D-\mu)e^-_\ell).
    \end{aligned}
\] Thus, the block \(B\in\Hom{\sh V_-}{\sh V_+}\) gives rise to an
operator of type \(\eps\,\eps\) creating particle-hole pairs, while the
block \(C\) (second-quantised as type \(\iota\,\iota\)) does the
opposite. If \(\mathsf H\) is taken to be exactly the Hartree-Fock
mean-field Hamiltonian, then \(B=C=0\) and \(A-\mu>0>D-\mu\) by the
definition of \(\sh V_+\) and \(\sh V_-\) from the chemical potential
\(\mu\). The \emph{charge operator} (representing the identity on
\(\sh V\)) is \[
\sh Q=\sum_j\eps(e^+_j)\iota(f^j_+)-\sum_\ell\eps(f^\ell_-)\iota(e^-_\ell).
\] Note that \(\sh Q|_{\wedge^{p,q}}=p-q\).

The full power of the formalism comes to play in the treatment of
superconducting systems. There, the \(\mathrm U(1)\) symmetry underlying
charge conservation is spontaneously broken (because of many-body
effects such as the interaction of the electrons with lattice vibrations
favouring the formation of so-called Cooper pairs). In the
Hartree--Fock--Bogoliubov mean-field approximation, one accounts for the
spontaneous breaking of \(\mathrm U(1)\) symmetry by augmenting the
free-fermion Hamiltonian \(\sh H\) with so-called \emph{pairing} terms
\(\Delta_+\) and \(\Delta_-\) that do \emph{not} conserve the charge.
Rather, they satisfy \[
    [\sh Q,\Delta_\pm]=\pm2\Delta_\pm,
\] and thus increase or decrease the charge by 2 units.

To summarise and simplify the notation, let us record here that any
free-fermion Hamiltonian \(\sh H\) (superconducting or not) acts on the
Nambu space \(\sh W\) of fields \(\psi\) as \[
    H\psi=i^{-1}d\rho(i\sh H)\psi=[\sh H,\psi].
\] Now, the ground state of a gapped free-fermion Hamiltonian \(\sh H\)
or, equivalently, of the induced Hamiltonian \(H\), is known to be
uniquely determined by the flat-band (or flattened) Hamiltonian \[
    J\defi-i\sign(H)
\] where a factor \(i= \sqrt{-1}\) was inserted to make \(J\) real
(\(\overline J=J\)). Note that \(\sign(H)=H\abs{H}^{-1}\) is
well-defined by the hypothesis of a non-vanishing energy gap. We then
infer the relation \(J^2=-1\), which makes \(J\) a complex structure of
the real vector space \(\sh W_\reals\). Recalling that \(H\) is skew
under transposition, we see that the same property holds for \(J\)
(\emph{i.e.} \(J^\intercal=-J\)). It then follows from
\(J^\intercal J=- J^2 =1\) that \(J\) preserves the CAR form: \[
    \Braces{J\psi,J\psi'}=\Braces{\psi,\psi'},
\] and is thus an orthogonal transformation. In summary,
\(J=-i\sign(H)\) is a CAR-preserving complex structure of the space of
Majorana fields \(\sh W_\reals\).

Let us briefly explain why the full information about the free-fermion
ground state is encoded precisely in the complex structure \(J\). The
operator \(J\) has two eigenspaces,
\(\ker(J\mp i)\subseteq\sh W_\reals\otimes_\reals\cplxs=\sh W\). These
eigenspaces provide a \emph{polarisation}: \[
    \sh W=\sh A^\mathrm c\oplus\sh A,\quad\sh A^\mathrm c=\ker(J+i),\quad\sh A=\ker(J-i),
\] and they determine \(J\). Because \(J\) preserves the CAR form, the
subspaces \(\sh A\) and \(\sh A^\mathrm c\) are isotropic: \[
    \Braces{\sh A,\sh A}=0=\Braces{\sh A^\mathrm c,\sh A^\mathrm c}.
\] For \(a\in\sh A\), \(c\in\sh A^\mathrm c\), and
\(\sh J=d\rho^{-1}(J)\), one has \[
    \Bracks{i\!{\sh J},a}=iJa=-a,\quad\Bracks{i\!{\sh J},c}=iJc=+c.
\] Thus the field operators in \(\sh A\) lower the energy of the
flattened Hamiltonian \(i\sh J\), while those in \(\sh A^\mathrm c\)
increase it. Clearly, the ground state as the state of lowest energy
must be annihilated by the field operators that lower the energy. It
follows that the isotropic subspace \(\sh A\) furnishes all the
(quasi-particle) annihilation operators. These, by a basic result of
many-body theory, precisely characterise the free-fermion ground state.
We mention in passing that the polarisation provided by the bare vacuum,
\[
    \sh W=\sh V\oplus\sh V^\ast,\quad\sh V=\sh V_+\oplus\sh V_-,
\] would be a poor starting point on which to build an effective theory
describing the low-energy physics that takes place near the chemical
potential.

Altogether, the following definition is now well-motivated.

\BeginKnitrBlock{definition}
\protect\hypertarget{def:qpv}{}{\label{def:qpv} }A free-fermion ground state
or \emph{quasi-particle vacuum} (QPV) is a complex structure \(J\) that
preserves the CAR form: \[
    J=-J^{-1}=-J^\intercal\in\blop(\sh W_\reals).
\] Here, \(\blop(-)\) denotes the set of bounded linear operators.
\EndKnitrBlock{definition}

\BeginKnitrBlock{remark}
{}For translation-invariant systems, \(J\) is
diagonal in the momentum \(k\). In that case, the orthogonality
condition \(J^\intercal=J^{-1}\) refines to
\(J(-k)^\intercal=J(k)^{-1}\). The latter is equivalent to a condition
on the spaces of quasi-particle annihilation operators:
\(\Braces{\sh A(k),\sh A(-k)}=0\) which is the \emph{Fermi constraint}
of Kennedy--Zirnbauer \autocite{rk-mrz}.
\EndKnitrBlock{remark}

\hypertarget{par:tight}{%
\subsubsection{\texorpdfstring{Hilbert space structure of
\(\sh W\)}{Hilbert space structure of \textbackslash sh W}}\label{par:tight}}

Looking at the bigger picture one realises that, up to the present
point, a similar development could have been carried out for bosons
instead of fermions. In fact, a free-boson ground state is precisely a
CCR-preserving complex structure of the real space of fields
\(\sh W_\reals\) viewed as a symplectic vector space (where CCR stands
for \textbf{c}anonical \textbf{c}ommutation \textbf{r}elations). In the
next step, however, we come to a structure specific to fermions in the
free-fermion limit: \(\sh W\) is a complex Hilbert space, as follows.

Recall that the CAR form \(\braces{\cdot,\cdot}\) on \(\sh W\) is
symmetric and complex bilinear, and restricts to a Euclidean structure
of \(\sh W_\reals\). For \(\psi\in\sh W\), let \[
    \psi=\psi_R+i\psi_I,\quad\psi_R\defi\tfrac12(\psi+\gamma\psi),\quad\psi_I\defi\tfrac1{2i}(\psi-\gamma\psi)
\] be the decomposition into real and imaginary parts. Then \[
    \Sdual{\psi}{\psi'}_\sh W\defi\Braces{\psi_R-i\psi_I,\psi_R'+i\psi_I'}
\] is an inner product turning \(\sh W\) into a Hermitian vector space.
In other words:

\BeginKnitrBlock{definition}[Inner product on Nambu space]
\protect\hypertarget{def:unnamed-chunk-2}{}{\label{def:unnamed-chunk-2}
\iffalse (Inner product on Nambu space) \fi{} }The Nambu space \(\sh W\)
of fermion field operators is equipped with an inner product
\(\sdual\cdot\cdot_\sh W\) by \begin{equation}
    \Sdual\psi{\psi'}_\sh W=\Braces{\gamma\psi,\psi'}\quad(\psi,\psi'\in\sh W).
    \label{eq:real-car}
\end{equation} Thus, \(\sh W\) acquires the structure of a complex
Hilbert space.
\EndKnitrBlock{definition}

\BeginKnitrBlock{remark}
{}The Hilbert space structure on \(\sh W\)
depends on our free-fermion assumption in two respects:

\begin{enumerate}
\def\labelenumi{(\roman{enumi})}
\item
  The same construction for bosons leads to an inner product with
  indefinite signature, and for this reason \emph{not} to a Hilbert
  space.
\item
  The invariance of \(\sdual\psi\psi_\sh W\sge 0\) under unitary time
  evolutions can be interpreted as a law of probability conservation for
  quasi-particles. That law is violated by electron-electron
  interactions, so \(\sdual\cdot\cdot_\sh W\) ceases to be relevant for
  interacting systems.
\end{enumerate}
\EndKnitrBlock{remark}

The presence of the inner product entails an operation
\(\endo{\sh W}\ni L\longmapsto L^*\) of taking the Hermitian adjoint. By
construction, this is the same as the operation of transpose combined
with that of \(\gamma\)-conjugate: \[
    L^\ast=\Parens{\overline{L}}^\intercal .
\] Now, recall from Definition \ref{def:qpv} the notion of a
free-fermion ground state or quasi-particle vacuum (QPV). Because any
QPV \(J\in\blop(\sh W_\reals)\) is skew, its complex linear extension
\(J\in\blop(\sh W)\) is skew-Hermitian (\(J^\ast=- J\)). It follows that
\(J\), being a complex structure \(J=-J^{-1}\) of \(\sh W_\reals\),
extends to a unitary operator \(J=(J^{-1})^\ast\) on \(\sh W\).

\BeginKnitrBlock{proposition}
\protect\hypertarget{prp:unnamed-chunk-4}{}{\label{prp:unnamed-chunk-4}
}Viewed as an operator on the complex Hilbert space \(\sh W\), any
quasi-particle vacuum \(J\) is skew-Hermitian and unitary: \[
    J^\ast=-J=J^{-1},
\] as well as real: \(\overline J=J\).
\EndKnitrBlock{proposition}

As we have stressed above, the Hilbert space structure, unlike the CAR
form and the real structure, is not a fundamental feature of the Nambu
space \(\sh W\). However, as long as we limit our attention to the
free-fermion case, it is permissible mathematically to upend the logic
and put the Hilbert space structure first. This is convenient for the
purely pragmatic purpose of connecting us to the mathematical
literature. Accordingly, we pose the following definition.

\BeginKnitrBlock{definition}[Real Hilbert spaces]
\protect\hypertarget{def:real-hilbert}{}{\label{def:real-hilbert}
\iffalse (Real Hilbert spaces) \fi{} }Let \(\sh W\) be a complex Hilbert
space with inner product \(\sdual\cdot\cdot\). A \emph{real structure}
on \(\sh W\) is an anti-unitary involution \(\gamma\). That is,
\(\gamma^2 = 1\) and \[ 
  \Sdual{\gamma\psi}{\gamma\psi'}=\Sdual{\psi'}{\psi}\quad(\psi,\psi'\in\sh W).
\] A \emph{real Hilbert space} is by definition a pair
\((\sh W,\gamma)\) of a Hilbert space \(\sh W\) and a real structure
\(\gamma\). A complex linear (or anti-linear) map
\(\phi:(\sh W_1,\gamma_1)\longrightarrow(\sh W_2,\gamma_2)\) is called
\emph{real} if it intertwines the real structures, \emph{i.e.} \[
  \phi\circ\gamma_1=\gamma_2\circ\phi.
\]
\EndKnitrBlock{definition}

In the framework of this definition, the Nambu space \(\sh W\), equipped
with its natural real structure, is a real Hilbert space, and any QPV
\(J\) is a real skew-Hermitian unitary.

\hypertarget{par:tba}{%
\subsubsection{Tight-binding approximation}\label{par:tba}}

A major benefit from developing the mathematical model from its proper
foundations in many-body quantum mechanics, is that we can now introduce
a \emph{tight-binding approximation} in a mathematically justified and
physically meaningful way. (This will be important for our treatment of
weak topological invariants; \emph{cf.} Section \ref{sec:discussion}. In
fact, without proper theory building, the weak invariants might be
criticised as artefacts due to the illegal step of replacing the spatial
continuum by a discrete approximation.)

Let us recall that for a many-fermion system at very low temperatures,
the active degrees of freedom reside near the chemical potential
\(\mu\). More precisely, the low-energy excitations are single particles
in the conduction bands just above \(\mu\) and single holes in the
valence bands just below \(\mu\). Hence, to capture the topological
properties of the ground state (and other low-temperature phenomena),
one may prune the locally infinite number of freedoms in the full
single-particle Hilbert space, by assuming a truncation to the
low-energy sector spanned by the conduction and valence bands near
\(\mu\). In this vein, we now make the following choice for our Hilbert
spaces \(\sh V_+\) and \(\sh V_-\): \[
    \sh V_\tau=\ell^2(\Lambda) \otimes V_\tau \quad (\tau=\pm)
\] with lattice \(\Lambda\cong\ints^d\). The factor \(\ell^2(\Lambda)\)
is the Hilbert space of a so-called \emph{tight-binding model}, where
the lattice \(\Lambda\cong\mathbb Z^d\) reflects the organisation of the
crystal by atomic sites. (In fact, the atomic sites may form a more
complex pattern or unit cell invariant under the translational action of
the \(\ints^d\) lattice; in this paper, we focus on systems homogeneous
w.r.t. translations, ignoring any further spatial symmetries.) The
finite-dimensional factors \(V_\pm\) account for the spin and orbital
degrees of freedom active in the conduction bands (\(V_+\)) and the
valence bands (\(V_-\)) near the chemical potential. The resulting
mathematical model is summarised as follows.

The Nambu space of field operators is \[
    \sh W=\sh V_+\oplus\sh V_-^*\oplus(\sh V_+\oplus\sh V_-^*)^*=\sh V\oplus\sh V^*,\quad\sh V\defi\sh V_+\oplus\sh V_-.
\] Here, we may identify \[
    \sh V\cong\ell^2(\Lambda)\otimes V,\quad V\defi V_+\oplus V_-.
\] By the procedure outlined above, \(\sh W\) acquires the structure of
a real Hilbert space. Using the inner product and the (complex)
conjugation on \(\ell^2(\Lambda)\), we may identify \[
    \sh W\cong\ell^2(\Lambda)\otimes W,\quad W\defi V\oplus V^*.
\]

The splitting \(\sh V\oplus\sh V^\ast\) of \(\sh W\) is characterised by
the (de-quantised) charge operator \[
    Q\defi 
    \begin{pmatrix}
        1&0\\0&-1
    \end{pmatrix}:\sh W=\sh V\oplus\sh V^*\longrightarrow\sh W,
\] resulting \emph{via} \(Q=i^{-1}d\rho(i\sh Q)\) from the
second-quantised charge operator \(\sh Q\) introduced earlier. Observe
that an arbitrary operator on \(\sh W\) leaves \(\sh V\) and
\(\sh V^\ast\) invariant if and only if it commutes with \(Q\). Since
\(iQ\) generates the \(\mathrm U(1)\) symmetry group underlying the
conservation of charge, we call such operators \emph{charge-conserving}.
Charge-conserving operators with the additional property of being real
take the general form \[
    O=
    \begin{pmatrix}
        O|_\sh V&0\\0&h(O|_\sh V)h^{-1}
    \end{pmatrix}
\] where \(h\equiv\gamma|_{\sh V}:\sh V\longrightarrow\sh V^*\) is the
Fréchet--Riesz isomorphism of the Hilbert space \(\sh V\). Thus, any
such operator is determined by its restriction to \(\sh V\).

\hypertarget{par:real-cst}{%
\subsubsection{\texorpdfstring{Real
C\(^*\)-algebras}{Real C\^{}*-algebras}}\label{par:real-cst}}

As derived above, a key feature of our mathematical model is the
presence of a real structure. Using a convenient mathematical short cut,
the definition of a free-fermion ground state or QPV \(J\) can be
formulated thus: \[
    \overline J=J,\quad J^\ast=-J,\quad J^2=-1;
\] that is, by reference only to the operations of forming the
\(\gamma\)-conjugate, the Hermitian adjoint, and the product of
operators.

Our aim, in the remainder of the current subsection, is to formulate an
algebraic framework for the study of free-fermion topological phases in
the presence of disorder. This will enable us to bring the
well-developed theory of operator \(K\)-theory to bear upon the problem
of bulk and boundary classification of such phases.

Defined algebraically as a (real) skew-Hermitian unitary, the
free-fermion QPV J resides in the algebra \(\blop(\sh W)\) of bounded
linear operators on \(\sh W\). By a classical theorem of Gel'fand and
Naimark (see \autocite[Corollary 3.7.5]{pedersen}), the closed
subalgebras of the algebra \(\blop(\sh H)\) of bounded operators on a
(complex) Hilbert space \(\sh H\) invariant under the formation of
Hermitian adjoints can be defined algebraically without reference to the
Hilbert space. The algebraic objects so defined are exactly the
so-called C\(^\ast\)-algebras.

In addition to the Hermitian adjoint, \emph{real C\(^\ast\)-algebras}
formalise the notion of a real structure; they are to real Hilbert
spaces what C\(^\ast\)-algebras are to Hilbert spaces. Abstractly, they
are defined as follows.

\BeginKnitrBlock{definition}[Real C$^*$-algebras]
\protect\hypertarget{def:unnamed-chunk-5}{}{\label{def:unnamed-chunk-5}
\iffalse (Real C\(^*\)-algebras) \fi{} }A \emph{real C\(^*\)-algebra} is
a pair \((A,\overline\cdot)\) consisting of a (complex) C\(^*\)-algebra
\(A\) and an anti-linear \(*\)-involution \(\overline\cdot\) called
\emph{conjugation}; \emph{viz.} \[
    \overline{xy}=\overline x\,\overline y,\quad\overline{x^*}=(\overline x)^*,\quad\overline{\lambda x+y}=\overline\lambda\overline x+\overline y\quad(x,y\in A,\lambda\in\cplxs).
\] An element \(x\in A\) is called \emph{real} if \(\overline x=x\) and
\emph{imaginary} if \(\overline x=-x\). A \(*\)-morphism
\(\phi:(A_1,\overline\cdot)\longrightarrow(A_2,\overline\cdot)\) of real
C\(^*\)-algebras is called \emph{real} if it intertwines the
conjugations, \emph{i.e.} \[
    \phi(\overline x)=\overline{\phi(x)}\quad(x\in A_1).
\]

The spatial tensor product \(A_1\otimes A_2\) of C\(^*\)-algebras, where
\((A_1,\overline\cdot)\) and \((A_2,\overline\cdot)\) are real
C\(^*\)-algebras, becomes a real C\(^*\)-algebra when equipped with the
conjugation defined by \[
    \overline{x_1\otimes x_2}\defi\overline{x_1}\otimes\overline{x_2}\quad(x_1\in A_1,x_2\in A_2).
\]
\EndKnitrBlock{definition}

\BeginKnitrBlock{remark}
{}Equivalently, a real C\(^*\)-algebra is
characterised as a C\(^*\)-algebra \(A\), together with a linear
anti-involution \(\mathsf T\) called \emph{transposition}. That is,
\(\mathsf T\) is complex linear and \[
    (xy)^\intercal=y^\intercal x^\intercal,\quad (x^*)^\intercal=(x^\intercal)^*,\quad (x^\intercal)^\intercal=x\quad(x,y\in A).
\] The two equivalent structures are related by the equation \[
    x^\intercal=\overline x^*\quad(x\in A).
\]
\EndKnitrBlock{remark}

In our setting based on the Nambu space of fields, \(\sh W\), the basic
example of a real C\(^*\)-algebra is a closed \(*\)-subalgebra
\(A\subseteq\blop(\sh W)\) that is invariant under conjugation (or
equivalently, under transposition with respect to the CAR form). Within
any such algebra, we will be able to define the notion of a QPV.

From an abstract point of view, all real C\(^*\)-algebras are realised
on a real Hilbert space, as the following remark implies.

\BeginKnitrBlock{remark}
{}For any real C\(^*\)-algebra
\((A,\overline\cdot)\), the real subalgebra \(B=A_\reals\) consisting of
all real elements is a real Banach \(*\)-algebra such that for every
\(x\in B\), \(\norm {x^*x}=\norm{x}^2\) and \(1+x^*x\) is invertible in
the unitisation of \(B\). Building on work of Arens \autocite{arens} and
Arens--Kaplansky \autocite{ak} in the commutative case, Ingelstam
\autocite{ingelstam} proved that conversely, if \(B\) is a real Banach
\(*\)-algebra satisfying these hypotheses, then its complexification
\(A=B_\cplxs\) is a real C\(^*\)-algebra for the conjugation given by \[
    \overline{x+iy}\defi x-iy\quad(x,y\in B).
\] Moreover, he proved that every real C\(^*\)-algebra
\((A,\overline\cdot)\) admits, for some real Hilbert space
\((\sh W,\gamma)\), a real isometric \(*\)-isomorphism onto a closed
subalgebra of \(\blop(\sh W)\) which is invariant under the adjoint and
the transpose.
\EndKnitrBlock{remark}

Although real C\(^\ast\)-algebras can thus always be realised on a real
Hilbert space, in practice, they also arise through quaternionic rather
than through real structures. This will be important for our treatment
of time reversal.

\BeginKnitrBlock{definition}[Quaternionic structures on Hilbert space]
\protect\hypertarget{def:quat-str}{}{\label{def:quat-str}
\iffalse (Quaternionic structures on Hilbert space) \fi{} }Let \(\sh V\)
be a Hilbert space (over the complex numbers). An anti-unitary operator
\(T\) on \(\sh V\) is called a \emph{quaternionic structure} if
\(T^2=-1\). That is, \(T\) is anti-linear and we have \[
    \Sdual{Tv_1}{Tv_2}=\Sdual{v_2}{v_1},\quad\Sdual{Tv_1}{v_2}=-\Sdual{Tv_2}{v_1}\quad(v_1,v_2\in\sh V).
\] A pair \((\sh V,T)\) consisting of a Hilbert space \(\sh V\) and a
quaternionic structure \(T\) is called a \emph{quaternionic Hilbert
space}. A complex linear map
\(\phi:(\sh W_1,T_1)\longrightarrow(\sh W_2,T_2)\) between quaternionic
Hilbert spaces is called \emph{quaternionic} if it intertwines the
quaternionic structures, \emph{viz.} \[
    \phi\circ T_1=T_2\circ\phi.
\]
\EndKnitrBlock{definition}

Let \((\sh V,T)\) be a quaternionic Hilbert space. Then \(\blop(\sh V)\)
is a \emph{real} C\(^*\)-algebra with the conjugation \[
    \overline A\defi T^*AT=-TAT\quad(A\in\blop(\sh V)). 
\] In the context of our applications, a quaternionic structure on
\(\sh V=\ell^2(\Lambda)\otimes V\) arises as \(\mathsf c\otimes T\),
where \(\mathsf c\) is complex conjugation on \(\ell^2(\Lambda)\) and
\(T\) is a quaternionic structure on \(V\), corresponding to the
operation of time reversal.

An example of a quaternionic structure is given for \(V=\cplxs^2\) by
\begin{equation}
    \ger c\defi
    \begin{pmatrix}
        0&\mathsf c\\-\mathsf c&0
    \end{pmatrix},
    \label{eq:quat-quat}
\end{equation} where \(\mathsf c\) is complex conjugation on \(\cplxs\).
This induces on \(M_2(\cplxs)=\endo{\cplxs^2}\) the conjugation
\(\overline\cdot=\Ad(\ger c)\), expressed explicitly by the formula \[
    \overline{
        \begin{pmatrix}
            a&b\\ c&d
        \end{pmatrix}
    }=
    \begin{pmatrix}
        \overline d&-\overline c\\-\overline b&\overline a
    \end{pmatrix}\quad(a,b,c,d\in\cplxs).
\] For the real C\(^\ast\)-algebra \((M_2(\cplxs),\Ad(\ger c))\), the
subalgebra of real elements is spanned (over the real numbers) by the
identity matrix and the matrices \(i\sigma_x\), \(i\sigma_y\),
\(i\sigma_z\), where \begin{equation}
    \sigma_x\defi
    \begin{pmatrix}
        0&1\\1&0
    \end{pmatrix},\quad
    \sigma_y\defi
    \begin{pmatrix}
        0&-i\\i&0
    \end{pmatrix},\quad
    \sigma_z\defi
    \begin{pmatrix}
        1&0\\0&-1
    \end{pmatrix}
    \label{eq:pauli}
\end{equation} are the usual Pauli matrices. This subalgebra is thus
isomorphic (as an algebra over the real numbers) to the quaternions
\(\mathbb H\). For this reason, we will denote the real C\(^*\)-algebra
\((M_2(\cplxs),\Ad(\ger c))\) by \(\mathbb H_\cplxs\).

\hypertarget{local-observables-as-controlled-operators}{%
\subsubsection{Local observables as controlled
operators}\label{local-observables-as-controlled-operators}}

Besides being Hermitian and preserving the CAR form, the Hamiltonians
\(H\) relevant to our considerations share a further common feature.
Namely, they are \emph{local} operators. By this we mean that
\((H\psi)(x)\) depends only on the values \(\psi(y)\) for \(y\) in a
neighbourhood of \(x\) whose diameter is bounded independently of \(x\).
Mathematically, this is captured by the notion of \emph{controlled
operators} defined below.

\BeginKnitrBlock{definition}[Controlled operators]
\protect\hypertarget{def:unnamed-chunk-8}{}{\label{def:unnamed-chunk-8}
\iffalse (Controlled operators) \fi{} }Let \(U\) be a finite-dimensional
Hilbert space, \(D\subseteq\Lambda\) any subset and \(O\) a bounded
linear operator on \(\ell^2(D)\otimes U\). Let \(O(x,y)\in\endo{W}\),
for \(x,y\in D\), denote the kernel function of \(O\), defined by \[
    O(x,y)\defi\Sdual x{Oy}_{\ell^2(D)}.
\] We say that \(O\) is \emph{controlled} if there is a finite constant
\(R>0\) such that \[
    O(x,y)=0\quad(\norm{x-y}>R).
\] The norm closure of the set of controlled operators shall be denoted
by \(C^*_u(D,U)\).
\EndKnitrBlock{definition}

\BeginKnitrBlock{remark}
{}The previous definition is due to Roe
\autocite[Definition 4.22]{roe}, in a much more general form. The
subscript \(u\) in \(C^*_u(D,U)\) refers to the word \enquote{uniform}.
Indeed, \(C^*_u(D,U)\cong C^*_u(D,\cplxs)\otimes \endo U\) where
\(C^*_u(D,\cplxs)\) is usually denoted by \(C^*_u(D)\) and called the
\emph{uniform Roe algebra} (of the metric space \(D\)) in the
literature. In the context of the topological classification of
solid-state systems in the tight-binding approximation, the uniform Roe
algebra was first used by Kubota \autocite{Kubota}.
\EndKnitrBlock{remark}

The following statement is immediate, see \autocite[Corollary 4.24]{roe}
for the first part.

\BeginKnitrBlock{proposition}
\protect\hypertarget{prp:unnamed-chunk-10}{}{\label{prp:unnamed-chunk-10}
}Let \(U\) be a finite-dimensional Hilbert space and
\(D\subseteq\Lambda\) a subset. Then \(C^*_u(D,U)\) is a closed
\(*\)-subalgebra of \(\blop(\ell^2(D)\otimes U)\) and hence a
C\(^*\)-algebra. If \(U\) has a real or a quaternionic structure, then
\(C^*_u(D,U)\) is invariant under the induced conjugation on
\(\blop\Parens{\ell^2(D)\otimes U}\), and thus, a real C\(^*\)-algebra.
\EndKnitrBlock{proposition}

\BeginKnitrBlock{proof}
{}Only the statement about real structure
requires proof. But \[
    \overline O(x,y)=\overline{O(x,y)}\quad(x,y\in D),
\] so that controlled operators are invariant under conjugation. Since
the latter is an isometry, the assertion follows.
\EndKnitrBlock{proof}

\hypertarget{the-algebra-of-bulk-observables}{%
\subsubsection{The algebra of bulk
observables}\label{the-algebra-of-bulk-observables}}

We now come to the final step of our construction of the bulk algebra.
Topological insulators and superconductors occur in solids with a
periodic, translation-invariant crystal structure in the bulk. This is
already reflected, in the tight-binding approximation, by the choice of
the lattice \(\Lambda\) of spatial translations leaving the atomic sites
invariant. As long as no disorder is present, the smallest
C\(^*\)-algebra containing all free-fermion Hamiltonians of interest on
\(\sh W\) is the algebra of \(\Lambda\)-invariants in
\(C^*_u(\Lambda,W)\). In this case, it would be fair to declare it to be
the algebra of bulk observables. (Incidentally, the Fourier transform
identifies \(\smash{C^*_u(\Lambda,W)^\Lambda}\) with the set of
continuous matrix functions \(H(k)\) of momenta \(k\) on the Brillouin
zone \(\reals^d/\Lambda\).)

However, we do not wish to ignore the effects of disorder. Indeed, as we
shall show, bulk-boundary correspondence holds in the presence of
disorder, both for the strong and the so-called weak invariants. Since
disorder breaks the translational invariance at a microscopic level, a
major step in building our model is to devise a mathematical counterpart
for the macroscopic homogeneity that is nonetheless preserved.

A general way to do this was suggested by Bellissard \autocite{b89}.
Starting from some basic Hamiltonian, he considers all
\(\Lambda\)-translates thereof and the closure \(\Omega\) of this
translational orbit in the space of operators (for a suitable topology).
If the Hamiltonian is macroscopically homogeneous, then, together with
its translates, it should \autocite{bn90} be an operator-valued
\emph{function} on \(\Omega\) translationally \emph{covariant} (or
\emph{equivariant}) in that it intertwines the action of the translation
group on \(\Omega\) and the space of operators.

As there is no way of knowing \emph{a priori} how to model the space
\(\Omega\) of disorder configurations, it is common practice to replace
this orbit closure by an arbitrary compact Hausdorff space \(\Omega\),
equipped with a continuous action of the group \(\Lambda\). For now, we
will leave the choice of \(\Omega\) open. We shall discuss a specific
choice later, in Subsection \ref{subs:disorder}.

\BeginKnitrBlock{definition}[Covariance algebra]
\protect\hypertarget{def:cov-alg}{}{\label{def:cov-alg} {} }Let \(\Omega\) be a compact Hausdorff topological space,
equipped with a right action of the group \(\Lambda\). Let \(U\) be a
finite-dimensional Hilbert space. Let \(u_x\), for \(x\in\Lambda\), be
defined by \[
    (u_x\psi)(y)\defi\psi(y-x)\quad(x,y\in\Lambda,\psi\in\ell^2(\Lambda)\otimes U).
\] We define \(A^U\) to be the set of all maps
\(O:\Omega\longrightarrow C^*_u(\Lambda, U)\) that are continuous in the
norm topology and \emph{covariant} (or equivariant) in the sense that
\begin{equation}
    O_{\omega\cdot x}=u_x^*O_\omega u_x\quad(x\in \Lambda,\omega\in\Omega).
    \label{eq:cov-cond}
\end{equation} In particular, referring to the notation of Paragraph
\ref{par:tba}, we set \(\mathbb A\defi A^W\) and \(\mathrm A\defi A^V\).
We sometimes speak of the former as the real case and the latter as the
complex case.
\EndKnitrBlock{definition}

\BeginKnitrBlock{lemma}
\protect\hypertarget{lem:unnamed-chunk-12}{}{\label{lem:unnamed-chunk-12}
}Retain the assumptions of Definition \ref{def:cov-alg}. With point-wise
operations \[
    (O_1+\lambda O_2)_\omega\defi (O_1)_\omega+\lambda(O_2)_\omega,\ (O^*)_\omega\defi (O_\omega)^*\quad(O,O_1,O_2\in A^U,\lambda\in\cplxs,\omega\in\Omega),
\] and the norm defined by \[
    \norm{O}\defi\sup_{\omega\in\Omega}\norm{O_\omega}\quad(O\in A^U),
\] the set \(A^U\) is a C\(^*\)-algebra. If \(U\) is real or
quaternionic, then \(A^U\) is a real C\(^*\)-algebra with conjugation
and transpose defined by \[
    (\overline O)_\omega\defi\overline{O_\omega},\quad(O^\intercal)_\omega\defi(O_\omega)^\intercal\quad(O\in A^{U},\omega\in\Omega).
\]
\EndKnitrBlock{lemma}

\BeginKnitrBlock{remark}[Magnetic fields]
{}The first one to study the
algebra from Definition \ref{def:cov-alg} in connection with solid-state
physics was Bellissard, in the context of the quantum Hall effect. (His
definition is phrased somewhat differently, using crossed products of
C\(^*\)-algebras; we shall show in Paragraph \ref{par:crossed} that this
is equivalent to our definition.) An important feature of the quantum
Hall effect considered by Bellissard is the presence of a homogeneous
external magnetic field. In building a tight-binding approximation for
such a system, one imposes a lattice spacing on the continuum of
positions in such a way that exactly one unit of magnetic flux passes
through every plaquette. This leads to a quantised magnetic co-cycle
\(\sigma(x,y)=\smash{e^{i\rscp x{\mathbf By}/2}}\in\mathrm U(1)\) with
\(B_{\mu\nu}\in\reals\).

Mathematically, there is no obstruction to incorporating such a co-cycle
into the definition of the algebra \(\mathrm A\) in our present context.
The situation is different for the case of the real algebra
\(\mathbb A\): One needs to replace \(\sigma\) by the real co-cycle \[
    \Sigma(x,y)\defi 
    \begin{pmatrix}
    e^{\frac i2\rscp x{\mathbf By}}&0\\0&e^{-\frac i2\rscp x{\mathbf By}}
    \end{pmatrix}.
\] In addition, this needs to be central, so one has to constrain the
units \(B_{\mu\nu}\) of magnetic flux per plaquette to integer multiples
of \(2\pi\), trivialising the co-cycle. Thus, it is not possible to
introduce a homogeneous magnetic field in the case of real symmetries.

This is not a serious limitation: Although magnetic fields can occur in
the real symmetry classes \(D\) (corresponding to gapped superconductors
or superfluids with no symmetries) and \(C\) (corresponding to such with
fermions of spin \(\tfrac12\) and \(\mathrm{SU}(2)\) spin-rotation
symmetry), they are far from homogeneous; in fact, in the so-called
vortex phase they admit magnetic fields as some pattern of magnetic flux
tubes.

Furthermore, a quantised magnetic co-cycle, even if it can be treated
mathematically, would only be of minor relevance for the physics of the
systems we consider. Indeed, in our setting, the lattice \(\Lambda\)
reflects the periodicity of the pattern of atomic sites. It is therefore
fixed by the crystalline structure of the solid; a homogeneous magnetic
field quantised perfectly in alignment with the crystalline lattice
could only be achieved by fine-tuning. Moreover, the absence of a
magnetic field should not affect the system's topological features in an
essential fashion. For instance, as was discovered by Haldane
\autocite{haldane-qahe}, the Hall conductance in the (anomalous) QHE may
still be quantised even when the total magnetic flux per unit plaquette
vanishes. See also \autocite[Corollary 5.7.2]{ps16}.
\EndKnitrBlock{remark}

Referring back to our previous discussion, the space \(\Omega\) in
Definition \ref{def:cov-alg} and the following lemma is the space of
\emph{disorder configurations}. The covariance condition in Equation
\eqref{eq:cov-cond} encodes the notion that the bulk observables of
interest are macroscopically homogeneous. We shall therefore call the
real C\(^*\)-algebra \(\mathbb A\) the \emph{algebra of bulk
observables} (or \emph{bulk algebra} for short). We call the
C\(^*\)-algebra \(\mathrm A\) the \emph{algebra of charge-conserving
bulk observables}. Note that if \(V\) is equipped with a quaternionic
structure \(T\), then \(\mathbb{A}^\sim\defi A^V\) is a real
C\(^*\)-algebra.

\BeginKnitrBlock{definition}[Disordered IQPV]
\protect\hypertarget{def:unnamed-chunk-14}{}{\label{def:unnamed-chunk-14}
\iffalse (Disordered IQPV) \fi{} }Let \(\Omega\) be a compact Hausdorff
topological space, equipped with a right action of the lattice
\(\Lambda\). Then a \emph{disordered macroscopically invariant
quasi-particle vacuum} (or \emph{disordered IQPV}) is a real unitary
\(J\in\mathbb A\) such that \(J^2=-1\).

In the complex case, a \emph{disordered charge-conserving IQPV} is
defined to be a unitary \(J\in\mathrm A=A^{V}\) such that \(J^2=-1\).
\EndKnitrBlock{definition}

The discussion at the end of Paragraph \ref{par:tight} shows that a
disordered charge-conserving IQPV is the same thing as a disordered IQPV
\(J\) such that \([Q,J]=0\).

\hypertarget{par:crossed}{%
\subsubsection{The bulk algebra as a crossed
product}\label{par:crossed}}

The definition of the bulk algebra \(\mathbb A\) \emph{via} controlled
operators given above is physically well-motivated. To derive, in
Section \ref{sec:bdy}, the sequence of C\(^\ast\)-algebras that lies at
the core of our derivation of the bulk-boundary correspondence, we shall
need an alternative description of \(\mathbb A\) by crossed products. In
the following, group elements \(g \in G\) are denoted as \(\tau_g\) when
they are to be re-interpreted as generators of a C\(^\ast\)-algebra.

\BeginKnitrBlock{definition}[Crossed product C$^*$-algebras]
\protect\hypertarget{def:unnamed-chunk-15}{}{\label{def:unnamed-chunk-15}
\iffalse (Crossed product C\(^*\)-algebras) \fi{} }Let \(G\) be a group
and \(A\) a C\(^*\)-algebra. An \emph{action} of \(G\) on \(A\) is a
group homomorphism \(\alpha:G\longrightarrow\auto A\); that is,
\(\alpha_g\) is a \(*\)-automorphism of \(A\) for every \(g\in G\), and
\[
    \alpha_g\circ\alpha_{g'}=\alpha_{gg'}\quad(g,g'\in G).
\]

The \emph{crossed product C\(^*\)-algebra} \(A\rtimes_\alpha G\) is, by
definition, the C\(^*\)-algebra defined by generators \(a\in A\) and
\(\tau_g\), for all \(g\in G\), and the following relations: the
C\(^*\)-algebraic relations of \(A\), together with \[
    \tau_g\tau_{g'}=\tau_{gg'},\quad \tau_g^*=\tau_{g^{-1}},\quad \tau_ga=\alpha_g(a)\tau_g\quad(g,g'\in G,a\in A).
\]

A tuple \((\pi,U)\) consisting of a non-degenerate \(*\)-representation
\(\pi:A\longrightarrow\blop(\sh H)\) and a unitary representation
\(U:G\longrightarrow\mathrm U(\sh H)\) on the same Hilbert space
\(\sh H\) is called a \emph{covariant pair} if \[
    \pi(\alpha_g(a))=U_g\pi(a)U_g^*\quad(g\in G,a\in A).
\] By definition, the C\(^*\)-algebra \(A\rtimes_\alpha G\) is universal
for covariant pairs. That is, any covariant pair \((\pi,U)\) determines
a unique \(*\)-representation
\(\pi\rtimes_\alpha U:A\rtimes_\alpha G\longrightarrow\blop(\sh H)\)
such that \[
    (\pi\rtimes_\alpha U)(a)=\pi(a),\quad(\pi\rtimes_\alpha U)(\tau_g)=U_g\quad(a\in A,g\in G).
\]
\EndKnitrBlock{definition}

\BeginKnitrBlock{remark}
{}C\(^*\)-algebras defined by generators and
relations are introduced and thoroughly discussed by Blackadar
\autocite{blackadar-shape}.

Twisting co-cycles can be included in the definition of the crossed
product, see \autocite{busby-smith}. The first two relations have to be
modified to \[
    \tau_g\tau_{g'}=\sigma(g,g')\tau_{gg'},\quad \tau_g^*=\tau_{g^{-1}}.
\] By \autocite[Theorem 3.3]{busby-smith}, the C\(^*\)-algebra defined
by the above relations is the enveloping C\(^*\)-algebra of the Banach
\(*\)-algebra \(L^1(A,G;\alpha,\sigma)\) introduced by Busby--Smith.
\EndKnitrBlock{remark}

Before we construct examples of group actions relevant to our
applications, let us see how to incorporate real structures.

\BeginKnitrBlock{definition}[Real structures on crossed products]
\protect\hypertarget{def:unnamed-chunk-17}{}{\label{def:unnamed-chunk-17}
\iffalse (Real structures on crossed products) \fi{} }Let \(G\) be a
group and \(\alpha\) an action on \(A\), where \((A,\overline\cdot)\) is
a real C\(^*\)-algebra. The action is \emph{real} if for every
\(g\in G\), \(\alpha_g\) is a real \(*\)-automorphism. In this case, we
may extend the conjugation \(\overline\cdot\) to one on
\(A\rtimes_\alpha G\) by setting \(\overline{\tau_g}\defi\tau_g\) for
all \(g\in G\). A covariant pair \((\pi,U)\) on the Hilbert space
\(\sh H\) is called \emph{real} if \(\sh H\) is a real Hilbert space,
\(\pi\) is real, and \(U_g\) is real for every \(g\in G\). The
\(*\)-representation \(\pi\rtimes_\alpha U\) induced by a real covariant
pair is real.
\EndKnitrBlock{definition}

As before, let \(\Omega\) be a compact Hausdorff space, equipped with a
right action of \(\Lambda\). We define a real action \(\alpha\) of
\(\Lambda\) on \(\sh C(\Omega)\otimes\endo W\) by \begin{equation}
    \alpha_x(f)(\omega)\defi f(\omega\cdot x)\quad(x\in\Lambda,f\in\sh C(\Omega)\otimes\endo W,\omega\in\Omega).
    \label{eq:alpha-def}
\end{equation} Here, \(\sh C(\Omega)\) is the C\(^*\)-algebra of
continuous functions on \(\Omega\), equipped with complex conjugation.
The same formula defines an action \(\alpha\) of \(\Lambda\) on
\(\sh C(\Omega)\otimes\endo V\).

\BeginKnitrBlock{theorem}
\protect\hypertarget{thm:covcrossed}{}{\label{thm:covcrossed} }The crossed
product C\(^*\)-algebra
\((\sh C(\Omega)\otimes\endo W)\rtimes_\alpha\Lambda\) is isomorphic to
\(\mathbb A\) as a real C\(^*\)-algebra. Similarly,
\(\mathrm A\cong(\sh C(\Omega)\otimes\endo V)\rtimes_\alpha\Lambda\) as
C\(^*\)-algebras.
\EndKnitrBlock{theorem}

\BeginKnitrBlock{remark}
{}Although our definition of the algebras
\(\mathbb A\) and \(\mathrm A\) is equivalent to a crossed product, we
believe that it is more natural to introduce these algebras in the
fashion we have chosen. We are especially motivated to do so by our
quest to develop the mathematical model from its foundations in
many-body quantum mechanics, thus providing it with a sound physical
meaning. Another benefit of this approach is that it can be extended to
include a group \(G\) of spatial symmetries. In this case, \(\Lambda\)
is the full set of atomic sites, invariant under the space group \(G\).
In this case, the resulting algebra is a corner of the crossed product
\((\sh C(\Omega)\otimes\endo W)\rtimes G\). This corner is Morita
equivalent to the crossed product itself under the assumptions that at
every point of \(\Lambda\), the on-site representations of the isotropy
group of the \(G\)-action contain all irreducible representations.

The statement of Theorem \ref{thm:covcrossed} is well-known to experts
and is implicit in the literature, see for example \autocite[Proposition
5]{b86}, although full proofs are not readily available. Therefore, and
because we need to refer back to it below, we give a detailed proof.
\EndKnitrBlock{remark}

\BeginKnitrBlock{proof}[of Theorem \ref{thm:covcrossed}]
{}We prove
only the statement for \(\mathbb A\); the other case is almost literally
the same. Abbreviate \(\tilde A\defi\sh C(\Omega)\otimes\endo W\). We
may equip \(\Omega\) with a fully supported Borel probability measure
\(\mathbb P\) (say), so that \(\mathbb A\) is realised in
\(\blop(L^2(\Omega,\mathbb P)\otimes\sh W)\). Define \[
    \left.
    \begin{aligned}
        \Parens{\pi(f)\psi}(\omega,y)&\defi f(\omega\cdot (-y))\psi(\omega,y)\\
        (R_x\psi)(\omega,y)&\defi\psi(\omega,y+x)
    \end{aligned}
    \right\}\quad(\psi\in L^2(\Omega,\mathbb P)\otimes\sh W,\omega\in\Omega,x,y\in\Lambda).
\] Then by \autocite[Section 7.7.1]{pedersen}, \((\pi,R)\) is a
covariant pair that is manifestly real. It thus defines a real
\(*\)-representation \(\Phi\defi\pi\rtimes_\alpha R\) of
\(\tilde A\rtimes_\alpha\Lambda\). Since \(\Lambda\) is amenable,
\autocite[Theorems 7.7.5 and 7.7.7]{pedersen} shows that \(\Phi\) is
faithful.

We claim that the image of \(\Phi\) lies in \(\mathbb A\). By the
definitions, \(\Phi(\tau_x)_\omega\) and \(\Phi(f)_\omega\) are
controlled for any \(x\in\Lambda\), \(f\in\sh C(\Omega)\otimes\endo W\),
and \(\omega\in\Omega\). Hence, \(\Phi(a)_\omega\in C^*_u(\Lambda,W)\)
for any \(a\in\tilde A\rtimes_\alpha\Lambda\). To check the covariance
condition, we compute \[
    u_g^*\Phi(\tau_x)_\omega u_g=u_{-g}u_{-x}u_g=u_{-x}=R_x=\Phi(\tau_x)_{\omega\cdot g}\quad(x,g\in\Lambda,\omega\in\Omega).
\] Similarly, we have \[
    (u_g)^*\Phi(f)_\omega u_g=\sum_{x\in\Lambda}f(\omega\cdot x)\ket{x-g}\bra{x-g}=\Phi(f)_{\omega\cdot g}\quad(f\in\tilde A,g\in\Lambda,\omega\in\Omega).
\] This shows that \(\im\Phi\subseteq\mathbb A\). We have to prove the
equality.

Let \(c_0(\Lambda)\) denote the C\(^*\)-algebra of null sequences
indexed by \(\Lambda\). Because \(\Lambda\) is an amenable group, there
is an approximate unit \((\phi_n)_{n\in\nats}\) for \(c_0(\Lambda)\)
consisting of finitely supported functions of positive type
\autocite[Lemma 11.19]{roe}. We may suppose that \(\phi_n(0)=1\). For
every \(n\), there is a finite-dimensional Hilbert space \(\sh H_n\), a
unitary representation \(\pi_n\) of \(\Lambda\) on \(\sh H_n\), and a
unit vector \(\xi_n\in\sh H_n\) such that
\(\phi_n(x)=\Sdual{\xi_n}{\pi_n(x)\xi_n}\) for \(x\in\Lambda\).

Let \((\xi_n^i)\) be an orthonormal basis of \(\sh H_n\). Set \[
    f_n^i(x)\defi\Sdual{\xi_n^i}{\pi_n(x)\xi_n}\quad(x\in\Lambda),
\] so that \(f_n^i\) is bounded on \(\Lambda\). For
\(T\in\blop(\sh W)\), we define \[
    S_n(T)\defi\sum_iM_{f_n^i}^*TM_{f_n^i}\in\blop(\sh W)
\] where \(M_f\) is multiplication by \(f\). Observe that \[
    S_n(T)=\sum_{x,y}\sum_i\Sdual{\pi_n(x)\xi_n}{\xi_n^i}\Sdual{\xi_n^i}{\pi_n(y)\xi_n}T(x,y)\ket x\bra y=\sum_{x,y}\phi_n(y-x)T(x,y)\ket x\bra y.
\] In particular, \(S_n(1)=1\), as \(\phi_n(0)=1\). It is clear that
\(S_n\) is completely positive, \emph{cf.} \autocite[Lemma 11.17]{roe},
and in particular positive. Hence \(\norm{S_n}\sle\norm{S_n(1)}=1\),
that is, \(S_n\) is a contraction. Since \(S_n\) leaves controlled
operators invariant, it leaves \(C^*_u(\Lambda,W)\) invariant.

Let \(T\in\mathbb A\). We set \(S_n(T)_\omega\defi S_n(T_\omega)\).
Since \(S_n\) commutes with \(u_x\) for any \(x\), this defines
\(S_n(T)\in\mathbb A\). Let \(\eps>0\) be arbitrary. Because \(\Omega\)
is compact, there exist a finite \(J\), points \(\omega_j\in\Omega\),
\(j\in J\), and open neighbourhoods \(U_j\subseteq\Omega\) of
\(\omega_j\) covering \(\Omega\) such that \[
    \Norm{T_\omega-T_{\omega_j}}\sle\frac\eps6\quad(\omega\in U_j,j\in J).
\] There exist a constant \(R>0\) and \(R\)-controlled operators
\(T_j\in C^*_u(\Lambda,W)\) such that \[
    \Norm{T_{\omega_j}-T_j}\sle\frac\eps6\quad(j\in J).
\] By \autocite[Lemma 4.27]{roe}, there is a constant \(C_R>0\)
depending only on \(R\) such that \[
    \Norm{T_j-S_n(T_j)}\sle C_R\cdot\sup\nolimits_{\norm x\sle R}\Abs{1-\phi_n(x)}\cdot\sup\nolimits_{\norm{x-y}\sle R}\norm{T_j(x,y)}\quad(n\in\nats,j\in J).
\] Since \(\phi_n\) converges to \(1\) uniformly on
\(\{\norm\cdot\sle R\}\), there is an \(N\in\nats\) such that for all
\(n\sge N\) and all \(j\in J\), the right-hand side is
\(\sle\frac\eps3\).

If now \(n\sge N\) and \(\omega\in\Omega\), then there is some \(j\)
such that \(\omega\in U_j\). It follows that \[
    \begin{split}
        \Norm{T_\omega-S_n(T_\omega)}&\sle
        \begin{aligned}[t]
            &\Norm{T_\omega-T_{\omega_j}}+\Norm{T_{\omega_j}-T_i}+\Norm{T_j-S_n(T_j)}\\
            &+\Norm{S_n(T_j-T_{\omega_j})}+\Norm{S_n(T_{\omega_j}-T_\omega)}
        \end{aligned}\\
        &\sle2\Norm{T_\omega-T_{\omega_j}}+2\Norm{T_{\omega_j}-T_i}+\Norm{T_j-S_n(T_j)}\sle\frac{4\eps}6+\frac\eps3=\eps,
    \end{split}
\] so \(S_n(T)\) converges to \(T\) in \(\mathbb A\). Since \(\im\Phi\)
is closed, it suffices to show that \(S_n(T)\in\im\Phi\).

For any \(n\in\nats\) and \(y\in{\supp}(\phi_n)\), define
\(f_{n,y}\in\sh C(\Omega)\otimes\endo W\) by \[
    f_{n,y}(\omega)\defi T_\omega(0,y)\quad(\omega\in\Omega).
\] Then we compute \[
    \begin{split}
        \im\Phi\ni\sum_{\phi_n(y)\neq0}\phi_n(y)\pi(f_{n,y})R_{-y}&=\sum_{\phi_n(y)\neq0,x\in\Lambda}\phi_n(y)T_{\omega\cdot x}(0,y)\ket x\bra x R_{-y}\\
        &=\sum_{\phi_n(y-x)\neq0}\phi_n(y-x)T_\omega(x,y)\ket x\bra y=S_n(T).
    \end{split}
\] Thus, \(\Phi\) is surjective. As we already know that \(\Phi\) is
injective, this proves the theorem.
\EndKnitrBlock{proof}

\hypertarget{sec:bulkinv}{%
\section{Symmetries, pseudo-symmetries, and bulk
invariants}\label{sec:bulkinv}}

\hypertarget{subs:symm}{%
\subsection{Quasi-particle vacua with symmetries}\label{subs:symm}}

Our mission is to study symmetry-protected topological (or SPT) phases
of disordered free fermions. While such phases constitute a rather large
class, a manageable and especially interesting subclass is the one given
by the classification scheme of the \emph{Tenfold Way} \autocite{hhz}.
In that scheme, one starts from the Nambu space of fields equipped with
the CAR form and at most two distinguished anti-unitary operations,
namely time-reversal and/or particle-hole symmetry. The setting also
allows for an arbitrary group \(G_0\) of unitary symmetries. The
classification statement of the Tenfold Way then is that, upon
projecting Nambu space to any isotypical component of irreducible
\(G_0\)-representations, one invariably ends up with a reduced space
that is of one of 10 types known as the AZ (or CAZ) classes
\autocite{az97}. The early work \autocite{kitaev,srfl} on free-fermion
SPT phases focused on precisely those 10 types or symmetry classes.

In trying to set up a concise description of all Tenfold-Way symmetry
classes in a unified \(\mathrm{C}^\ast\)-algebraic framework, one faces
the complication that several real structures co-exist. Indeed, there is
the Majorana or \(\gamma\)-real structure and one may also have the
anti-linear operations of time-reversal and/or particle-hole symmetry in
play (see also the discussion in Paragraph \ref{diii-iqpv} and
following). How can one then avoid a cumbersome case-by-case treatment?
It was Kitaev \autocite{kitaev} who first put forth the good scheme to
use: adopting the universal real structure offered by the real space of
Majorana fields, he implements the anti-linear symmetries as part of a
Clifford algebra of complex linear (!) generators. That scheme was
worked out in systematic detail by Kennedy--Zirnbauer \autocite{rk-mrz},
and our first step here is to adapt their Bloch-bundle picture to the
\(\mathrm{C}^\ast\)-algebraic setting for systems with disorder.

\hypertarget{symm}{%
\subsubsection{On-site symmetries}\label{symm}}

Our bare algebraic framework is enriched by symmetry operations that act
on the Nambu space \(\sh W=\ell^2(\Lambda)\otimes W\) or rather, as
on-site symmetries, on \(W\). These operations are linear or anti-linear
endomorphisms of \(W\) that, if present, constrain our Hamiltonians.
With the exception of particle-hole symmetry, all of them are initially
defined on the single-particle Hilbert space
\(\sh V=\ell^2(\Lambda)\otimes V\), or actually, on the on-site Hilbert
space \(V\). Given that origin, they all turn into operators on \(W\) by
a canonical procedure: second quantisation takes an operator on
\(\sh V\) to an operator on the Fock space
\(\bigwedge(\sh V_+\oplus \sh V_-^\ast)\). The latter then determines a
(\enquote{de-quantised}) operator on \(\sh W\) by the commutator or
adjoint action, \emph{cf.} Section \ref{subs:nambu-bulk}.

The first symmetry we wish to consider is \emph{time reversal}, \(T\).
On electrons (or any other fundamental fermions), this is an
anti-unitary operator such that \(T^2=-1\). Put differently, time
reversal is simply a \emph{quaternionic structure} on \(V\), see
Definition \ref{def:quat-str}. The afore-mentioned functor of second
quantisation followed by de-quantisation extends \(T\) to a real
anti-linear endomorphism of \(W\) by \(T|_{V^*} \defi hTh^{-1}\).

The next set of symmetries that occur on \(V\) are related to the
\emph{spin rotation generators} \(S_1,S_2,S_3\). For electrons (or any
other fundamental fermions) carrying spin \(\tfrac12\) these satisfy the
relations \[
    S_1S_2=-S_2S_1=iS_3,\quad S_\mu^2=1\quad(\mu=1,2,3).
\] They anti-commute with \(T\) (physically speaking, this is because
the operation of inverting the time direction makes any motion, also
that of spin, run backwards). The operators representing spin rotations
are obtained by exponentiation:
\(g=\mathrm{exp} (i\sum\omega_\mu S_\mu)\in\mathrm{SU}(2)=\mathrm{Spin}(3)\),
with real parameters \(\omega_\mu\). To generate \(\mathrm{SU}(2)\) over
the real numbers, we set \(j_\mu\defi iS_\mu\), so that \[
    j_1j_2=-j_2j_1=-j_3,\quad j_\mu^2=-1\quad(\mu=1,2,3).
\] Overriding physics conventions, we still refer to the \(j_\mu\) as
spin rotation generators. Our quantisation functor yields
\(\smash{j_\mu|_{V^*}\defi hj_\mu h^{-1}}\). Note that the \(j_\mu\)
commute with \(T\).

There are two remaining symmetries in our picture. The first of them is
\(iQ\), which we met earlier as the charge operator
\(Q= i^{-1}d\rho(i\sh Q)\). The \(\gamma\)-real operator \(iQ\)
generates the \(\mathrm U(1)\) symmetry group underlying the law of
charge conservation. Recall our above observation, in Paragraph
\ref{par:tight}, that an operator on \(W\) commutes with \(Q\) if and
only if it leaves \(V\) and \(V^\ast\) invariant. It follows that \(T\)
anti-commutes with \(iQ\), and that the \(j_\mu\), \(\mu=1,2,3\),
commute with \(iQ\).

Finally, we consider the \emph{particle-hole symmetry} \(C\) on \(W\).
Recall from Subsection \ref{subs:nambu-bulk} that the Fock space
\(\bigwedge\) of the many-particle system was built from the Hilbert
space \(\sh V_+\oplus\sh V_-^\ast\), where
\(\sh V_+=\ell^2(\Lambda)\otimes V_+\) and
\(\sh V_-=\ell^2(\Lambda)\otimes V_-\) are the subspaces harbouring the
conduction and the valence states, respectively. A special situation
that may occur is that the conduction and valence spaces are in
bijection \emph{via} an operator \[
    S:\sh V_+\oplus\sh V_-\longrightarrow\sh V_-\oplus\sh V_+
\] which exchanges the conduction and valence states and anti-commutes
with the first-quantised Hamiltonian \(\mathsf H\). In principle, \(S\)
may be linear or anti-linear. For example, for Dirac fermions one knows
the operation of \emph{charge conjugation}
\(\psi\longmapsto i\gamma^2\overline\psi\) (using the standard physics
notation for \(\gamma\)-matrices, not to be confused with our choice of
real structure). Charge conjugation is complex anti-linear in first
quantisation (and, perhaps confusingly, complex linear after second
quantisation) and anti-commutes with the Dirac Hamiltonian in the
absence of electromagnetic fields. Yet, in the condensed matter physics
of non-relativistic electrons, there is no such thing as the
relativistic charge conjugation operator. However, what may exist in
that setting is a sub-lattice (or \enquote{chiral}) operation
\(S:\psi_A+\psi_B\longmapsto\psi_A-\psi_B\), which is complex linear.

In keeping with our use of the tight-binding approximation, we assume
the operator \(S\) to be realised on \(V=V_+\oplus V_-\), so that
\(S\in\mathrm U(V)\), \(S^2=1\), \([S,T]=0\), and \([S,j_\mu]=0\),
\(\mu=1,2,3\). As before, \(S\) extends to a real linear endomorphism of
\(W\) by \(\smash{S|_{V^*}}\defi hSh^{-1}\). In particular, \([S,Q]=0\)
for this extension. The \emph{particle-hole symmetry} \(C\) is then
defined by \(C\defi\gamma S\); indeed, if the free-fermion Hamiltonian
\(H=-\gamma H\gamma\) anti-commutes with the sub-lattice operation
\(S\), then \(C=\gamma S\) commutes with \(H\) and is thus a true
physical symmetry. (Moreover, the many-particle Fock vacuum
\(\smash{\bigwedge^0(\sh V_+ \otimes \sh V_-^*)}\) is particle-hole
symmetric.)

It should be stressed here that our notion of particle-hole symmetry
continues to make sense in the realm of interacting systems beyond the
free-fermion approximation. As a matter of fact, the anti-linear
operator \(C : \sh W\to\sh W\) lifts to a real or quaternionic structure
on the many-particle Fock space \(\bigwedge\), and the latter operation
induces a real structure on the algebra of bounded many-body operators.
An important example of an interacting Hamiltonian with particle-hole
symmetry is that of the Hubbard model.

\hypertarget{symmetry-classes}{%
\subsubsection{Symmetry classes}\label{symmetry-classes}}

We shall say that a Hermitian and \(\gamma\)-imaginary Hamiltonian
\(H\in\blop(\sh W)\) is in the \emph{(real) symmetry class} \(s\) (where
\(s=0,\dotsc,7\)) if it commutes with the symmetries indicated in Table
\ref{tab:symm-class}. There, the symbol \(\Uc\) stands for the
\(\mathrm{U}(1)\) group generated by the charge operator \(iQ\), and
\(\Usp\) denotes the \(\mathrm{SU}(2)\) group of spin rotations
generated by the \(j_\mu\). We say that \(H\) is in the \emph{complex
symmetry class} \(s\) (where \(s=0,1\)) if it commutes with the
symmetries indicated in Table \ref{tab:complex-symm-class}.

Note that if the Hamiltonian preserves charge, \emph{i.e.} commutes with
\(Q\), then we may reduce it to its first-quantised form on \(\sh V\).

\begin{longtable}[]{@{}lllll@{}}
\caption{\label{tab:symm-class} Symmetry classes}\tabularnewline
\toprule
\begin{minipage}[b]{0.06\columnwidth}\raggedright
\(s\)\strut
\end{minipage} & \begin{minipage}[b]{0.11\columnwidth}\raggedright
class\strut
\end{minipage} & \begin{minipage}[b]{0.32\columnwidth}\raggedright
symmetry group\strut
\end{minipage} & \begin{minipage}[b]{0.16\columnwidth}\raggedright
generators\strut
\end{minipage} & \begin{minipage}[b]{0.20\columnwidth}\raggedright
comments\strut
\end{minipage}\tabularnewline
\midrule
\endfirsthead
\toprule
\begin{minipage}[b]{0.06\columnwidth}\raggedright
\(s\)\strut
\end{minipage} & \begin{minipage}[b]{0.11\columnwidth}\raggedright
class\strut
\end{minipage} & \begin{minipage}[b]{0.32\columnwidth}\raggedright
symmetry group\strut
\end{minipage} & \begin{minipage}[b]{0.16\columnwidth}\raggedright
generators\strut
\end{minipage} & \begin{minipage}[b]{0.20\columnwidth}\raggedright
comments\strut
\end{minipage}\tabularnewline
\midrule
\endhead
\begin{minipage}[t]{0.06\columnwidth}\raggedright
\(0\)\strut
\end{minipage} & \begin{minipage}[t]{0.11\columnwidth}\raggedright
\(D\)\strut
\end{minipage} & \begin{minipage}[t]{0.32\columnwidth}\raggedright
trivial\strut
\end{minipage} & \begin{minipage}[t]{0.16\columnwidth}\raggedright
-\strut
\end{minipage} & \begin{minipage}[t]{0.20\columnwidth}\raggedright
\strut
\end{minipage}\tabularnewline
\begin{minipage}[t]{0.06\columnwidth}\raggedright
\(1\)\strut
\end{minipage} & \begin{minipage}[t]{0.11\columnwidth}\raggedright
\(\diii\)\strut
\end{minipage} & \begin{minipage}[t]{0.32\columnwidth}\raggedright
\(\trs\)\strut
\end{minipage} & \begin{minipage}[t]{0.16\columnwidth}\raggedright
\(T\)\strut
\end{minipage} & \begin{minipage}[t]{0.20\columnwidth}\raggedright
time reversal\strut
\end{minipage}\tabularnewline
\begin{minipage}[t]{0.06\columnwidth}\raggedright
\(2\)\strut
\end{minipage} & \begin{minipage}[t]{0.11\columnwidth}\raggedright
\(\aii\)\strut
\end{minipage} & \begin{minipage}[t]{0.32\columnwidth}\raggedright
\(\trs\ltimes\Uc\)\strut
\end{minipage} & \begin{minipage}[t]{0.16\columnwidth}\raggedright
\(T,iQ\)\strut
\end{minipage} & \begin{minipage}[t]{0.20\columnwidth}\raggedright
charge\strut
\end{minipage}\tabularnewline
\begin{minipage}[t]{0.06\columnwidth}\raggedright
\(3\)\strut
\end{minipage} & \begin{minipage}[t]{0.11\columnwidth}\raggedright
\(\cii\)\strut
\end{minipage} & \begin{minipage}[t]{0.32\columnwidth}\raggedright
\(\trs\ltimes\Uc\times\phs\).\strut
\end{minipage} & \begin{minipage}[t]{0.16\columnwidth}\raggedright
\(T,iQ,C\)\strut
\end{minipage} & \begin{minipage}[t]{0.20\columnwidth}\raggedright
particle-hole\strut
\end{minipage}\tabularnewline
\begin{minipage}[t]{0.06\columnwidth}\raggedright
\(4\)\strut
\end{minipage} & \begin{minipage}[t]{0.11\columnwidth}\raggedright
\(C\)\strut
\end{minipage} & \begin{minipage}[t]{0.32\columnwidth}\raggedright
\(\Usp\)\strut
\end{minipage} & \begin{minipage}[t]{0.16\columnwidth}\raggedright
\(j_\mu\)\strut
\end{minipage} & \begin{minipage}[t]{0.20\columnwidth}\raggedright
spin\strut
\end{minipage}\tabularnewline
\begin{minipage}[t]{0.06\columnwidth}\raggedright
\(5\)\strut
\end{minipage} & \begin{minipage}[t]{0.11\columnwidth}\raggedright
\(C\mathrm I\)\strut
\end{minipage} & \begin{minipage}[t]{0.32\columnwidth}\raggedright
\(\Usp\times\trs\)\strut
\end{minipage} & \begin{minipage}[t]{0.16\columnwidth}\raggedright
\(j_\mu,T\)\strut
\end{minipage} & \begin{minipage}[t]{0.20\columnwidth}\raggedright
\strut
\end{minipage}\tabularnewline
\begin{minipage}[t]{0.06\columnwidth}\raggedright
\(6\)\strut
\end{minipage} & \begin{minipage}[t]{0.11\columnwidth}\raggedright
\(A\mathrm I\)\strut
\end{minipage} & \begin{minipage}[t]{0.32\columnwidth}\raggedright
\(\Usp\times(\trs\ltimes\Uc)\)\strut
\end{minipage} & \begin{minipage}[t]{0.16\columnwidth}\raggedright
\(j_\mu,T,iQ\)\strut
\end{minipage} & \begin{minipage}[t]{0.20\columnwidth}\raggedright
\strut
\end{minipage}\tabularnewline
\begin{minipage}[t]{0.06\columnwidth}\raggedright
\(7\)\strut
\end{minipage} & \begin{minipage}[t]{0.11\columnwidth}\raggedright
\(BD\mathrm I\)\strut
\end{minipage} & \begin{minipage}[t]{0.32\columnwidth}\raggedright
\(\Usp\times\trs\ltimes\Uc\times\phs\)\strut
\end{minipage} & \begin{minipage}[t]{0.16\columnwidth}\raggedright
\(j_\mu,T,iQ,C\)\strut
\end{minipage} & \begin{minipage}[t]{0.20\columnwidth}\raggedright
\strut
\end{minipage}\tabularnewline
\bottomrule
\end{longtable}

\begin{longtable}[]{@{}lllll@{}}
\caption{\label{tab:complex-symm-class} Complex symmetry
classes}\tabularnewline
\toprule
\begin{minipage}[b]{0.08\columnwidth}\raggedright
\(s\)\strut
\end{minipage} & \begin{minipage}[b]{0.09\columnwidth}\raggedright
class\strut
\end{minipage} & \begin{minipage}[b]{0.20\columnwidth}\raggedright
symmetry group\strut
\end{minipage} & \begin{minipage}[b]{0.21\columnwidth}\raggedright
generators\strut
\end{minipage} & \begin{minipage}[b]{0.28\columnwidth}\raggedright
comments\strut
\end{minipage}\tabularnewline
\midrule
\endfirsthead
\toprule
\begin{minipage}[b]{0.08\columnwidth}\raggedright
\(s\)\strut
\end{minipage} & \begin{minipage}[b]{0.09\columnwidth}\raggedright
class\strut
\end{minipage} & \begin{minipage}[b]{0.20\columnwidth}\raggedright
symmetry group\strut
\end{minipage} & \begin{minipage}[b]{0.21\columnwidth}\raggedright
generators\strut
\end{minipage} & \begin{minipage}[b]{0.28\columnwidth}\raggedright
comments\strut
\end{minipage}\tabularnewline
\midrule
\endhead
\begin{minipage}[t]{0.08\columnwidth}\raggedright
\(0\)\strut
\end{minipage} & \begin{minipage}[t]{0.09\columnwidth}\raggedright
\(A\)\strut
\end{minipage} & \begin{minipage}[t]{0.20\columnwidth}\raggedright
\(\Uc\)\strut
\end{minipage} & \begin{minipage}[t]{0.21\columnwidth}\raggedright
\(iQ\)\strut
\end{minipage} & \begin{minipage}[t]{0.28\columnwidth}\raggedright
charge\strut
\end{minipage}\tabularnewline
\begin{minipage}[t]{0.08\columnwidth}\raggedright
\(1\)\strut
\end{minipage} & \begin{minipage}[t]{0.09\columnwidth}\raggedright
\(\aiii\)\strut
\end{minipage} & \begin{minipage}[t]{0.20\columnwidth}\raggedright
\(\Uc\times\phs\)\strut
\end{minipage} & \begin{minipage}[t]{0.21\columnwidth}\raggedright
\(iQ,C\)\strut
\end{minipage} & \begin{minipage}[t]{0.28\columnwidth}\raggedright
particle-hole\strut
\end{minipage}\tabularnewline
\bottomrule
\end{longtable}

Although our symmetry-class tables look different from those widely
cited and used in the literature \autocite{kitaev,srfl}, they lead to
mathematically equivalent results. What we follow here is the assessment
of Kennedy--Zirnbauer \autocite{rk-mrz} that, at the level of setting
the foundations, all declarations of symmetry should conform to the
principle that true physical symmetries \emph{commute} with the
Hamiltonian (never do they anti-commute). One benefit from adhering to
that strict notion of symmetry is that our framework directly extends to
systems with electron-electron interactions. Indeed, all our symmetries
\(X\) make immediate sense at the interacting many-particle level
(actually, as we have indicated, they \emph{descend} from their
second-quantised ancestors \(\sh X\) by the transfer to Nambu space),
and they constrain the symmetry-allowed Hamiltonians by
\(\sh X\sh H=+\sh H\sh X\) (never by \(\sh X\sh H=-\sh H\sh X\)). The
relation of the present scheme to that of
Schnyder--Ryu--Furusaki--Ludwig \autocite{srfl} is reviewed in the
appendix to this section.

\BeginKnitrBlock{remark}
{}The terminology for symmetry classes was
introduced by Altland--Zirnbauer \autocite{az97} in their work on
disordered free fermions in mesoscopic systems, especially
superconductors. The labels (\(D\), \(D\mathrm{I\!I\!I}\), \emph{etc.})
used in that context refer to the Cartan--Killing classification of
Riemannian symmetric spaces. Let us emphasise that we use the (C)AZ
terminology precisely as intended by its pioneers, namely for the case
\emph{with disorder}, and please be warned that much of the recent
literature has taken to (ab-)using that terminology for
translation-invariant systems without disorder.
\EndKnitrBlock{remark}

\hypertarget{pseudo-symmetries}{%
\subsubsection{Pseudo-symmetries}\label{pseudo-symmetries}}

Some of the physical symmetries we consider are linear, while others are
anti-linear. Given a symmetry \(g\) that commutes with the gapped
Hamiltonian \(H\), whether \(g\) commutes or anti-commutes with the
corresponding IQPV \(J=-i\mathrm{sign}(H)\) depends on whether \(g\) is
linear or anti-linear. Since we are interested primarily in ground
states rather than Hamiltonians, our description of symmetry classes of
IQPV would not be smooth if done in terms of the original symmetries.
This situation is resolved by the concept of \emph{pseudo-symmetries}
advertised by Kennedy--Zirnbauer \autocite{rk-mrz}.

Following Kitaev \autocite{kitaev} they organise the symmetry classes of
translationally invariant free-fermion ground states by assigning, to
each of the sets of symmetries in Table \ref{tab:symm-class}, which
commute with \(H\), a set of \(s\) operators \(J_1,\dotsc,J_s\), called
\emph{pseudo-symmetries}, which \emph{anti-commute} with the IQPV \(J\).
Here, \(s\) denotes the numerical label in the table. Besides
anti-commuting with \(J\), the pseudo-symmetries define a real
\(\ast\)-representation of a Clifford algebra. To make this precise, we
recall the following standard definition.

\BeginKnitrBlock{definition}[Clifford algebras]
\protect\hypertarget{def:cliff}{}{\label{def:cliff} {} }Let \(\mathbb C\ell_{r,s}\) be the universal
C\(^*\)-algebra with unitary generators
\(k_1,\dotsc,k_r,j_1,\dotsc,j_s\) and relations \[
    \left.
    \begin{aligned}
        &k_ak_b+k_bk_a=2\delta_{ab}\\
        &j_\alpha j_\beta+j_\beta j_\alpha=-2\delta_{\alpha\beta}\\ 
        &k_aj_\alpha+j_\alpha k_a=0
    \end{aligned}\right\}\quad(a,b=1,\dotsc,r,\alpha,\beta=1,\dotsc,s).
\] This is the familiar \emph{Clifford algebra} with \(r\) positive and
\(s\) negative generators.

We define an anti-linear involution \(\overline\cdot\) on
\(\mathbb C\ell_{r,s}\) by declaring all of the generators
\(k_a,j_\alpha\) to be fixed by it. The resulting real C\(^*\)-algebra
\((\mathbb C\ell_{r,s},\overline\cdot)\) will be denoted by
\(C\ell_{r,s}\).
\EndKnitrBlock{definition}

\BeginKnitrBlock{remark}
{}With the terminology introduced in Definition
\ref{def:cliff}, any collection of real unitaries
\(K_1,\dotsc,K_r,J_1,\dotsc,J_s\) on \(W\) such that \begin{equation}
    \left.
    \begin{aligned}
        &K_aK_b+K_bK_a=2\delta_{ab}\\
        &J_\alpha J_\beta+J_\beta J_\alpha=-2\delta_{\alpha\beta}\\ 
        &K_aJ_\alpha+J_\alpha K_a=0
    \end{aligned}\right\}\quad(a,b=1,\dotsc,r,\alpha,\beta=1,\dotsc,s).
    \label{eq:rs-pseudo}
\end{equation} determines a unique real \(*\)-morphism
\(C\ell_{r,s}\longrightarrow\endo W\) \emph{via} \(k_a\longmapsto K_a\)
and \(j_a\longmapsto J_\alpha\), and \emph{vice versa}. Similarly, a set
of unitaries on \(V\) satisfying the relations in Equation
\eqref{eq:rs-pseudo} amount to the same data as a \(*\)-morphism
\(\mathbb C\ell_{r,s}\longrightarrow\endo V\).
\EndKnitrBlock{remark}

Let us now review the Kennedy--Zirnbauer construction briefly. When
\(0\sle s\sle 3\), we may define a set \(J_1,\dotsc,J_s\) of
anti-commuting real unitaries on \(W\) by \begin{equation}
    J_1\defi J_T\defi\gamma T,\quad J_2\defi J_Q\defi i\gamma QT,\quad J_3\defi J_C\defi i\gamma QC.
    \label{eq:j03}
\end{equation} When \(s\sge 4\), we amplify \(W\) to
\(W\otimes\cplxs^2\), giving \(\cplxs^2\) its canonical real structure
\(\mathbb{R}^2 \subset \cplxs^2\), and we define \begin{align}
    J_\mu&\defi j_\mu\otimes\sigma_z\quad(\mu=1,\dotsc,3),
    \label{eq:jspin}\\
    J_4&\defi 1\otimes i\sigma_y,\quad J_5\defi J_T\otimes\sigma_x,\quad J_6\defi J_Q\otimes\sigma_x,\quad J_7\defi J_C\otimes\sigma_x.
    \label{eq:j4plus}
\end{align} In particular, this redefines \(J_1,\dotsc, J_3\). We obtain
a set \(J_1,\dotsc,J_s\) of \(s\) anti-commuting real unitaries on
\(W\otimes\cplxs^2\).

For \(0\sle s\sle 3\), \(H\) commutes with the corresponding symmetries
in Table \ref{tab:symm-class} if and only if \(J\) anti-commutes with
\(J_1,\dotsc,J_s\). For \(s\sge4\), \(H\) commutes with the
corresponding symmetries if and only if \(J\otimes\sigma_x\)
anti-commutes with \(J_1,\dotsc,J_s\). Therefore, and motivated by
\autocite[Definition 2.4]{rk-mrz}, we pose the following definition.

\BeginKnitrBlock{definition}[Disordered IQPV with symmetries]
\protect\hypertarget{def:unnamed-chunk-3}{}{\label{def:unnamed-chunk-3}
\iffalse (Disordered IQPV with symmetries) \fi{} }A \emph{disordered
IQPV of symmetry index} \((r,s)\) (or simply \emph{of index} \((r,s)\))
is a tuple \((J;K_1,\dotsc,K_r,J_1,\dotsc,J_s)\) where \(J\in\mathbb A\)
is a disordered IQPV and \((K_1,\dotsc,K_r,J_1,\dotsc,J_s)\) define a
real \(*\)-representation \(\phi:C\ell_{r,s}\longrightarrow\endo{W}\)
such that\\
\begin{equation}
    JK_a+K_aJ=JJ_\alpha+J_\alpha J=0\quad(a=1,\dotsc,r,\alpha=1,\dotsc,s).
\label{eq:iqpv-sym}
\end{equation} If \(r=0\) and \(0\sle s\sle 7\), then we also call this
a disordered IQPV of \emph{symmetry class \(s\)}. We shall also refer to
it by the Cartan--Killing label in Table \ref{tab:symm-class}. (For
instance, \enquote{\(J\) is a disordered IQPV of class \(D\).}) We call
the \(J_\alpha\) \emph{negative} and the \(K_a\) \emph{positive
pseudo-symmetries} of \(J\).

Similarly, a \emph{disordered IQPV of complex symmetry index } \((r,s)\)
consists of a disordered charge-preserving IQPV \(J\in\mathrm A\) and
\(K_a,J_\alpha\in\endo V\) defining a \(*\)-representation of
\(\cplxs\ell_{r,s}\) such that Equation \eqref{eq:iqpv-sym} holds; we call
these \(K_a,J_\alpha\) \emph{complex pseudo-symmetries}. For \(r=0\) and
\(s=0,1\), we say that \((J;J_\alpha)\) are in \emph{complex symmetry
class \(s\)}.
\EndKnitrBlock{definition}

\BeginKnitrBlock{remark}
{}In view of the correspondence between
operators \(K_a,J_\alpha\) and Clifford representations \(\phi\), we
will interchangeably write a disordered IQPV of a given (complex)
symmetry index as \((J;\phi)\) where \(\phi\) is the \(*\)-morphism
representing the (complex) pseudo-symmetries.

The discussion at the end of Paragraph \ref{par:tight} shows that a
disordered IQPV of complex symmetry index \((r,s)\) is the same as a
disordered IQPV of symmetry index \((r,s)\) with the additional property
that both \(J\) and the \(K_a,J_\alpha\) commute with \(Q\).
\EndKnitrBlock{remark}

\hypertarget{background-on-clifford-algebras}{%
\subsubsection{Background on Clifford
algebras}\label{background-on-clifford-algebras}}

In this section, we collect some background results on Clifford
algebras. These will explain several delicate but important points in
our classification scheme for IQPV with pseudo-symmetries.

One background fact is the \(\mathrm{mod}\,8\) \emph{periodicity} of
Clifford algebras (resp. the \(\mathrm{mod}\,2\) periodicity of complex
Clifford algebras); this explains why we consider \enquote{real}
symmetry classes \(s\) where \(s=0,\dotsc,7\) (resp. \enquote{complex}
symmetry classes indexed by \(s=0,1\)). Another one is the quaternion
periodicity of Clifford algebras, which explains the symmetry between
the \enquote{real} cases \(0\sle s\sle 3\) and \(4\sle s\sle 7\). Yet
another point is that up to equivalence and multiplicity, the symmetry
class labelled by \(s\) is determined by \(s\) and does not depend on
the choice of the operators \(J_\alpha\). This is related to the
classification of Clifford representations. Finally, we will review the
Clifford \((1,1)\) periodicity, which is not only at the basis of the
other periodicity results, but also fundamental for our construction of
boundary invariants. For the remaining discussion, we introduce the
notion of a \emph{grading}.

\BeginKnitrBlock{definition}[Graded real C$^*$-algebras]
\protect\hypertarget{def:unnamed-chunk-5}{}{\label{def:unnamed-chunk-5}
\iffalse (Graded real C\(^*\)-algebras) \fi{} }A \emph{grading} on a
\(*\)-algebra \(A\) is a decomposition \(A=A_\ev\oplus A_\odd\), where
\(\ints/2\ints=\{\ev,\odd\}\), into closed subspaces such that \[
    A_i\cdot A_j\subseteq A_{i+j},\ *(A_i)\subseteq A_i\quad(i,j\in\ints/2\ints).
\] When equipped with a grading, \(A\) is called a \emph{graded
\(*\)-algebra}. Elements of \(A_\ev\) are called \emph{even}, while
those of \(A_\odd\) are called \emph{odd}. Non-zero elements \(a\) in
either of these summands are called \emph{homogeneous}; in this case,
\(\abs a\) denotes the \emph{parity}, \(\ev\) if \(a\) is even and
\(\odd\) if \(a\) is odd. If \(A\) is a C\(^*\)-algebra, then we suppose
in addition that \(A_i\), \(i=\ev,\odd\), are closed; \(A\) is then
called a \emph{graded C\(^*\)-algebra}. A map of graded \(*\)-algebras
is called \emph{even} if it preserves the grading. If
\((A,\overline\cdot)\) is a real \(*\)-algebra equipped with a grading,
then it is called a \emph{graded real \(*\)-algebra} if the conjugation
is even. A graded real \(*\)-algebra that is also a graded
C\(^*\)-algebra is called a \emph{graded real C\(^*\)-algebra}.

For graded \(*\)-algebras \(A\) and \(B\), the algebraic tensor product
\(A\odot B\) is graded by \[
    (A\odot B)_k\defi\bigoplus_{i+j=k}A_i\odot B_j\quad(k\in\ints/2\ints).
\] If \(A\) and \(B\) are real, then so is \(A\odot B\). If \(A\) and
\(B\) are graded (real) C\(^*\)-algebras, then the grading on
\(A\odot B\) extends to the spatial tensor product \(A\otimes B\),
turning it into a graded (real) C\(^*\)-algebra.

A different object altogether is the \emph{graded tensor product}. If
\(A\) and \(B\) are graded \(*\)-algebras, then the algebraic tensor
product \(A\odot B\) carries a new algebra structure and involution
induced by \[
    (a\otimes b)\cdot(a'\otimes b')\defi(-1)^{\abs{a'}\abs b}aa'\otimes bb',\quad(a\otimes b)^*=(-1)^{\abs a\abs b}a^*\otimes b^*,
\] for homogeneous \(a,a'\in A\), \(b,b'\in B\), defining the
\emph{graded algebraic tensor product} \(A\mathop{\widehat{\odot}}B\),
not isomorphic to \(A\odot B\) in general. It is real if so are \(A\)
and \(B\). If \(A\) and \(B\) are graded (real) C\(^*\)-algebras, then
there is a natural C\(^*\)-norm on \(A\mathop{\widehat{\odot}}B\). By
completing \(A\mathop{\widehat{\odot}}B\) with respect to that norm, one
obtains a graded (real) C\(^*\)-algebra \(A\mathop{\widehat\otimes}B\)
\autocite[§ 14.4]{blackadar-kthyopalg} called the \emph{graded tensor
product}. We will denote the image of an elementary tensor
\(a\otimes b\) in \(A\mathop{\widehat{\otimes}}B\) by the same symbol
\(a \otimes b\).
\EndKnitrBlock{definition}

Any (real) C\(^*\)-algebra can be considered as a graded (real)
C\(^*\)-algebra with the trivial grading, in which every element is
even. In particular, we will usually consider the matrix algebras
\(M_n(\cplxs)\) as trivially graded and with the component-wise
conjugation. The real C\(^*\)-algebra \(\mathbb H_\cplxs\) will also be
considered as ungraded. On the Clifford algebra \(C\ell_{r,s}\), one
considers the grading that is determined by letting the generators
\(k_a,j_\alpha\) be odd.

We now collect some periodicity results on Clifford algebras,
\emph{cf.}~\autocite[§4]{abs}. Basic to all of the relations between
Clifford algebras explained here is the fact that \begin{equation}
    C\ell_{p+r,q+s}\cong C\ell_{p,q}\mathop{\widehat\otimes} C\ell_{r,s}
    \label{eq:cliff-tensor}
\end{equation} as graded real C\(^*\)-algebras. The following is
straightforward.

\BeginKnitrBlock{proposition}[Clifford (1,1) periodicity]
\protect\hypertarget{prp:cliff11}{}{\label{prp:cliff11} {} }There is an isomorphism of real
C\(^*\)-algebras \(C\ell_{1,1}\longrightarrow M_2(\cplxs)\), defined on
generators by \[
    k_1\longmapsto\sigma_x,\quad j_1\longmapsto-i\sigma_y.
\] If we grade \(M_2(\cplxs)\) by \[
    M_2(\cplxs)_\ev\defi\BBBraces{\begin{pmatrix}*&0\\0&*\end{pmatrix}},\quad M_2(\cplxs)_\odd\defi\BBBraces{\begin{pmatrix}0&*\\ *&0\end{pmatrix}},
\] then this isomorphism is even. In particular, we have
\emph{(1,1)-periodicity}: \[
    C\ell_{r+1,s+1}\cong C\ell_{r,s}\mathop{\widehat\otimes} M_2(\cplxs).
\]
\EndKnitrBlock{proposition}

Recall the following result \autocite[Proposition 4.2]{abs}.

\BeginKnitrBlock{proposition}[Atiyah–Bott–Shapiro]
\protect\hypertarget{prp:cliff2-shift}{}{\label{prp:cliff2-shift}
\iffalse (Atiyah--Bott--Shapiro) \fi{} }There are isomorphisms of graded
real C\(^*\)-algebras \[
    C\ell_{0,r+2}\cong C\ell_{0,2}\otimes C\ell_{r,0},\quad C\ell_{r+2,0}\cong C\ell_{2,0}\otimes C\ell_{0,r}
\] induced respectively by \[
    \begin{cases}
        j_\alpha\longmapsto j_\alpha\otimes 1, &\text{ if }\alpha=1,2,\\
        j_\alpha\longmapsto j_1j_2\otimes k_{\alpha-2}, &\text{ if }\alpha=3,\dotsc,r+2,
    \end{cases}
\] and \[
    \begin{cases}
        k_a\longmapsto k_a\otimes 1, &\text{ if }a=1,2,\\
        k_a\longmapsto k_1k_2\otimes j_{a-2}, &\text{ if }a=3,\dotsc,r+2.
    \end{cases}
\] Note that above, we consider the \emph{ungraded} tensor product.
\EndKnitrBlock{proposition}

From Equation \eqref{eq:pauli}, it is easy to see that
\(C\ell_{0,2}\cong\mathbb H_\cplxs\) as ungraded real C\(^*\)-algebras.
Similarly, there is an isomorphism \(C\ell_{2,0}\cong M_2(\cplxs)\) of
ungraded real C\(^*\)-algebras, induced by the assignments
\(k_1\longmapsto\sigma_z\) and \(k_2\longmapsto\sigma_x\). From this,
one deduces (\emph{cf.} \autocite[§4]{abs}) isomorphisms of ungraded
real C\(^*\)-algebras \[
    C\ell_{0,8}\cong C\ell_{0,2}\otimes C\ell_{0,2}\otimes C\ell_{2,0}\otimes C\ell_{2,0}\cong \mathbb H_\cplxs\otimes\mathbb H_\cplxs\otimes M_4(\cplxs).
\] Since \(\mathbb H_\cplxs\otimes\mathbb H_\cplxs\cong M_4(\cplxs)\) as
real C\(^*\)-algebras, it follows that \[
    C\ell_{0,8}\cong M_{16}(\cplxs)
\] as ungraded real C\(^*\)-algebras. A similar argument also shows
\(C\ell_{8,0}\cong M_{16}(\cplxs)\) as ungraded real C\(^*\)-algebras.
Together, these isomorphisms are the famous \(\mathrm{mod}\,8\)
periodicity of Clifford algebras \autocite[Table 1]{abs}. Disregarding
the real structure, we have \(M_2(\cplxs)=\mathbb H_\cplxs\); hence, we
obtain the isomorphism of (ungraded) C\(^*\)-algebras \[
    \cplxs\ell_{0,2}\cong\cplxs\ell_{2,0}\cong M_2(\cplxs).
\] This is the \(\mathrm{mod}\,2\) periodicity of complex Clifford
algebras. Summarising, the 8 (real) symmetry classes and 2 complex
symmetry classes of the Tenfold Way are reflected in the classification
of Clifford algebras.

From the above, we also deduce the following isomorphism: \[
    \begin{split}
        \cplxs\ell_{0,s+4}&\cong\cplxs\ell_{0,2}\otimes\cplxs\ell_{s+2,0}
        \cong\cplxs\ell_{0,2}\otimes\cplxs\ell_{2,0}\otimes\cplxs\ell_{0,s}\\
        &\cong\mathbb H_\cplxs\otimes M_2(\cplxs)\otimes\cplxs\ell_{0,s} 
        \cong M_2(\mathbb H_\cplxs)\otimes\cplxs\ell_{0,s},
    \end{split}
\] which reflects the symmetry between the symmetry classes indexed by
\(0\sle s\sle 3\) and \(4\sle s\sle 7\), respectively. Using
periodicity, it is possible to compute \(C\ell_{0,s}\) in all cases. We
reproduce the result from \autocite[Table 2]{abs}.

\begin{longtable}[]{@{}ll@{}}
\caption{\label{tab:clifford} Negative Clifford algebras}\tabularnewline
\toprule
\begin{minipage}[b]{0.10\columnwidth}\raggedright
\(s\)\strut
\end{minipage} & \begin{minipage}[b]{0.34\columnwidth}\raggedright
\(C\ell_{0,s}\)\strut
\end{minipage}\tabularnewline
\midrule
\endfirsthead
\toprule
\begin{minipage}[b]{0.10\columnwidth}\raggedright
\(s\)\strut
\end{minipage} & \begin{minipage}[b]{0.34\columnwidth}\raggedright
\(C\ell_{0,s}\)\strut
\end{minipage}\tabularnewline
\midrule
\endhead
\begin{minipage}[t]{0.10\columnwidth}\raggedright
\(0\)\strut
\end{minipage} & \begin{minipage}[t]{0.34\columnwidth}\raggedright
\(\cplxs=\reals_\cplxs\)\strut
\end{minipage}\tabularnewline
\begin{minipage}[t]{0.10\columnwidth}\raggedright
\(1\)\strut
\end{minipage} & \begin{minipage}[t]{0.34\columnwidth}\raggedright
\(\cplxs_\cplxs\)\strut
\end{minipage}\tabularnewline
\begin{minipage}[t]{0.10\columnwidth}\raggedright
\(2\)\strut
\end{minipage} & \begin{minipage}[t]{0.34\columnwidth}\raggedright
\(\mathbb H_\cplxs\)\strut
\end{minipage}\tabularnewline
\begin{minipage}[t]{0.10\columnwidth}\raggedright
\(3\)\strut
\end{minipage} & \begin{minipage}[t]{0.34\columnwidth}\raggedright
\(\mathbb H_\cplxs\oplus\mathbb H_\cplxs\)\strut
\end{minipage}\tabularnewline
\begin{minipage}[t]{0.10\columnwidth}\raggedright
\(4\)\strut
\end{minipage} & \begin{minipage}[t]{0.34\columnwidth}\raggedright
\(M_2(\mathbb H_\cplxs)\)\strut
\end{minipage}\tabularnewline
\begin{minipage}[t]{0.10\columnwidth}\raggedright
\(5\)\strut
\end{minipage} & \begin{minipage}[t]{0.34\columnwidth}\raggedright
\(M_4(\cplxs_\cplxs)\)\strut
\end{minipage}\tabularnewline
\begin{minipage}[t]{0.10\columnwidth}\raggedright
\(6\)\strut
\end{minipage} & \begin{minipage}[t]{0.34\columnwidth}\raggedright
\(M_8(\cplxs)=M_8(\reals_\cplxs)\)\strut
\end{minipage}\tabularnewline
\begin{minipage}[t]{0.10\columnwidth}\raggedright
\(7\)\strut
\end{minipage} & \begin{minipage}[t]{0.34\columnwidth}\raggedright
\(M_8(\cplxs)\oplus M_8(\cplxs)\)\strut
\end{minipage}\tabularnewline
\bottomrule
\end{longtable}

\BeginKnitrBlock{proposition}
\protect\hypertarget{prp:cliffsimple}{}{\label{prp:cliffsimple} }If
\(s\not\equiv3\,(4)\), then \(C\ell_{0,s}\) is simple as a real
C\(^*\)-algebra. Moreover, any two finite-dimensional real
\(*\)-representations are unitarily equivalent, provided only that they
have the same dimension.
\EndKnitrBlock{proposition}

\BeginKnitrBlock{proof}
{}Simplicity is clear from Table
\ref{tab:clifford}. Any finite-dimensional real \(*\)-representation is
the direct sum of irreducible representations, and hence, a multiple of
the unique isomorphism class of irreducible real \(*\)-representation.
\EndKnitrBlock{proof}

The situation for \(s=3\,(4)\) is different in that there exists a
central element \(\omega=j_1\dotsm j_s\) with \(\omega^2=+1\). The
following statement may be found in \autocite[Chapter 1, Propositions
5.9 and 5.10]{lawson-michelsohn} in a slightly different form.

\BeginKnitrBlock{proposition}
\protect\hypertarget{prp:cliffnonsimple}{}{\label{prp:cliffnonsimple} }Let
\(s\equiv3\,(4)\) and set \(\omega\defi j_1\dotsm j_s\). There are up to
unitary equivalence exactly two irreducible real \(*\)-representations
\(\phi\) of \(C\ell_{0,s}\), and they are distinguished by
\(\phi(\omega)=1\) and \(\phi(\omega)=-1\), respectively.

Any finite-dimensional real \(*\)-representation \((W,\phi)\) of
\(C\ell_{0,s}\) extends to a real \(*\)-representation of
\(C\ell_{0,s+1}\) if and only if the eigenvalues \(\pm1\) of
\(\phi(\omega)\) have equal multiplicity. In this case, the unitary
equivalence class of \((W,\phi)\) and indeed of its extension to
\(C\ell_{0,s+1}\) is uniquely determined by the dimension of \(W\).
\EndKnitrBlock{proposition}

\BeginKnitrBlock{proof}
{}The first part is immediate from Schur's
lemma. In particular, the unitary equivalence class of any
finite-dimensional real \(*\)-representation \((W,\phi)\) of
\(C\ell_{0,s}\) is determined by the multiplicity of the eigenvalues
\(\pm1\) of \(\phi(\omega)\).

Assume that in the representation \((W,\phi)\),
\(W_\pm\defi\ker(\phi(\omega)\mp1)\) have equal dimension. There is an
automorphism \(\Gamma\) of \(C\ell_{0,s}\), defined by \[
    \Gamma(j_\alpha)\defi -j_\alpha\quad(\alpha=1,\dotsc,s).
\] Then \(\Gamma(\omega)=-\omega\). It follows that
\(\phi_+\circ\Gamma\) and \(\phi_-\) are unitarily equivalent, where
\(\phi_\pm\) is the restriction of \(\phi\) to \(W_\pm\). Hence, there
is a real unitary isomorphism \(u:W_+\longrightarrow W_-\) such that \[
    -u\phi_+(j_\alpha)u^*=u\phi_+(\Gamma(j_\alpha))u^*=\phi_-(j_\alpha)\quad(\alpha=1,\dotsc,s).
\] We may define \[
    \phi(j_{s+1})\defi
    \begin{pmatrix}
        0&-u^*\\u&0
    \end{pmatrix}.
\] This defines the required extension.

Conversely, let \((W,\phi)\) be a real \(*\)-representation of
\(C\ell_{0,s+1}\). Because \(\phi(\omega_s)\) anti-commutes with
\(\phi(j_{s+1})\), the eigenvalues of \(\phi(\omega_s)\) have equal
multiplicity. The statement about uniqueness is obvious from Proposition
\ref{prp:cliffsimple}, as \(s+1\equiv0\,(4)\).
\EndKnitrBlock{proof}

\BeginKnitrBlock{remark}
{}In any given symmetry class labelled by
\(s\), we shall assume a reference disordered IQPV
\(J_\mathrm{ref}\in\mathbb A\) with \(s\) negative pseudo-symmetries
\(J_1,\dotsc,J_s\in\endo{W}\). Physically speaking, \(J_\mathrm{ref}\)
represents the \enquote{trivial} topological phase. If
\(J_\mathrm{ref}\) is \enquote{atomic} or local in the strong sense that
\(J_\mathrm{ref}\in\endo{W}\), then by Propositions
\ref{prp:cliffsimple} and \ref{prp:cliffnonsimple}, the real Clifford
\(*\)-representation defined by \(J_1,\dotsc,J_s\) is determined up to
unitary equivalence by \(\dim W\).
\EndKnitrBlock{remark}

We now discuss case-by-case what it means for \(J\) to be in one of the
symmetry classes \(s=0,\dotsc,7\), in terms of the physical symmetries
introduced in Paragraph \ref{symm}. We will in particular use the
notation from Equations (\ref{eq:j03}--\ref{eq:j4plus}). However, the
construction of both the bulk and the boundary classes, given in
Subsection \ref{subs:bulkinv} and in Section \ref{sec:bdy}, will be
independent of this case-by-case discussion and will rely solely on the
description in terms of pseudo-symmetries.

\hypertarget{diii-iqpv}{%
\subsubsection{\texorpdfstring{Class \(D\mathrm{I\!I\!I}\),
\(s=1\)}{Class D\textbackslash mathrm\{I\textbackslash!I\textbackslash!I\}, s=1}}\label{diii-iqpv}}

In this case, \(V\) carries a time-reversal operator or
\emph{quaternionic structure} \(T\) (see Definition \ref{def:quat-str}).
That is, \(T\) is anti-unitary and \(T^2=-1\). We assign to \(T\) the
real unitary \(J_1=J_T=\gamma T\).

Recall that the skew-Hermitian unitary \(J\) anti-commutes with the
pseudo-symmetry \(J_T\). Thus \(J\) exchanges the two eigenspaces
\(W_\pm = \mathrm{ker}(J_T \mp i)\) and takes the form \[
    J=\begin{pmatrix} 0&-u^\ast\\u&0\end{pmatrix}, \quad u^\ast=u^{-1},
\] with respect to the decomposition \(W=W_+\oplus W_-\). Now
\(J=-J^\ast\) is also real or, equivalently, skew (w.r.t. the CAR form).
By the algebraic properties of \(J_T\), the CAR form restricts to a
non-degenerate bilinear form \(W_+\otimes W_-\longrightarrow\cplxs\).
The resulting identification \(W_-\cong W_+^\ast\) lets us interpret
\(u\in\mathrm{Hom}(W_+,W_-)\cong\mathrm{Hom}(W_+,W_+^\ast)\) as a
bilinear form on \(W_+\). The skewness of \(J\) then implies skewness of
\(u\). Thus, any disordered IQPV of class \(\diii\) corresponds to a
unitary and skew-symmetric bilinear form on \(W_+\). Such forms
constitute a symmetric space \(\mathrm{U}/\mathrm{Sp}\), which is of
Cartan--Killing type \(\aii\).

\hypertarget{aii-iqpv}{%
\subsubsection{\texorpdfstring{Class \(A\mathrm{I\!I}\),
\(s=2\)}{Class A\textbackslash mathrm\{I\textbackslash!I\}, s=2}}\label{aii-iqpv}}

Here, in addition to the time reversal \(T\), we consider the charge
operator \(Q\in\endo W\) and \(J_2\defi J_Q\defi iJ_TQ\). The following
lemma will help to describe the situation.

\BeginKnitrBlock{lemma}
\protect\hypertarget{lem:twoPS}{}{\label{lem:twoPS} }Let
\(x_1,x_2\in\endo W\subseteq\mathbb A\) be two anti-commuting real
skew-Hermitian unitaries. Then \((W^\sim,T)\), where
\(W^\sim\defi\ker(x_2x_1-i)\) and \(T\defi-\gamma x_2\), is a
quaternionic vector space. There is an isomorphism
\((W,\gamma)\cong(W^\sim\otimes\cplxs^2,T\otimes\ger c)\) of real vector
spaces such that under the induced isomorphism of real C\(^*\)-algebras
\[
    \mathbb A\cong\mathbb A^\sim\otimes\mathbb H_\cplxs,\quad\mathbb A^\sim\defi A^{W^\sim},
\] we have \(x_1\longmapsto 1\otimes i\sigma_x\) and
\(x_2\longmapsto 1\otimes i\sigma_y\). Under the isomorphism, real
unitaries \(J\in\mathbb A\) anti-commuting with \(x_1\) correspond to
operators \(y_1\otimes i\sigma_y+y_2\otimes i\sigma_z\) where
\(y_1,y_2\in\mathbb A^\sim\) are real Hermitian elements such that \[
    y_1^2+y_2^2=1,\quad y_1y_2=y_2y_1.
\]
\EndKnitrBlock{lemma}

\BeginKnitrBlock{proof}
{}Let \(x_3\defi x_2x_1\in\endo W\). Then
\(x_3\) is a real skew-Hermitian unitary that anti-commutes with \(x_1\)
and \(x_2\). Setting \(W^\pm\defi\ker(x_3\mp i)\), we obtain a splitting
\(W=W^+\oplus W^-\) with respect to which \[
    x_1=\begin{pmatrix}0&iu^*\\ iu&0\end{pmatrix},\quad
    x_2=\begin{pmatrix}0&u^*\\ -u&0\end{pmatrix},\quad
    x_3=\begin{pmatrix}i&0\\0&-i\end{pmatrix}.
\] The map \(u:W^+\longrightarrow W^-\) is a unitary isomorphism, and so
is \[
    \phi\defi
    \begin{pmatrix}
        1&0\\0&u
    \end{pmatrix}\colon W^+\oplus W^+=W^+\otimes\cplxs\longrightarrow W^+\oplus W^-.
\] We compute \[
    \phi^*x_1\phi=1\otimes i\sigma_x,\quad\phi^*x_2\phi=1\otimes i\sigma_y.
\] Since the anti-unitary map \(\gamma x_2\) anti-commutes with \(x_3\),
it leaves \(W^+\) invariant. We let \(T\defi-\gamma x_2|_{W^+}\). Then
\(T\) is a quaternionic structure on \(W^\sim\defi W^+\) and \[
    -uT=-x_2\gamma x_2|_{W^+}=-\gamma x_2^2|_{W^+}=\gamma|_{W^+},
\] so that \[
    \phi(T\otimes\ger c)\phi^*=
    \phi
    \begin{pmatrix}
        0&T\\-T&0
    \end{pmatrix}
    \phi^*=
    \begin{pmatrix}
        0&Tu^*\\-uT&0
    \end{pmatrix}=\gamma.
\] This proves the claim.
\EndKnitrBlock{proof}

With the lemma above in hand, we can finish the story as follows. Under
the isomorphism \(\phi:V\otimes\cplxs^2\longrightarrow W\) the
pseudo-symmetry \(J_Q\) corresponds to \[
    \phi^\ast J_Q\phi={\id}_V\otimes i\sigma_y.
\] Real unitaries \(J\in\mathbb A\) anti-commuting with \(J_1,J_2\)
correspond under
\(\mathbb A\cong\mathbb{A}^\sim \otimes\mathbb H_\cplxs\) exactly to
elements of the form \(x \otimes i\sigma_z\), where
\(x\in\mathbb A^\sim\) is a real Hermitian unitary.

Here, the condition that \(x\) be real is equivalent to \(Tx=xT\). Such
elements \(x\) constitute a symplectic Grassmann manifold
\(\mathrm{Sp}/(\mathrm{Sp}\times\mathrm{Sp})\), which is a symmetric
space of Cartan--Killing type \(C\mathrm{I\!I}\).

\hypertarget{cii-iqpv}{%
\subsubsection{\texorpdfstring{Class \(C\mathrm{I\!I}\),
\(s=3\)}{Class C\textbackslash mathrm\{I\textbackslash!I\}, s=3}}\label{cii-iqpv}}

Besides preserving the time-reversal and charge symmetries, ground
states in this symmetry class are invariant under \emph{particle-hole
symmetry}. We have \(J_3=J_C=i\gamma CQ\). The isomorphism \(\phi\) from
Lemma \ref{lem:twoPS} gives \[
    \phi^*J_C\phi=
    \begin{pmatrix}
        iS&0\\0&-iS
    \end{pmatrix}
    =S\otimes i\sigma_z.
\] Here, we apply the decomposition \(V=V_+\oplus V_-\) for
\(V_\pm\defi\ker(S\mp1)\).

Similarly, a skew-Hermitian unitary \(J\in\mathbb A\) anti-commuting
with \(J_1,J_2,J_3\) acquires the shape \(x\otimes i\sigma_z\) where
\(x\) is a real Hermitian unitary commuting with \(T\) and
anti-commuting with \(S\). Thus, \(x\) takes the form \[
    x=
    \begin{pmatrix}
        0&y\\ y^*&0
    \end{pmatrix}
\] where \(y_\omega\) is a unitary symplectic operator for every
\(\omega\in\Omega\), corresponding to the symmetric space
\(\mathrm{USp}\) of type \(C\). In terms of the decomposition
\(\mathbb A^\sim\cong\mathbb A^\approx\otimes M_2(\cplxs)\), where
\(\mathbb A^\approx\defi A^{V_+}\) is equipped by the real structure
induced by \(T|_{V_+}\), \(y\) is a real unitary in
\(\mathbb A^\approx\).

\hypertarget{s4-iqpv}{%
\subsubsection{\texorpdfstring{Classes
\(s\sge4\)}{Classes s\textbackslash sge4}}\label{s4-iqpv}}

In this situation, we consider real unitaries \(J\in\mathbb A\) such
that \(J\otimes\sigma_x\) anti-commutes with \(J_1,\dotsc, J_s\),
defined as in Equations \eqref{eq:jspin} and \eqref{eq:j4plus}. This is
equivalent to \(J\) anti-commuting with the real spin rotation
generators \(S_1,\dotsc, S_3\) and with the first \(s-4\) elements of
the sequence \(J_T,J_Q,J_C\). Applying Lemma \ref{lem:twoPS} with
\(x_1=j_1\) and \(x_2=j_2\), we obtain \[
    \endo{W}\cong\endo{W^\sim}\otimes\mathbb H_\cplxs,\quad\mathbb A\cong\mathbb A^\sim\otimes\mathbb H_\cplxs
\] where \(W^\sim=\ker(j_3-i)=\ker(S_3-1)\) and
\(\mathbb A^\sim=A^{W^\sim}\). Under the isomorphism, \[
    J_T\longmapsto j_T\otimes1,\quad J_Q\longmapsto j_Q\otimes1,\quad J_C\longmapsto j_C\otimes1,
\] where \(j_T,j_Q,j_C\in\endo{W^\sim}\) are anti-commuting real
skew-Hermitian unitaries. The operator \(J\) corresponds to a real
skew-Hermitian unitary \(j\in\mathbb A^\sim\) anti-commuting with the
first \(s-4\) of \(j_T,j_Q,j_C\). In this sense, the symmetry classes
for \(s\sge 4\) reduce to the symmetry classes for \(0\sle s\sle 3\).
The only difference is that the real structure on \(\mathbb A^\sim\)
comes from a quaternionic structure on \(W^\sim\) rather than from a
real structure.

\hypertarget{the-srfl-scheme}{%
\subsection{The SRFL scheme}\label{the-srfl-scheme}}

In the literature, symmetries are often organised according to a scheme
introduced by Schnyder--Ryu--Furusaki--Ludwig \autocite{srfl}, which is
different from the approach of Kennedy and Zirnbauer that we follow. We
speak of the SRFL scheme. It is based on two anti-unitary endomorphisms
\(\Theta,\Xi:V\longrightarrow V\) and a unitary operator \(\Pi\)
proportional to \(\Xi\Theta\). The following relations are assumed: \[
    \Theta^2=\pm1,\quad\Xi^2=\pm1,\quad\Pi^2=1.
\] In the literature, \(\Theta\) is called \emph{time reversal}, \(\Xi\)
is called \emph{particle-hole conjugation}, and \(\Pi\) the
\emph{chiral} or \emph{sub-lattice symmetry}. As we have already
reserved these expressions for certain physical symmetries with
potentially different commutation relations, we will not use this
terminology here.

According to the SRFL scheme, a Hamiltonian is called
\emph{\(\Theta\)-symmetric}, \emph{\(\Xi\)-symmetric}, or
\emph{\(\Pi\)-symmetric}, if respectively, \[
    \Theta H\Theta^*=H,\quad \Xi H\Xi^*=-H,\quad\Pi H\Pi^*=-H.
\] All possible combinations of the presence or absence of these
symmetries can be summarised in the following table. Here, an entry
\(0\) means that the corresponding symmetry is absent, whereas a
non-zero entry \(\pm1\) indicates that the symmetry is present; in this
case, the entry equals the square of the corresponding symmetry
operator.

\begin{longtable}[]{@{}llll@{}}
\caption{\label{tab:srfl} Symmetry classes according to
Schnyder--Ryu--Furusaki--Ludwig}\tabularnewline
\toprule
\begin{minipage}[b]{0.25\columnwidth}\raggedright
class\strut
\end{minipage} & \begin{minipage}[b]{0.15\columnwidth}\raggedright
\(\Theta\)\strut
\end{minipage} & \begin{minipage}[b]{0.10\columnwidth}\raggedright
\(\Xi\)\strut
\end{minipage} & \begin{minipage}[b]{0.10\columnwidth}\raggedright
\(\Pi\)\strut
\end{minipage}\tabularnewline
\midrule
\endfirsthead
\toprule
\begin{minipage}[b]{0.25\columnwidth}\raggedright
class\strut
\end{minipage} & \begin{minipage}[b]{0.15\columnwidth}\raggedright
\(\Theta\)\strut
\end{minipage} & \begin{minipage}[b]{0.10\columnwidth}\raggedright
\(\Xi\)\strut
\end{minipage} & \begin{minipage}[b]{0.10\columnwidth}\raggedright
\(\Pi\)\strut
\end{minipage}\tabularnewline
\midrule
\endhead
\begin{minipage}[t]{0.25\columnwidth}\raggedright
\(A\)\strut
\end{minipage} & \begin{minipage}[t]{0.15\columnwidth}\raggedright
\(0\)\strut
\end{minipage} & \begin{minipage}[t]{0.10\columnwidth}\raggedright
\(0\)\strut
\end{minipage} & \begin{minipage}[t]{0.10\columnwidth}\raggedright
\(0\)\strut
\end{minipage}\tabularnewline
\begin{minipage}[t]{0.25\columnwidth}\raggedright
\(A\mathrm{I\!I\!I}\)\strut
\end{minipage} & \begin{minipage}[t]{0.15\columnwidth}\raggedright
\(0\)\strut
\end{minipage} & \begin{minipage}[t]{0.10\columnwidth}\raggedright
\(0\)\strut
\end{minipage} & \begin{minipage}[t]{0.10\columnwidth}\raggedright
\(1\)\strut
\end{minipage}\tabularnewline
\begin{minipage}[t]{0.25\columnwidth}\raggedright
\(D\)\strut
\end{minipage} & \begin{minipage}[t]{0.15\columnwidth}\raggedright
\(0\)\strut
\end{minipage} & \begin{minipage}[t]{0.10\columnwidth}\raggedright
\(1\)\strut
\end{minipage} & \begin{minipage}[t]{0.10\columnwidth}\raggedright
\(0\)\strut
\end{minipage}\tabularnewline
\begin{minipage}[t]{0.25\columnwidth}\raggedright
\(D\mathrm{I\!I\!I}\)\strut
\end{minipage} & \begin{minipage}[t]{0.15\columnwidth}\raggedright
\(-1\)\strut
\end{minipage} & \begin{minipage}[t]{0.10\columnwidth}\raggedright
\(1\)\strut
\end{minipage} & \begin{minipage}[t]{0.10\columnwidth}\raggedright
\(1\)\strut
\end{minipage}\tabularnewline
\begin{minipage}[t]{0.25\columnwidth}\raggedright
\(A\mathrm{I\!I}\)\strut
\end{minipage} & \begin{minipage}[t]{0.15\columnwidth}\raggedright
\(-1\)\strut
\end{minipage} & \begin{minipage}[t]{0.10\columnwidth}\raggedright
\(0\)\strut
\end{minipage} & \begin{minipage}[t]{0.10\columnwidth}\raggedright
\(0\)\strut
\end{minipage}\tabularnewline
\begin{minipage}[t]{0.25\columnwidth}\raggedright
\(C\mathrm{I\!I}\)\strut
\end{minipage} & \begin{minipage}[t]{0.15\columnwidth}\raggedright
\(-1\)\strut
\end{minipage} & \begin{minipage}[t]{0.10\columnwidth}\raggedright
\(-1\)\strut
\end{minipage} & \begin{minipage}[t]{0.10\columnwidth}\raggedright
\(1\)\strut
\end{minipage}\tabularnewline
\begin{minipage}[t]{0.25\columnwidth}\raggedright
\(C\)\strut
\end{minipage} & \begin{minipage}[t]{0.15\columnwidth}\raggedright
\(0\)\strut
\end{minipage} & \begin{minipage}[t]{0.10\columnwidth}\raggedright
\(-1\)\strut
\end{minipage} & \begin{minipage}[t]{0.10\columnwidth}\raggedright
\(0\)\strut
\end{minipage}\tabularnewline
\begin{minipage}[t]{0.25\columnwidth}\raggedright
\(C\mathrm I\)\strut
\end{minipage} & \begin{minipage}[t]{0.15\columnwidth}\raggedright
\(1\)\strut
\end{minipage} & \begin{minipage}[t]{0.10\columnwidth}\raggedright
\(-1\)\strut
\end{minipage} & \begin{minipage}[t]{0.10\columnwidth}\raggedright
\(1\)\strut
\end{minipage}\tabularnewline
\begin{minipage}[t]{0.25\columnwidth}\raggedright
\(A\mathrm I\)\strut
\end{minipage} & \begin{minipage}[t]{0.15\columnwidth}\raggedright
\(1\)\strut
\end{minipage} & \begin{minipage}[t]{0.10\columnwidth}\raggedright
\(0\)\strut
\end{minipage} & \begin{minipage}[t]{0.10\columnwidth}\raggedright
\(0\)\strut
\end{minipage}\tabularnewline
\begin{minipage}[t]{0.25\columnwidth}\raggedright
\(BD\mathrm I\)\strut
\end{minipage} & \begin{minipage}[t]{0.15\columnwidth}\raggedright
\(1\)\strut
\end{minipage} & \begin{minipage}[t]{0.10\columnwidth}\raggedright
\(1\)\strut
\end{minipage} & \begin{minipage}[t]{0.10\columnwidth}\raggedright
\(1\)\strut
\end{minipage}\tabularnewline
\bottomrule
\end{longtable}

\hypertarget{class-d}{%
\subsubsection{\texorpdfstring{Class \(D\)}{Class D}}\label{class-d}}

The only restriction here is that a flat-band Hamiltonian \(H\)
anti-commutes with the real structure \(\gamma\) on \(W\). Hence, if we
set \(\Xi\defi\gamma\), then \(\Xi^2=1\) and any \(\Xi\)-symmetric
Hamiltonian gives rise to an IQPV of class \(D\).

\hypertarget{class-dmathrmiii}{%
\subsubsection{\texorpdfstring{Class
\(D\mathrm{I\!I\!I}\)}{Class D\textbackslash mathrm\{I\textbackslash!I\textbackslash!I\}}}\label{class-dmathrmiii}}

Here, a time-reversal symmetry \(T\) on \(V\), \(T^2=-1\), is present,
and extended to \(W\) as usual. We may set \(\Xi\defi\gamma\),
\(\Theta\defi T\), and \(\Pi\defi i\Theta\Xi\). Then \(\Theta^2=-1\),
\(\Xi^2=1\), and \(\Pi^2=1\), and any \(H\) that is symmetric for
\(\Theta\), \(\Xi\), and \(\Pi\) gives rise to an IQPV of class
\(D\mathrm{I\!I\!I}\).

\hypertarget{class-amathrmii}{%
\subsubsection{\texorpdfstring{Class
\(A\mathrm{I\!I}\)}{Class A\textbackslash mathrm\{I\textbackslash!I\}}}\label{class-amathrmii}}

In this class we have the symmetries \(T\) and \(iQ\). As explained in
\ref{aii-iqpv}, a IQPV of class \(A\mathrm{I\!I}\) is the same as an
operator \(x\) on \(V\), \(x^2=1\), which commutes with \(T\). Hence, if
we set \(\Theta\defi T\), then \(\Theta^2=-1\) and any
\(\Theta\)-symmetric flat-band Hamiltonian \(H\) determines an IQPV of
class \(A\mathrm{I\!I}\) \emph{via} \(x\defi H\).

\hypertarget{class-cmathrmii}{%
\subsubsection{\texorpdfstring{Class
\(C\mathrm{I\!I}\)}{Class C\textbackslash mathrm\{I\textbackslash!I\}}}\label{class-cmathrmii}}

In addition to the previous symmetries, the particle-hole symmetry \(C\)
is present and given by \(C=\gamma S\) where \(S\) is linear, \(S^2=1\),
and \([S,T]=0\). As explained in \ref{cii-iqpv}, an IQPV of class
\(C\mathrm{I\!I}\) is the same as an operator \(x\), \(x^2=1\), on
\(V\), such that \(Tx=xT\) and \(Sx=-xS\). We may set \(\Theta\defi T\),
\(\Pi\defi S\), \(\Xi\defi\Theta\Pi\). Then \(\Theta^2=-1\),
\(\Pi^2=1\), \(\Xi^2=-1\), and any flat-band Hamiltonian \(H\) symmetric
for \(\Theta\), \(\Xi\), and \(\Pi\) determines an IQPV of class
\(C\mathrm{I\!I}\) \emph{via} \(x\defi H\).

\hypertarget{classes-ssge-4}{%
\subsubsection{\texorpdfstring{Classes
\(s\sge 4\)}{Classes s\textbackslash sge 4}}\label{classes-ssge-4}}

For \(s\sge 4\), the generators \(S_1,S_2,S_3\) of spin rotation induce
an isomorphism \(W\cong W^\sim\otimes\cplxs^2\) and a splitting
\(\mathbb A=A^{\smash{W^\sim}}\otimes \mathbb{H}_\cplxs\), as explained
in \ref{s4-iqpv}. In this splitting, \(\gamma\) corresponds to
\(T\otimes\ger c\) where \(T\) is a quaternionic structure on \(W^\sim\)
and \(\ger c\) the standard quaternionic structure on \(\cplxs^2\).

As we have seen, IQPV of class \(s\sge 4\) correspond to operators
\(j\in A^{W^\sim}\) commuting with \(T\) such that \(j^2=-1\) and \(j\)
anti-commutes with the first \(s-4\) of the operators \(j_T\), \(j_Q\),
and \(j_C\), corresponding to \(J_T\), \(J_Q\), and \(J_C\),
respectively. We may proceed as above, with \(T\) playing the role of
\(\gamma\), to establish the correspondence in all real symmetry
classes.

\hypertarget{complex-class-a}{%
\subsubsection{\texorpdfstring{Complex class
\(A\)}{Complex class A}}\label{complex-class-a}}

The complex classes are those where both \(J\) and the group of
symmetries commute with \(Q\). As we have noted, this implies that \(J\)
and any other symmetries present leave \(\sh V\) invariant and are
determined by their restriction to this space. In class \(A\), there are
no further symmetries present, so the Hamiltonians \(H\) on \(\sh V\)
determine IQPV in the complex symmetry class \(A\).

\hypertarget{complex-class-amathrmiii}{%
\subsubsection{\texorpdfstring{Complex class
\(A\mathrm{I\!I\!I}\)}{Complex class A\textbackslash mathrm\{I\textbackslash!I\textbackslash!I\}}}\label{complex-class-amathrmiii}}

In complex class \(A\mathrm{I\!I\!I}\), we have the additional symmetry
\(C=\gamma S\). IQPV in this class correspond to operators on \(\sh V\)
anti-commuting with \(S\). If we set \(\Pi\defi S\), then \(\Pi^2=1\),
and \(\Pi\)-symmetric Hamiltonians determine IQPV in the complex
symmetry class \(A\mathrm{I\!I\!I}\).

\hypertarget{subs:bulkinv}{%
\subsection{From quasi-particle vacua to bulk
invariants}\label{subs:bulkinv}}

\hypertarget{van-daeles-picture-of-k-theory}{%
\subsubsection{\texorpdfstring{Van Daele's picture of
\(K\)-theory}{Van Daele's picture of K-theory}}\label{van-daeles-picture-of-k-theory}}

There exists a plethora of different pictures for \(K\)-theory, so there
is a choice to be made in the construction of \(K\)-theory classes
attached to free-fermion topological phases. We follow the guiding
principle to stay as close as possible to the physical model. This leads
to different pictures for \(K\)-theory in the bulk and at the boundary.
At the boundary, Kasparov's Fredholm picture is the most natural option,
as we shall see in Section \ref{sec:bdy}.

In the bulk, the most natural choice is the picture due to Van Daele
\autocite{van_Daele1,van_Daele2}, as was first observed by Kellendonk
\autocite{kellendonk-VD}. This picture is pleasantly simple, and it
works for graded Banach algebras over the real and complex fields. Since
Van Daele's picture appears to be little known, we give a brief
exposition for the case of (real) C\(^*\)-algebras.

\BeginKnitrBlock{definition}[Van Daele's picture for real K-theory]
\protect\hypertarget{def:unnamed-chunk-10}{}{\label{def:unnamed-chunk-10}
\iffalse (Van Daele's picture for real K-theory) \fi{} }Let \(A\) be a
unital graded real C\(^*\)-algebra. An odd real Hermitian unitary in
\(A\) will be called an \emph{ORHU}. Let \(\sh F(A)\) be the set of ORHU
in \(A\) and \(F(A) \defi\pi_0(\sh F(A))\) the set of path-connected
components of \(\sh F(A)\) for the topology induced by the norm. We say
that two elements of \(\sh F(A)\) are \emph{homotopic} if they define
the same element of \(F(A)\).

For any integer \(n\sge 1\), we consider the entry-wise grading on
\(M_n(A)=A\otimes M_n(\cplxs)\). We let \(\sh F_n(A)\defi\sh F(M_n(A))\)
and \(F_n(A)\defi F(M_n(A))\) where \(M_n(A)\) has entry-wise grading.
For \(x\in\sh F_m(A)\) and \(y\in\sh F_n(A)\), we define \[
x\oplus y\defi
\begin{pmatrix}
    x&0\\0&y
\end{pmatrix}\in\sh F_{m+n}(A)
\] and \([x]\oplus[y]\defi[x\oplus y]\in F_{m+n}(A)\).

Suppose that \(\sh F(A)\neq\vvoid\) and choose \(e\in\sh F(A)\). Then
\(F_n(A)\) becomes an inductive system of sets with the connecting maps
\(F_n(A)\longrightarrow F_{n+1}(A):[x]\longmapsto[x]\oplus[e]\).
Consider the inductive limit \[
F_\infty(A)\defi\varinjlim\nolimits_nF_n(A).
\] Then \(F_\infty(A)\) is an Abelian monoid with the addition induced
by \(\oplus\) and the unit \([e]\) \autocite[Proposition
2.7]{van_Daele1}. We define \emph{Van Daele's \(KR\)-group} \[
    DKR(A)\defi DKR_e(A)\defi G(F_\infty(A)).
\] Here, \(G(-)\) denotes the Grothendieck group of an Abelian monoid.
In case \(A\) is a unital graded C\(^*\)-algebra, we define
\(\sh F_n(A)\), \(F_n(A)\) and \(\sh F_\infty(A)\) by dropping all the
reality constraints, and we set \(DK(A)\defi G(F_\infty(A))\).
\EndKnitrBlock{definition}

\BeginKnitrBlock{remark}
{}Note that up to canonical isomorphism,
\(DKR(A)\) is independent of \(e\). Van Daele \autocite[Definition
2.1]{van_Daele1} does not impose the condition that the elements
defining \(\sh F(A)\) be Hermitian; however, this can be assumed by
\autocite[Proposition 2.5]{van_Daele1}. If \(e\) and \(-e\) are
homotopic, then \(F_\infty(A)\) is an Abelian group and the canonical
map \(F_\infty(A)\longrightarrow DKR(A)\) is an isomorphism
\autocite[Proposition 2.11]{van_Daele1}. The picture of \(DKR(A)\) we
introduce is due to Roe \autocite{roe-paschke-duality}; Van Daele's
original definition of \(DKR(A)\) coincides with it by
\autocite[Proposition 3.3]{van_Daele1}. The independence of \(DKR_e(A)\)
of \(e\) \autocite[Proposition 2.12]{van_Daele1} can be used to show
that \(DKR(A)\) is functorial for unital (real) \(*\)-morphisms. The
definition of \(DKR(A)\) can be extended to the cases where
\(\sh F(A)=\vvoid\) and where \(A\) is non-unital \autocite[Section
3]{van_Daele1}.
\EndKnitrBlock{remark}

\BeginKnitrBlock{lemma}
\protect\hypertarget{lem:dkr-stable}{}{\label{lem:dkr-stable} }Let \(A\) be
a unital graded real C\(^*\)-algebra and \(e\in\sh F(A)\) an ORHU. Let
\(m\sge1\) be an integer. Then the maps \[
    \sh F_n(A)\longrightarrow\sh F_n(M_m(A)):x\longmapsto E_{11}\otimes x\oplus (1-E_{11})\otimes e_n,
\] where \(E_{ij}\), \(1\sle i,j\sle m\), are the standard matrix units
and \(e_k=e\oplus\dotsm\oplus e\) is the sum of \(k\) copies of \(e\),
induce an isomorphism of Abelian groups
\(DKR_e(A)\longrightarrow DKR_{e_m}(M_m(A))\).
\EndKnitrBlock{lemma}

For us, the importance of Van Daele's group \(DKR(A)\) lies in its
relation to (real) \(K\)-theory, as detailed in the following theorem.

\BeginKnitrBlock{theorem}[Van Daele's picture of real K-theory]
\protect\hypertarget{thm:DKR-KR}{}{\label{thm:DKR-KR} {} }Let \(A\) be a unital graded real
C\(^*\)-algebra resp. a unital graded C\(^*\)-algebra. Then
\begin{equation}
  DKR(A\mathop{\widehat\otimes} C\ell_{r,s}) \cong KR_{s-r+1}(A),\quad\text{resp. }DK(A\mathop{\widehat\otimes}\cplxs\ell_{r,s}) \cong K_{s-r+1}(A).
  \label{eq:dkr-kr}
\end{equation}
\EndKnitrBlock{theorem}

Theorem \ref{thm:DKR-KR} implies that up to isomorphism,
\(DKR(A\mathop{\widehat\otimes} C\ell_{r,s})\) depends only on
\(s-r\pmod8\) and
\(\smash{DK(A\mathop{\widehat\otimes}\mathbb C\ell_{r,s})}\) depends
only on \(s-r\pmod2\). The theorem is proved in Ref. \autocite[Section
2]{roe-paschke-duality}. We give a sketch, recalling Roe's construction,
as we shall later refer to it. We will make use of Kasparov's bivariant
real \(KKR\)-theory; see
\autocite{blackadar-kthyopalg,kasparov-operatork,schroder1993k}.

\BeginKnitrBlock{proof}[of Theorem \ref{thm:DKR-KR}, sketch]
{}Applying
\((1,1)\)-periodicity of \(DKR\) \autocite[Lemma
2.4]{roe-paschke-duality}, we may assume that \(r\sge1\). Possibly
replacing \(A\) by a matrix algebra, fix a reference ORHU
\(e\in\sh F(Q^s(A)\mathop{\widehat{\otimes}}C\ell_{r-1,s})\), where
\(Q^s(A)\defi\mathrm M^s(A)/(A\mathop{\widehat{\otimes}}\knums)\) and
\(\knums\) denotes the algebra of compact operators on a separable
graded real Hilbert space.

It is well-known that
\(KR_{s-r+1}(A)=KKR(\cplxs,A\mathop{\widehat{\otimes}}C\ell_{r-1,s})\),
see \autocite[Theorem 2.3.8 and remarks following Corollary
2.5.2]{schroder1993k}. It also standard \autocite[Proposition
2.3.5]{schroder1993k} that any class in the latter group can be
represented by a cycle of the form
\(\Parens{\sh H_A\mathop{\widehat{\otimes}}C\ell_{r-1,s},1_\cplxs,x}\),
where \(\sh H_A\) is the standard graded real Hilbert \(A\)-module and
\(x\in\mathrm M^s(A)\mathop{\widehat{\otimes}}C\ell_{r-1,s}\) maps to an
ORHU in \(Q^s(A)\mathop{\widehat{\otimes}}C\ell_{r-1,s}\) \emph{via} the
canonical map \(\pi:\mathrm M^s(A)\longrightarrow Q^s(A)\).

Let \(\partial_Q\) be the connecting map for \(DKR\), defined in
\autocite{van_Daele2}, for the short exact sequence \[
\begin{tikzcd}
    0\rar{}&A\mathop{\widehat{\otimes}}\knums\rar{}&\mathrm M^s(A)\rar{}&Q^s(A)\rar{}&0.
\end{tikzcd}
\] Because \(\mathrm M^s(A)\) is \(KR\)-trivial, it is an isomorphism \[
    \partial_Q:DKR_e\Parens{Q^s(A)\mathop{\widehat{\otimes}}C\ell_{r-1,s}}\longrightarrow DKR\Parens{A\mathop{\widehat{\otimes}}C\ell_{r,s}}.
\] The desired isomorphism \[
    \alpha:KKR\Parens{\cplxs,A\mathop{\widehat{\otimes}}C\ell_{r-1,s}}\longrightarrow DKR\Parens{A\mathop{\widehat{\otimes}}C\ell_{r,s}}
\] is defined by mapping the \(KKR\) class corresponding to \(x\) to
\(\partial_Q([x]-[e])\).
\EndKnitrBlock{proof}

\hypertarget{bulk-invariants-attached-to-iqpv}{%
\subsubsection{Bulk invariants attached to
IQPV}\label{bulk-invariants-attached-to-iqpv}}

We will now attach classes in the \(KR\)-theory of \(\mathbb{A}\) to any
disordered IQPV with symmetries. The construction is purely algebraic
and in terms of Van Daele's picture of (real) \(K\)-theory, and does not
depend on the real C\(^*\)-algebra \(\mathbb A\). In fact, we will show
that \(KR\)-theory can be described entirely in terms of the algebraic
relations of a disordered IQPV with pseudo-symmetries.

In what follows, let \(B\) be an (ungraded) unital real C\(^*\)-algebra.
Suppose \(r,s\sge0\) and that \((K_1,\dotsc,K_r,J_1,\dotsc,J_s)\)
represent a unital real \(*\)-morphism
\(\phi:C\ell_{r,s}\longrightarrow B\).

\BeginKnitrBlock{definition}
\protect\hypertarget{def:psym-proj}{}{\label{def:psym-proj} }Define
commuting projections \(Q_a,P_\alpha\in B\otimes C\ell_{r,s+1}\) by
\begin{equation}
    2Q_a-1=(-1)^sK_a\otimes k_aj_1,\quad 2P_\alpha-1=J_\alpha\otimes j_1j_{\alpha+1} \quad(a=1,\dotsc,r,\alpha=1,\dotsc,s).
\label{eq:comm-proj}
\end{equation} The product over all of these defines a further
projection \[
    P^{r,s}\defi Q_1\dotsm Q_rP_1\dotsm P_s.
\]
\EndKnitrBlock{definition}

Let \(\FF^{r,s}(B)\) be the set of all \(J\in B\) satisfying the
algebraic relations of an IQPV of symmetry index \((r,s)\), that is \[
    \overline J=J=-J^*,\quad J^2=-1,\quad JK_a+K_aJ=JJ_\alpha+J_\alpha J=0\quad(a=1,\dotsc,r,\alpha=1,\dotsc,s).
\] In case \(B\) is not assumed to be real, we make the same definition,
dropping the reality constraints on the \(K_a\), \(J_\alpha\) and on
\(J\).

\BeginKnitrBlock{proposition}
\protect\hypertarget{prp:iqpv-orhu}{}{\label{prp:iqpv-orhu} }The assignment
\[
    J\longmapsto(J\otimes j_1)P^{r,s}
\] defines a bijective map \[
    \FF^{r,s}(B)\longrightarrow\sh F\Parens{P^{r,s}(B\otimes C\ell_{r,s+1})P^{r,s}}.
\] If \(B\) and the \(K_a,J_\alpha\) are not assumed to be real, one has
a corresponding bijection \[
    \FF^{r,s}(B)\longrightarrow\sh F\Parens{P^{r,s}(B\otimes\mathbb C\ell_{r,s+1})P^{r,s}}.
\]
\EndKnitrBlock{proposition}

We only give the details of the proof in the real case. It relies on the
following lemma.

\BeginKnitrBlock{lemma}
\protect\hypertarget{lem:corner-orhu}{}{\label{lem:corner-orhu} }Let \(C\)
be a unital graded real C\(^*\)-algebra and \(x\in C\) be an ORHU.
Consider the projection \(q\in C\mathop{\widehat\otimes}C\ell_{0,1}\)
defined by \(2q-1=x\otimes j_1\). (Note that \(x \otimes j_1\) is
Hermitian due to the appearance of the graded tensor product.) Then \[
    \sh F(q(C\mathop{\widehat\otimes}C\ell_{0,1})q)=\Set{(y\otimes1)q}{y\in\sh F(C),xy+yx=0}.
\] Moreover, the ORHU \(y\in C\) representing an ORHU
\((y\otimes1)q\in q(C\mathop{\widehat\otimes}C\ell_{0,1})q\) is unique.

Similarly, let \(x\in C\) be an odd real skew-Hermitian unitary and
\(p\in C\mathop{\widehat\otimes}C\ell_{1,0}\) be the projection defined
by \(2p-1=x\otimes k_1\). Then \[
    \sh F(p(C\mathop{\widehat\otimes}C\ell_{1,0})p)=\Set{(y\otimes1)p}{y\in\sh F(C),xy+yx=0},
\] where once again, the representatives \(y\) are uniquely determined.
\EndKnitrBlock{lemma}

\BeginKnitrBlock{proof}
{}We only prove the first statement, the proof
of the second one being nearly identical. The odd elements of
\(C\mathop{\widehat\otimes}C\ell_{0,1}\) are exactly those of the form
\[
    z=\tfrac12(y\otimes1)+\tfrac12(y'\otimes j_1)
\] where \(y\) is odd and \(y'\) is even. Then \(z\) lies in the corner
\(q(C\mathop{\widehat\otimes}C\ell_{0,1})q\) if and only if \[
    (x\otimes j_1)z=z=z(x\otimes j_1),
\] that is, if and only if \(y'=-xy=yx\). If this is the case, then
\(z=(y\otimes1)q\) and \(z\) is Hermitian if and only if \(y\) is.
Finally, in that case, \(z\) is unitary in the corner if and only if \[
    \tfrac12(1\otimes1)+\tfrac12(x\otimes j_1)=q=z^2=(y\otimes1)^2q=\tfrac12(y^2\otimes1)+\tfrac12(y^2x\otimes j_1),
\] which in turn is equivalent to \(y^2=1\). By construction, \(y\) is
uniquely determined by \(z\).
\EndKnitrBlock{proof}

\BeginKnitrBlock{proof}[of Proposition \ref{prp:iqpv-orhu}]
{}It is
obvious that the \(Q_a\) and \(P_\alpha\) are indeed commuting
projections. We will prove the statement by two separate inductions with
respect to \(r\) and \(s\). In case \(r=s=0\), we have \(P^{0,0}=1\),
and there is nothing to prove. Suppose that \(r=0\), \(s\sge1\), and
that the statement has been proved for \((0,s-1)\) and any unital real
C\(^*\)-algebra \(B\). Since \(1\otimes j_{s+1}\) commutes with
\(P^{0,s-1}\), Equation \eqref{eq:cliff-tensor} implies that \[
    P^{0,s}\Parens{B\otimes C\ell_{0,s+1}}P^{0,s}\cong P_s\Parens{B'\mathop{\widehat\otimes}C\ell_{0,1}}P_s
\] where \[
    B'\defi P^{0,s-1}\Parens{B\otimes C\ell_{0,s}}P^{0,s-1}.
\] Here, the \(C\ell_{0,1}\) factor on the right-hand side of the former
equation is generated by \(j_{s+1}\), and the \(C\ell_{0,s}\) factor on
the right-hand side of the latter is generated by \(j_1,\dotsc,j_s\).

Under the isomorphism in the former equation, \(J_s\otimes j_1j_{s+1}\)
is mapped to \(x\otimes j_{s+1}\), where \(x\defi J_s\otimes j_1\).
Hence, we may apply the first part of Lemma \ref{lem:corner-orhu} with
\(C=B'\) to the right-hand side of the former equation. We conclude that
there is a bijection, defined by the equation \(a=bP_s\), between the \[
    a\in\sh F(P^{0,s}\Parens{B\otimes C\ell_{0,s+1}}P^{0,s}),
\] and the
\(b\in\sh F(\smash{P^{0,s-1}\Parens{B\otimes C\ell_{0,s}}P^{0,s-1}})\)
anti-commuting with \(J_s\otimes j_1\). By the inductive assumption, the
\(b\in\sh F(\smash{P^{0,s-1}\Parens{B\otimes C\ell_{0,s}}P^{0,s-1}})\)
are in bijection \emph{via} \(b=(J\otimes j_1)P^{0,s-1}\) with the real
Hermitian unitaries \(J\in B\) anti-commuting with
\(J_1,\dotsc,J_{s-1}\). The \(b\) anti-commuting with \(J_s\otimes j_1\)
are in bijection with the \(J\) anti-commuting with \(J_1,\dotsc,J_s\).

This proves the statement for \((0,s)\). If now \(r\sge1\), then
similarly \[
    P^{r,s}\Parens{B\otimes C\ell_{r,s+1}}P^{r,s}\cong Q_r\Parens{B''\mathop{\widehat\otimes}C\ell_{1,0}}Q_r
\] where \[
    B''\defi P^{r-1,s}\Parens{B\otimes C\ell_{r-1,s+1}}P^{r-1,s}.
\] Here, the \(C\ell_{1,0}\) factor on the right-hand side of the former
equation is generated by \(k_r\), whereas the \(C\ell_{r-1,s+1}\) factor
in the latter is generated by \(k_1,\dotsc,k_{r-1},j_1,\dotsc,j_{s+1}\).
The assertion now follows by induction on \(r\), applying the second
part of Lemma \ref{lem:corner-orhu} to \(C=B''\) and
\(x=(-1)^sK_r\otimes j_1\).
\EndKnitrBlock{proof}

In order to see that the ORHU defined by Proposition \ref{prp:iqpv-orhu}
give classes in the correct \(K\)-theory, we need the following lemma.

\BeginKnitrBlock{lemma}
\protect\hypertarget{lem:corner-morita}{}{\label{lem:corner-morita} }As
before, assume given a unital real \(*\)-morphism
\(\phi:C\ell_{r,s}\longrightarrow B\). In addition, suppose there is
some \(J_{\mathrm{ref}}\in\FF^{r,s}(B)\). Then the graded real
C\(^*\)-algebra \(B\otimes C\ell_{r,s+1}\) is isomorphic to the graded
real C\(^*\)-algebra of \(2^{r+s}\times 2^{r+s}\) matrices over
\(P^{r,s}(B\otimes C\ell_{r,s+1})P^{r,s}\) with entry-wise grading.
\EndKnitrBlock{lemma}

\BeginKnitrBlock{proof}
{}We will make a specific choice of isomorphism
that will allow us to compute its effect on \(K\)-theory. We shall index
the rows and columns of \(2^{r+s}\times2^{r+s}\) matrices by indices
\(\eps\) running over \(\{\pm\}^{r+s}\). Set
\(Q_\alpha^+\defi Q_\alpha\) and \(Q_\alpha^-\defi1-Q_\alpha\), and
similarly for \(P^a\). This enables us to define \[
    P^{r,s}_\eps\defi Q_1^{\eps_1}\dotsm Q_r^{\eps_r}P_1^{\eps_{r+1}}\dotsm P_s^{\eps_{r+s}}.
\] The projections \(P^{r,s}_\eps\) satisfy \begin{equation}
    P^{r,s}_{\eps^{\phantom\prime}} P^{r,s}_{\eps'}=\delta_{\eps\eps'}P^{r,s}_\eps,\quad\sum_{\eps\in\{\pm\}^{r+s}}P^{r,s}_\eps=1.
    \label{eq:prs-rel}
\end{equation} The projections \(P^{r,s}_\eps\) are all unitarily
equivalent. We make a careful choice of unitaries implementing the
equivalence. Observe first that \[
    \left.
    \begin{aligned}
        \Ad(K_aJ_\mathrm{ref}\otimes 1)(P^{r,s})&=P^{r,s}_{\eps-2e_a}\\
        \Ad(J_\alpha J_\mathrm{ref}\otimes 1)(P^{r,s})&=P^{r,s}_{\eps-2e_{r+\alpha}}
    \end{aligned}
    \right\}\quad(a=1,\dotsc,r,\alpha=1,\dotsc,s).
\] Hence, setting \[
    u_\eps\defi\prod_{1\sle a\sle r,\eps_a=-}(K_aJ_\mathrm{ref}\otimes 1)\prod_{1\sle\alpha\sle s,\eps_{r+\alpha}=-}(J_\alpha J_\mathrm{ref}\otimes 1),
\] we obtain \[
    u_\eps P^{r,s}=P^{r,s}_\eps u_\eps,\quad u_\eps(J_\mathrm{ref}\otimes j_1)=\eps(J_\mathrm{ref}\otimes j_1)u_\eps
\] where we follow the convention \(\eps\equiv\sgn\eps\). Define a map
\[
    \Psi:B\otimes C\ell_{r,s+1}\longrightarrow M_{2^{r+s}}(\cplxs)\otimes P^{r,s}(B\otimes C\ell_{r,s+1})P^{r,s}
\] by \[
    \Psi(x)\defi\sum_{\eps,\eps'\in\{\pm\}^{r+s}}E_{\eps\eps'}\otimes P^{r,s}u_{\eps^{\phantom\prime}}^*xu_{\eps'}^{\phantom*}P^{r,s}=\sum_{\eps,\eps'\in\{\pm\}^{r+s}}E_{\eps\eps'}\otimes u_{\eps^{\phantom\prime}}^*P^{r,s}_{\eps^{\phantom\prime}} xP^{r,s}_{\eps'}u_{\eps'}^{\phantom*}.
\] Here, \(E_{\eps\eps'}\) are the standard matrix units. It is clear
that \(\Psi\) commutes with \(*\) and \(\overline\cdot\). A computation
using Equation \eqref{eq:prs-rel} shows that \(\Psi\) is in fact a real
\(*\)-morphism. Since the matrix units form a basis of
\(M_{2^{r+s}}(\cplxs)\), it is easy to see that \(\Psi\) is bijective.
\EndKnitrBlock{proof}

\BeginKnitrBlock{remark}
{}The result of Lemma \ref{lem:corner-morita}
is not surprising, as by \autocite[Example 3.6 and Proposition
3.28]{raeburn-williams}, the corner
\(P^{r,s}(B\otimes C\ell_{r,s+1})P^{r,s}\) is Morita equivalent to the
closed two-sided ideal generated by \(P^{r,s}\). One can show that this
ideal is already all of \(B\otimes C\ell_{r,s+1}\). Since \(B\) is
unital by hypothesis, the Brown--Green--Rieffel Theorem
\autocite[Theorem 5.55]{raeburn-williams} applies, and the Morita
equivalence implies stable isomorphism.

However, the more precise statement of the lemma will be useful in order
to express the bulk classes that we will presently construct in an
explicit form.

Circumstantial evidence suggests that Lemma \ref{lem:corner-morita}
holds in the absence of the assumption that there is an element
\(J_{\mathrm{ref}}\), although our proof does not work in this case.
Indeed, if \(\phi\) is injective, then the minimal choice for \(B\) is
\(B=C\ell_{r,s}\). For \(r=0\), \(B=C\ell_{0,s}\), one has \[
    B\otimes C\ell_{0,s+1}\cong M_{2^s}(E),\quad E_\reals=
    \begin{cases}
        \cplxs,&\text{if }s\equiv 0,1\,(4),\\
        \reals\oplus\reals,&\text{if }s\equiv 2,3\,(4).
    \end{cases}
\] For the cases of \(s-r\equiv 3\,(4)\), \(\phi\) may fail to be
injective. (Otherwise, \(C\ell_{r,s}\) is simple.) Setting \(r=0\), we
have \(C\ell_{0,s}=C\ell_{0,s-1}\oplus C\ell_{0,s-1}\), so the minimal
choice for \(B\) (without the assumption that \(\phi\) be injective) is
\(C\ell_{0,s-1}\), which is \(\mathbb H_\cplxs\) for \(s=3\) and
\(M_8(\cplxs)\) for \(s=7\). In both cases, one has
\(B\otimes C\ell_{0,s+1}\cong M_{2^s}(\cplxs)\), and \(\cplxs\) is a
corner of the tensor product.
\EndKnitrBlock{remark}

\BeginKnitrBlock{definition}[Bulk class of a disordered IQPV with symmetries]
\protect\hypertarget{def:bulk-class}{}{\label{def:bulk-class} {} }We assume that
\(\FF^{r,s}(B)\neq\vvoid\) and select
\(J_{\mathrm{ref}}\in\FF^{r,s}(B)\). Let
\(e^{r,s}\defi(J_{\mathrm{ref}}\otimes j_1)P^{r,s}\) be the ORHU
corresponding to \(J_{\mathrm{ref}}\) \emph{via} Proposition
\ref{prp:iqpv-orhu}. To any \(J\in\FF^{r,s}(B)\), we assign the class \[
    [(J;\phi)]\defi\Bracks{(J\otimes j_1)P^{r,s}}-\Bracks{e^{r,s}}\in DKR_{e^{r,s}}(P^{r,s}(B\otimes C\ell_{r,s+1})P^{r,s}). 
\] By Lemma \ref{lem:corner-morita}, Lemma \ref{lem:dkr-stable}, and
Theorem \ref{thm:DKR-KR}, we may consider this as a class in \[
    DKR(B\otimes C\ell_{r,s+1})=KR_{s-r+2}(B).
\] In case \(B=\mathbb A\) or \(B=\mathrm A\) and the \(K_a,J_\alpha\)
are operators on \(W\) resp. on \(V\), we say that
\([(J;\phi)]=[(J;K_1,\dotsc,K_r,J_1,\dotsc,J_r)]\) is the \emph{bulk
class} associated with the disordered IQPV \((J;\phi)\) of symmetry
index \((r,s)\).
\EndKnitrBlock{definition}

By inspecting the proof of Lemma \ref{lem:corner-morita}, we may
determine the class in \(DKR(B\otimes C\ell_{r,s+1})\) corresponding to
\([(J;\phi)]\) explicitly.

\BeginKnitrBlock{lemma}
\protect\hypertarget{lem:bulk-explicit}{}{\label{lem:bulk-explicit} }Suppose
there is some \(J_{\mathrm{ref}}\in\FF^{r,s}(B)\). Then the class in
\(DKR_{J_{\mathrm{ref}}\otimes j_1}(B\otimes C\ell_{r,s+1})\)
corresponding to \([(J;\phi)]\) under the isomorphisms from Lemmas
\ref{lem:dkr-stable} and \ref{lem:corner-morita} is \begin{equation}
    [(J\otimes j_1)P^{r,s}+(J_{\mathrm{ref}}\otimes j_1)(1-P^{r,s})]-[J_{\mathrm{ref}}\otimes j_1].
\label{eq:bulk-explicit}
\end{equation}
\EndKnitrBlock{lemma}

\BeginKnitrBlock{proof}
{}Unless \(r+s>0\), there is nothing to prove.
Set \(n\defi 2^{r+s-1}\). We abbreviate \(e_0\defi e^{r,s}\),
\(p\defi P^{r,s}\), and \(C\defi B\otimes C\ell_{r,s+1}\). Then
\(e\defi e_0\oplus -e_0\) is homotopic to \(-e_0\oplus e_0\) in
\(\sh F_2(pCp)\), by \autocite[p.~191]{van_Daele1}, and hence,
\(F_{\infty}(M_2(pCp))\) is an Abelian group isomorphic to \(DKR(pCp)\)
\autocite[Proposition 2.11]{van_Daele1}. Under the isomorphism,
\([(J\otimes j_1)P]-[e_0]\) corresponds to the class \([x]\) of the
element \(x\defi (J\otimes j_1)P\oplus(-e_0)\).

The map from Lemma \ref{lem:dkr-stable} sends \(x\) to
\(x\oplus e_{n-1}\in\sh F_{2n}(pBp)\). This is the \(2n\times2n\) matrix
\[
    E_{++}\otimes(J\otimes j_1)p+\sum_{\eps\neq(+,\dotsc,+)}\eps E_{\eps\eps}\otimes e_0,
\] where \(E_{++}\) is the matrix unit corresponding to
\(\eps=(+,\dotsc,+)\).

Let \(\Psi\) be the isomorphism from Lemma \ref{lem:corner-morita}. As
\(J_\mathrm{ref}\otimes j_1\) commutes with \(p=P^{r,s}\) and
\(\Ad(u_\eps)(J_\mathrm{ref}\otimes j_1)=\eps(J_\mathrm{ref}\otimes j_1)\),
we find \[
    \Psi(J_\mathrm{ref}\otimes j_1)=\sum_{\eps,\eps'\in\{\pm\}^{r+s}}\eps'E_{\eps\eps'}\otimes(J_\mathrm{ref}\otimes j_1)u_{\eps^{\phantom\prime}}^*P^{r,s}_{\eps^{\phantom\prime}} P^{r,s}_{\eps'}u_{\eps'}^{\phantom*}=\sum_{\eps\in\{\pm\}^{r+s}}\eps E_{\eps\eps}\otimes(J_\mathrm{ref}\otimes j_1)P^{r,s}.
\] Thus we see that \(\Psi(J_{\mathrm{ref}}\otimes j_1)=e_n\). A similar
computation shows that \[
    \Psi\Parens{(J\otimes j_1)p+(J_{\mathrm{ref}}\otimes j_1)(1-p)}=x\oplus e_{n-1}.
\] Therefore, \(x\oplus e_{n-1}\) is identified with the desired element
in \(M_{2n}(C)\). Because the class
\([e_n]=[J_{\mathrm{ref}}\otimes j_1]\) is zero in the Abelian group
\(F_\infty(M_{2n}(C))\), the assertion follows.
\EndKnitrBlock{proof}

The assignment to \(J\in\FF^{r,s}(B)\) of the \(DKR\)-class of the ORHU
\((J\otimes j_1)P^{r,s}\) in fact gives an alternative definition of
real \(K\)-theory. To that end, let \[
    \FF^{r,s}_n(B)\defi\FF^{r,s}(M_n(B)),
\] where the right-hand side is defined in terms of the elements
\(1_n\otimes K_a\), \(a=1,\dotsc,r\), and \(1_n\otimes J_\alpha\),
\(\alpha=1,\dotsc,s\). The sets \(\FF^{r,s}_n(B)\) form an inductive
system \emph{via} \[
    \FF^{r,s}_n(B)\longrightarrow\FF^{r,s}_{n+1}(B):J\longmapsto J\oplus J_{\mathrm{ref}}\defi
    \begin{pmatrix}
        J&0\\0&J_{\mathrm{ref}}
    \end{pmatrix}.
\] Denote the set of path-connected components in the norm topology of
\(\FF^{r,s}_n(B)\) by \[
    FF^{r,s}_n(B)\defi\pi_0\Parens{\FF^{r,s}_n(B)}.
\] For the induced maps, the sets \(FF^{r,s}_n(B)\) form an inductive
system of sets.

\BeginKnitrBlock{theorem}[Real K-theory via IQPV]
\protect\hypertarget{thm:iqpv-kthy}{}{\label{thm:iqpv-kthy} {} }Let \(B\) be an ungraded unital real
C\(^*\)-algebra. Let \(r,s\sge0\) and
\((K_1,\dotsc,K_r,J_1,\dotsc,J_s)\) represent a unital real
\(*\)-morphism \(C\ell_{r,s}\longrightarrow B\). Suppose that
\(\FF^{r,s}(B)\neq\vvoid\) and select
\(J_{\mathrm{ref}}\in\FF^{r,s}(B)\).

Then the inductive limit \[
    FF^{r,s}_\infty(B)\defi\varinjlim\nolimits_nFF^{r,s}_n(B)
\] acquires the structure of an Abelian monoid with addition induced by
the maps \begin{equation}
    \FF^{r,s}_m(B)\times\FF^{r,s}_n(B)\longrightarrow\FF^{r,s}_{m+n}(B):(J,J')\longmapsto J\oplus J'=
    \begin{pmatrix}
        J&0\\0&J'
    \end{pmatrix}
    \label{eq:ff-monoid}
\end{equation} and unit element induced by \(J_{\mathrm{ref}}\).
Moreover, \(J\mapsto(J\otimes j_1)P^{r,s}\) induces an isomorphism of
Abelian groups \[
    G\Parens{FF^{r,s}_\infty(B)}\longrightarrow DKR_{e^{r,s}}(P^{r,s}(B\otimes C\ell_{r,s+1})P^{r,s})=KR_{s-r+2}(B),
\] where we consider the reference ORHU
\(e^{r,s}=(J_{\mathrm{ref}}\otimes j_1)P^{r,s}\).

In case \(B\) is not assumed real, we have a corresponding isomorphism\\
\[
    G\Parens{FF^{r,s}_\infty(B)}\longrightarrow DK_{e^{r,s}}(P^{r,s}(B\otimes C\ell_{r,s+1})P^{r,s})=K_{s-r+2}(B).
\]
\EndKnitrBlock{theorem}

\BeginKnitrBlock{proof}
{}Note that \[
    M_n\Parens{P^{r,s}(B\otimes C\ell_{r,s+1})P^{r,s}}=(1_n\otimes P^{r,s})(M_n(B)\otimes C\ell_{r,s+1})(1_n\otimes P^{r,s}).
\] We compute for all \(J\in\FF^{r,s}_n(B)\) that \[
    \Parens{(J\oplus J_{\mathrm{ref}})\otimes j_1}(1_{n+1}\otimes P^{r,s})=(J\otimes j_1)(1_n\otimes P^{r,s})\oplus e^{r,s}.
\] Hence, the map from Proposition \ref{prp:iqpv-orhu} defines an
isomorphism of the inductive systems \((\FF^{r,s}_n(B))_{n\sge 1}\) and
\((\sh F_n(P^{r,s}(B\otimes C\ell_{r,s+1})P^{r,s}))_{n\sge1}\), where
the latter is defined by the reference ORHU \(e^{r,s}\).

Inspecting the proof of Proposition \ref{prp:iqpv-orhu}, the map defined
there is a homeomorphism with respect to the topologies induced by the
norm. Indeed, in Lemma \ref{lem:corner-orhu}, we have \[
    C\mathop{\widehat\otimes}C\ell_{0,1}=C\oplus Cj_1
\] as normed vector spaces, and the projection onto the first factor,
which is continuous, defines the inverse of the first bijection
described there. The same holds for \(C\ell_{1,0}\) in place of
\(C\ell_{0,1}\). Thus, the map from Proposition \ref{prp:iqpv-orhu} is a
homeomorphism.

Hence, the inductive systems \((FF^{r,s}_n(B))_{n\sge1}\) and
\((F_n(P^{r,s}(B\otimes C\ell_{r,s+1})P^{r,s}))_{n\sge1}\) are
isomorphic, so that we obtain a bijection \[
    FF^{r,s}_\infty(B)\longrightarrow F_\infty\Parens{P^{r,s}(B\otimes C\ell_{r,s+1})P^{r,s}}. 
\]

In order to see that the left-hand side is an Abelian monoid and this
bijection is an isomorphism of such objects, we study the effect of the
map from Proposition \ref{prp:iqpv-orhu} on the \(\oplus\) operation
from Equation \eqref{eq:ff-monoid}. We compute for \(J\in\FF^{r,s}_m(B)\)
and \(J'\in\FF^{r,s}_n(B)\): \[
    \Parens{(J\oplus J')\otimes j_1}(1_{m+n}\otimes P^{r,s})=(J\otimes j_1)(1_m\otimes P^{r,s})\oplus(J'\otimes j_1)(1_n\otimes P^{r,s}).
\] Since \(F_\infty(P^{r,s}(B\otimes C\ell_{r,s+1})P^{r,s})\) is an
Abelian monoid and its Grothendieck group is
\(DKR_{e^{r,s}}(P^{r,s}(B\otimes C\ell_{r,s+1})P^{r,s})\), we have
proved the claim.
\EndKnitrBlock{proof}

\hypertarget{sec:bdy}{%
\section{The view from the boundary}\label{sec:bdy}}

In this section, we will construct a framework for the topological
classification of boundaries of topological insulators and
superconductors in terms of \(KR\)-theory. At an interface between bulk
systems with different topological invariants, the gap closes, thus
defining a gapless boundary system. (Indeed, as we shall see in Section
\ref{sec:discussion}, this is in a mathematically precise sense a
necessary condition for topological non-triviality.)

For this reason, the construction of bulk invariants in Van Daele's
picture of real \(K\)-theory, as developed in the previous section,
cannot work at the boundary. The solution for this problem is two-fold:
Gapless boundary systems need a gap in the bulk in order to be
topologically protected, so we will always consider \emph{the boundary
together with the bulk}. This shall be accomplished by introducing a
\emph{half-space algebra}, as first suggested by
Kellendonk--Richter--Schulz-Baldes \autocite{skr00,krs02}. Secondly, we
shall apply an alternative picture of real \(K\)-theory which allows for
the closing of the gap at the boundary.

\hypertarget{the-sequence-connecting-bulk-and-boundary}{%
\subsection{The sequence connecting bulk and
boundary}\label{the-sequence-connecting-bulk-and-boundary}}

We introduce a boundary into our system by considering the half-space
semilattice \[
  \hat\Lambda\defi\Lambda_\partial\times\nats,\quad\Lambda_\partial\defi\ints^{d-1},\quad\nats=\{0,1,2,\dotsc\},
\] in which the translational symmetry is broken in the \(d\)th
coordinate direction. The aim of this section is to exhibit a short
exact sequence of real C\(^*\)-algebras connecting bulk and boundary, as
follows: \begin{equation}
\begin{tikzcd}
  0\rar{}&\mathbb A_\partial\otimes\mathbb K(\ell^2(\mathbb N))\rar{\iota}&\widehat{\mathbb A}\rar{\varrho}&\mathbb A\rar{}&0.
\end{tikzcd}
\label{eq:bb-ses}
\end{equation} Here, \(\mathbb A_\partial\) is defined just as
\(\mathbb A\), with \(\Lambda_\partial\) replacing \(\Lambda\), and
\(\widehat{\mathbb A}\) is the so-called \emph{half-space algebra}, we
which now proceed to construct.

We define Hilbert spaces on \(\hat\Lambda\) by setting
\(\widehat{\sh V}\defi\ell^2(\hat\Lambda)\otimes V\) and \[
  \widehat{\sh W}\defi\widehat{\sh V}\oplus\smash{\widehat{\sh V}}^*\cong\ell^2(\hat\Lambda)\otimes W.
\] Of basic importance is the orthogonal projection (called the
\emph{Szegő projection}) \[
  q:\ell^2(\Lambda)\longrightarrow\ell^2(\hat\Lambda)\subseteq\ell^2(\Lambda).
\] Here, we identify \(\ell^2(\hat\Lambda)\) with the set of all
square-summable sequences \(\psi\in\ell^2(\Lambda)\) which vanish
outside \(\smash{\hat\Lambda}\). Implicitly tensoring with \({\id}_V\)
or \({\id}_W\), we may consider \(q\) as an orthogonal projection
defined on \(\sh V\) or \(\sh W\).

The translation \(u_i\defi u_{e_i}\), where \(e_i\) is the \(i\)th
standard basis vector, defined by \[
  \Parens{u_x\psi}(y)\defi\psi(y-x)\quad(x,y\in\Lambda,\psi\in\ell^2(\Lambda)),
\] commutes with \(q\) if \(i<d\). For \(i=d\), we obtain the partial
isometry \[
  \hat u_d\defi qu_dq\in\blop(\ell^2(\hat\Lambda)).
\] Then \(\hat u_d\) is characterised by the fact that
\(\hat u_d^*\hat u_d^{\phantom*}=1\) and that
\(1-\hat u_d^{\phantom*}\hat u_d^*\) is the projection onto the space
\(\ell^2(\Lambda_\partial)\) of states supported at the boundary. In
other words, \(\hat u_d\) is the \emph{unilateral shift} in the \(d\)th
coordinate direction. We are now able to give a definition for the
half-space algebra.

\BeginKnitrBlock{definition}[Half-space covariance algebra]
\protect\hypertarget{def:half-covalg}{}{\label{def:half-covalg}
\iffalse (Half-space covariance algebra) \fi{} }Let \(\Omega\) be a
compact Hausdorff space, equipped with a continuous right action of the
group \(\Lambda\). Let \(U\) be a finite-dimensional Hilbert space and
let \(\hat u_x\), for \(x\in\hat\Lambda\), be defined by \[
  (\hat u_x\psi)(y)\defi\Parens{(qu_xq\otimes{\id}_U)\psi}(y)\quad( x,y\in\hat\Lambda,\psi\in\ell^2(\hat\Lambda)\otimes U).
\] Define \(\smash{\widehat A}^U\) to be the closed subalgebra of the
set \(\sh C(\Omega,C^*_u(\hat\Lambda,U))\) of norm continuous maps
\(\Omega\longrightarrow C^*_u(\hat\Lambda,U)\) generated by those maps
\(O\) that are \emph{covariant} in the sense that \[
  O_{\omega\cdot x}=\hat u_x^*O_\omega\hat u_x^{\phantom*}\quad(x\in \hat\Lambda,\omega\in\Omega). 
\]
\EndKnitrBlock{definition}

Again, \(\smash{\widehat A}^U\) is a C\(^*\)-algebra with point-wise
operations and the sup norm. Whenever \(U\) is real or quaternionic,
then \(\smash{\widehat A}^U\) is real. We define \[
  \widehat{\mathbb A}\defi\smash{\widehat A}^W,\quad\widehat{\mathrm A}\defi\smash{\widehat A}^V.
\]

In order to exhibit \(\mathbb A\) as a quotient of
\(\widehat{\mathbb A}\), we need to prove that the latter has a
crossed-product-like structure. To that end, we pose the following
definition.

\BeginKnitrBlock{definition}
\protect\hypertarget{def:half-covpair}{}{\label{def:half-covpair} }Let
\(B\defi \sh C(\Omega)\otimes\endo W\). Consider a tuple
\((\phi,\hat V)\), where \(\phi:B\longrightarrow C\) is a \(*\)-morphism
to a real C\(^*\)-algebra \(C\) and \(\hat V_x\in\mathrm M(C)\) (the
multiplier algebra), \(x\in\hat\Lambda\), are real elements. We call
\((\phi,\hat V)\) a \emph{real covariant pair} if \[
  \hat V^{\phantom*}_x\hat V_x^*=1,\quad \hat V_x\hat V_y=\hat V_{x+y},\quad\phi(\alpha_x(f))\hat V_x=\hat V_x\phi(f)\quad(x,y\in\hat\Lambda,f\in B).
\] (Recall from Equation \eqref{eq:alpha-def} that
\(\alpha_x(f)(\omega)=f(\omega\cdot x)\).) For
\(B=\sh C(\Omega)\otimes\endo V\), we define \emph{covariant pairs} by
dropping the reality constraints from the above definitions.
\EndKnitrBlock{definition}

\BeginKnitrBlock{remark}
{}Observe that the equation
\(\hat V\hat V^*=1\) for the partial isometries \(\hat V=\hat V_x\) in
the definition of (real) covariant pairs differs from the equation
\(\hat u^*\hat u=1\) for the unilateral shift \(\hat u=\hat u_d\). The
rationale here is the unilateral shift is a truncated left translation,
whereas the \(\hat V\) are truncated right translations: Operators that
\emph{commute} with (truncated) left translations are \emph{generated}
by (truncated) right translations.
\EndKnitrBlock{remark}

\BeginKnitrBlock{proposition}
\protect\hypertarget{prp:halfsp-covcrossed}{}{\label{prp:halfsp-covcrossed}
}Define \(\widehat R_x\defi qR_xq\) for \(x\in\hat\Lambda\) (where
\(R_x\) is right translation by \(x\), as in the proof of Theorem
\ref{thm:covcrossed}) and \[
  (\hat\pi(f)_\omega\psi)(x)=f(\omega\cdot(-x))\psi(x)\quad(f\in\sh C(\Omega)\otimes\endo W,x\in\hat\Lambda,\psi\in\widehat{\sh W}).
\] Then \((\hat\pi,\widehat R)\) is a real covariant pair, and together
with it, the real C\(^*\)-algebra \(\widehat{\mathbb A}\) is universal
for real covariant pairs. That is, for any real covariant pair
\((\phi,\hat V)\), there is a unique real \(*\)-morphism
\(\smash{\widehat{\mathbb A}}\longrightarrow C\), denoted by
\(\phi\rtimes_{\alpha}\hat V\), such that \[
  (\phi\rtimes_{\alpha}\hat V)(\hat\pi(f))=\phi(f)\quad(\phi\rtimes_\alpha\hat V)(\smash{\widehat R}_x)=\hat V_x\quad(f\in\sh C(\Omega)\otimes\endo W,x\in\hat\Lambda).
\] Similarly, the C\(^*\)-algebra \(\widehat{\mathrm A}\) is universal
for covariant pairs.
\EndKnitrBlock{proposition}

\BeginKnitrBlock{proof}
{}We limit ourselves to the real case. Let the
C\(^*\)-algebra \(\tilde A\) be defined by the generators
\(f\in\sh C(\Omega)\otimes\endo W\) and \(\hat\tau_x\), for
\(x\in\hat\Lambda\), by the \(*\)-algebraic relations of
\(\sh C(\Omega)\otimes\endo W\), by the algebraic relations of
\(\smash{\hat\Lambda}\), and by the relations \[
  \hat\tau_x^{\phantom*}\hat\tau_x^*=1,\quad \hat\tau_x\hat\tau_y=\hat\tau_{x+y},\quad \alpha_x(f)\hat\tau_x=\hat\tau_xf\quad(x,y\in\hat\Lambda,f\in\sh C(\Omega)\otimes\endo W).
\] Then \(\tilde A\) is a real C\(^*\)-algebra if we require
\(\sh C(\Omega)\otimes\endo W\) to be a real subalgebra and the
generators \(\hat\tau_x\), \(x\in\smash{\hat\Lambda}\), to be real. By
definition, \(\tilde A\) is universal for real covariant pairs.

By restriction to \(\hat\Lambda\) of the real covariant pair \((\pi,R)\)
for the action of the group \(\Lambda\) on
\(\sh C(\Omega)\otimes\endo W\), we already know that
\((\hat\pi,\smash{\widehat R})\) is a real covariant pair. Hence, we
obtain a unique real \(*\)-morphism \[
  \hat\Phi\defi\hat\pi\rtimes_{\alpha}\smash{\widehat R}:\tilde A\longrightarrow\blop(L^2(\Omega,P)\otimes\widehat{\sh W}),
\] where \(P\) is the probability measure on \(\Omega\) from the proof
of Theorem \ref{thm:covcrossed}.

Let \(\tilde A_\partial\) denote the universal C\(^*\)-algebra generated
by the \(f\in\sh C(\Omega)\otimes\endo W\) and \(\tau_x\), for
\(x\in\Lambda_\partial\), subject to the C\(^*\)-algebraic relations of
\(\sh C(\Omega)\otimes\endo W\), the algebraic relations of
\(\Lambda_\partial\), and the relations \[
  \tau_x^*\tau_x^{\phantom*}=\tau_x^{\phantom*}\tau_x^*=1,\quad\tau_x\tau_y=\tau_{x+y},\quad\alpha_x(f)\tau_x=\tau_xf\quad(x,y\in\Lambda_\partial,f\in\sh C(\Omega)\otimes\endo W).
\] There is a canonical \(*\)-morphism
\(\Phi_\partial:\tilde A_\partial\longrightarrow \tilde A\), defined by
\(\tau_x\longmapsto\hat\tau_x\) and \(f\longmapsto f\). By Theorem
\ref{thm:covcrossed}, applied to \(\Lambda_\partial\) instead of
\(\Lambda\), \(\smash{\widehat\Phi\circ\Phi_\partial}\) is injective.
Moreover, \(\tilde A_\partial\) is a closed \(*\)-subalgebra of
\(\tilde A\).

We now apply Murphy's results on crossed products by ordered groups
\autocite{murphy-orderedcrossed}. By definition, \(\tilde A\) satisfies
the universality for covariant pairs stated in \autocite[Proposition
1.1]{murphy-orderedcrossed} for Murphy's crossed product
\(\tilde A_\partial\rtimes\nats\), and hence, \(\tilde A\) is
canonically isomorphic to that crossed product. Since
\(\smash{\widehat\Phi\circ\Phi_\partial}\) is faithful,
\autocite[Theorem 4.4]{murphy-orderedcrossed} applies, and it follows
that the real \(*\)-morphism \(\smash{\widehat\Phi}\) is injective.

It remains to be shown that \(\im\widehat\Phi=\widehat{\mathbb A}\). To
that end, we refer back to the proof of Theorem \ref{thm:covcrossed}.
The statement that
\(\im\smash{\widehat\Phi}\subseteq\smash{\widehat{\mathbb A}}\) follows
in much the same way as the corresponding statement for \(\mathbb A\).
For the converse inclusion, we begin with some preparations which are
largely similar to the corresponding steps in the bulk: Recall the
definition of the maps \(S_n\) and the functions \(f^i_n\); since we may
consider \(\blop(\smash{\widehat{\sh W}})\) as a subset of
\(\blop(\sh W)\) by the use of the projection \(q\), we may restrict
\(S_n\) to \(\blop(\smash{\widehat{\sh W}})\). For
\(T\in\blop(\smash{\widehat{\sh W}})\), we find that \[
  \widehat S_n(T)\defi S_n(T)=\sum_iM_{\smash{\hat f}_n^i}^*TM_{\smash{\hat f}_n^i}^{\phantom*},\quad\smash{\hat f}_n^i\defi f_n^i|_{\hat\Lambda}\quad(T\in\blop(\widehat{\sh W})),
\] where \(M_f\) is the operator of multiplication by \(f\). In
particular, \(\widehat S_n\) is a completely positive endomorphism of
\(\blop(\smash{\widehat{\sh W}})\). Hence, as for \(S_n\),
\(\smash{\widehat S_n}\) is a contraction. We see as before that \[
  \widehat S_n(T)=\sum_{x,y\in\hat\Lambda}\phi_n(x-y)T(x,y)\ket x\bra y\quad(T\in\blop(\widehat{\sh W})).
\] Then \(\widehat S_n\) leaves controlled operators on
\(\ell^2(\hat\Lambda)\) invariant, and therefore also the algebra
\(C^*_u(\hat\Lambda,W)\). Defining, for \(T\in\widehat{\mathbb A}\),
\(\widehat S_n(T)\) by
\(\widehat S_n(T)_\omega\defi\widehat S_n(T_\omega)\) for all
\(\omega\in\Omega\), we obtain a contractive, completely positive
endomorphism of \(\smash{\widehat{\mathbb A}}\). As previously, it
follows that \(S_n(T)\) converges to \(T\) for any
\(T\in\smash{\widehat{\mathbb A}}\).

The remaining steps differ from our previous considerations. Let
\(T:\Omega\longrightarrow C^*_u(\hat\Lambda,W)\) be norm continuous and
satisfy the covariance condition in Definition \ref{def:half-covalg}.
Setting \[
  x\sle y\quad:\Longleftrightarrow\quad y-x\in\hat\Lambda
\] defines a total preorder on \(\hat\Lambda\). Therefore, we may write
\(\widehat S_n(T)=\mathrm I+\mathrm{I\!I}\) where \[
  \mathrm I_\omega=\sum_{x,y\in\hat\Lambda,x\sle y}\phi_n(x-y)T_\omega(x,y)\ket x\bra y
\] and the sum defining \(\mathrm{I\!I}\) extends over all
\(x,y\in\hat\Lambda\), \(x>y\). For \(x\sle y\), we have \[
  \bra x\smash{\widehat R}_{y-x}=\bra y,
\] so that \[
  \mathrm I_\omega=\sum_{x,y\in\hat\Lambda,x\sle y}\phi_n(x-y)T_{\omega\cdot x}(0,y-x)\ket x\bra y=\sum_{g\in\hat\Lambda\cap\supp\phi_n}\pi(f^\mathrm I_{n,g})_\omega\smash{\widehat R}_g
\] where \[
  f^\mathrm I_{n,g}(\omega)\defi\phi_n(g)T_\omega(0,g)\quad(n\sge 0,g\in\hat\Lambda,\omega\in\Omega).
\] Similarly, for \(x>y\), we have \[
  \Parens{\smash{\widehat R}_{x-y}}^*w\ket y=w\ket x\quad(w\in W),
\] so that \[
  \mathrm{I\!I}_\omega=\sum_{x,y\in\hat\Lambda,x>y}\phi_n(x-y)T_{\omega\cdot y}(x-y,0)\ket x\bra y=\sum_{g\in(\hat\Lambda\setminus\Lambda_\partial)\cap(-\supp\phi_n)}\Parens{\smash{\widehat R}_g}^*\pi(f^{\mathrm{I\!I}}_{n,g})_\omega
\] where \[
  f^{\mathrm{I\!I}}_{n,g}(\omega)\defi\phi_n(-g)T_\omega(g,0)\quad(n\sge 0,g\in\hat\Lambda,\omega\in\Omega).
\] This proves that \(\widehat S_n(T)\in\im\widehat\Phi\); since
\(\im\widehat\Phi\) is closed, it also contains \(T\), proving that
\(\widehat\Phi\) is surjective, and hence, the assertion, as
\(\tilde A\) satisfies the claimed universality.
\EndKnitrBlock{proof}

The following theorem constructs the desired sequence \eqref{eq:bb-ses}.

\BeginKnitrBlock{theorem}
\protect\hypertarget{thm:bb-ses}{}{\label{thm:bb-ses} }There is a unique
surjective real \(*\)-morphism
\(\vrho:\widehat{\mathbb A}\longrightarrow\mathbb A\) such that \[
  \vrho(\hat\pi(f))=\pi(f),\quad \vrho(\smash{\widehat R}_x)=R_x\quad(f\in\sh C(\Omega)\otimes\endo W,x\in\hat\Lambda).
\] The kernel of \(\vrho\) is the closed ideal \((e)\) generated by the
projection \(e\defi 1-(\smash{\widehat R}_d)^*\smash{\widehat R}_d\),
where \(R_d\defi R_{e_d}\). It is isomorphic as a real C\(^*\)-algebra
to \(\mathbb A_\partial\otimes\knums(\ell^2(\nats))\). The same holds
for \(\smash{\widehat{\mathrm A}}\), with \(W\) replaced by \(V\), and
without the reality constraints.
\EndKnitrBlock{theorem}

\BeginKnitrBlock{proof}
{}As before, we limit our detailed proof to the
real case. The unique existence and the surjectivity of \(\vrho\) follow
directly from Proposition \ref{prp:halfsp-covcrossed} and Theorem
\ref{thm:covcrossed}. It is also clear from the definitions that the
projection \(e\) is contained in the kernel of \(\vrho\).

We abbreviate \(\widehat S\defi\hat u_d\) and
\(\hat e\defi 1-\widehat S\smash{\widehat S}^*\), the orthogonal
projection onto \(\ell^2(\Lambda_\partial)\). By Theorem
\ref{thm:covcrossed}, there is a unique real \(*\)-morphism
\(\phi:\mathbb A\longrightarrow\mathbb A\otimes M_2(\blop(\ell^2(\nats)))\),
such that \[
  \left.
  \begin{aligned}
    \phi(\pi(f))&\defi\pi(f)\otimes\begin{pmatrix}1&0\\0&1\end{pmatrix}\\
    \phi\bigl(R_{(x,n)}\bigr)&\defi R_{(x,n)}\otimes\begin{pmatrix}\smash{\widehat S}^*&0\\\hat e&\widehat S\end{pmatrix}^n
  \end{aligned}
  \right\}\quad(f\in\sh C(\Omega)\otimes\endo W,x\in\ints^{d-1},n\in\ints).
\] In fact, observe that \[
  \begin{pmatrix}\smash{\widehat S}^*&0\\\hat e&\widehat S\end{pmatrix}\begin{pmatrix}\widehat S&\hat e\\0&\smash{\widehat S}^*\end{pmatrix}=\begin{pmatrix}1&0\\0&1\end{pmatrix}=\begin{pmatrix}\widehat S&\hat e\\0&\smash{\widehat S}^*\end{pmatrix}\begin{pmatrix}\smash{\widehat S}^*&0\\\hat e&\widehat S\end{pmatrix}.
\] By Proposition \ref{prp:halfsp-covcrossed}, we have a unique real
\(*\)-morphism
\(\hat\phi:\widehat{\mathbb A}\longrightarrow\mathbb A\otimes\blop(\ell^2(\nats))\),
\[
  \left.
  \begin{aligned}
    \hat\phi(\hat\pi(f))&\defi\pi(f)\otimes1\\
    \hat\phi\bigl(\smash{\widehat R}_{(x,n)}\bigr)&\defi R_{(x,n)}\otimes(\smash{\widehat S}^*)^n
  \end{aligned}
  \right\}\quad(f\in\sh C(\Omega)\otimes\endo W,x\in\ints^{d-1},n\in\nats).
\] As in the proof of Proposition \ref{prp:halfsp-covcrossed}, using the
obvious action of \(\reals/\ints\) on \(\blop(\ell^2(\nats))\), we see
that \(\hat\phi\) is injective, and hence defines a real
\(*\)-isomorphism onto its image.

By the definition of \(\phi\), the image of the upper left component
\(\phi_{11}\) is contained in the image of \(\hat\phi\). We may
therefore define \[
  \sigma:\mathbb A\longrightarrow\widehat{\mathbb A},\quad\sigma\defi(\hat\phi)^{-1}\circ\phi_{11}. 
\] This map is not a \(*\)-morphism, but it is a completely positive
contraction intertwining the \(*\) operations. Moreover, we have \[
  \phi_{11}(xy)=\phi_{11}(x)\phi_{11}(y)+\phi_{12}(x)\phi_{21}(y)\quad(x,y\in\mathbb A),
\] where \(\phi_{12}(x)\phi_{21}(y)\) is contained in the ideal
generated by \(1\otimes\hat e\). By \(\hat\phi^{-1}\), this ideal is
mapped onto the ideal generated by \(e\). It follows that \(\sigma\) is
a section of \(\vrho\), as it is sufficient to check this statement on
generators. We claim that \(x-\sigma(\vrho(x))\) is contained in the
closed ideal generated by \(e\). By the same argument as in the previous
paragraph, it is sufficient to check on generators \(x\) that
\(\hat\phi(x)-\phi_{11}(\vrho(x))\) is contained in the closed ideal
generated by \(1\otimes\hat e\). But this expression actually vanishes
on generators.

Now we prove \(\ker\vrho\subseteq(e)\). Thus, let
\(x\in\widehat{\mathbb A}\), \(\vrho(x)=0\). Then we have \[
  x=x-\sigma(\vrho(x))\in(e),
\] proving that indeed, \(\ker\vrho=(e)\).

It remains to prove that the ideal \((e)\) is isomorphic to
\(\mathbb A_\partial\otimes\knums(\ell^2(\nats))\). Define \[
  \phi_\partial(a\otimes\ket m\bra n)\defi (R^*_d)^ma(R^{\phantom*}_d)^n\otimes\ket m\bra n\in\mathbb A\otimes\blop(\ell^2(\nats))\quad(a\in\mathbb A_\partial).
\] Then \(\phi_\partial\) is a real \(*\)-morphism on
\(\mathbb A_\partial\otimes\blop_{\mathrm{fin}}(\ell^2(\nats))\), where
\(\blop_{\mathrm{fin}}\) denotes the finite rank operators. It is easy
to see that \(\phi_\partial\) is an isometry on
\(\mathbb A_\partial\otimes\blop_{\mathrm{fin}}(\ell^2(\nats))\), and
hence it extends to \(\mathbb A_\partial\otimes\knums(\ell^2(\nats))\)
as an isometric real \(*\)-morphism. Now, observe that \[
    (R_d^*)^ma(R^{\phantom*}_d)^n\otimes\ket m\bra n=\hat\phi\Parens{\smash{\widehat R}_{e_d}^*}^m(a\otimes 1)(1\otimes\hat e)\hat\phi\Parens{\smash{\widehat R}_{e_d}^{\phantom*}}^{n}.
\] Thus, it is clear that the image of \(\phi_\partial\) equals the
closed ideal of \(\im\hat\phi\) generated by \(1\otimes\hat e\), proving
the claim, and thereby, the theorem.
\EndKnitrBlock{proof}

\hypertarget{real-k-theory-kasparovs-fredholm-picture}{%
\subsection{Real K-theory: Kasparov's Fredholm
picture}\label{real-k-theory-kasparovs-fredholm-picture}}

In the previous subsection, we have constructed the bulk-to-boundary
sequence \eqref{eq:bb-ses}. Our next aim, which we will address in
Subsection \ref{subs:bdy-class}, is to attach boundary \(K\)-theory
classes to every disordered IQPV of a given symmetry index. In order to
do that, we will need a different picture of real \(K\)-theory, due to
Kasparov \autocite{kasparov-topell1}. This we introduce in the present
section.

Kasparov's picture of \(K\)-theory is best known in the context of
Kasparov's bivariant \(KK\)-theory, which unifies both \(K\)-theory and
its dual theory, \(K\)-homology. We have encountered the bivariant
theory before, in the proof of Theorem \ref{thm:DKR-KR} in Section
\ref{sec:bulkinv}. It continues to be an important technical tool for
us, and we shall employ it in the present section, in the proof of
Proposition \ref{prp:kr-kasparov}, and in the proof of our main result
in Section \ref{sec:bb}. For the purpose of constructing boundary
invariants and stating bulk-boundary correspondence, it will be
unnecessary to give a the definition of \(KK\)-theory, and we will
eschew it because of its considerable technical difficulty. For more
details, see
\autocite{blackadar-kthyopalg,kasparov-operatork,kasparov-topell1,schroder1993k}.

The following definition of Kasparov's picture of \(K\)-theory, which
does not appeal to the bivariant theory, is taken from
\autocite{kasparov-topell1}.

\BeginKnitrBlock{definition}[Kasparov's Fredholm picture of real K-theory]
\protect\hypertarget{def:kasp-kr}{}{\label{def:kasp-kr} {} }Let \(A\) be a trivially
graded real C\(^*\)-algebra. Let \(\knums\) denote the real
C\(^*\)-algebra of compact operators on a separable real Hilbert space.
Let \(\mathrm M^s(A)\defi\mathrm M(A\otimes\knums)\) be the stable
multiplier algebra.

Suppose we are given a pair \((F,\psi)\) consisting of an ungraded real
\(\ast\)-morphism \(\psi:C\ell_{p,q-1} \longrightarrow\mathrm M^s(A)\)
and a real operator \(F\in\mathrm M^s(A)\) such that all of the
expressions \[
  \psi(k_a)F+F\psi(k_a),\quad\psi(j_\alpha)F+F\psi(j_\alpha),\quad F^*+F,\quad1+F^2,
\] where \(a=1,\dotsc,p\) and \(\alpha=1,\dotsc,q-1\), are contained in
\(A\otimes\knums\). Such a pair is called a \emph{Kasparov \(KR_{p,q}\)
cycle}. It is called \emph{degenerate} if all the above expressions
vanish. Let \(\mathbb E_{p,q}(A)\) be the set of Kasparov \(KR_{p,q}\)
cycles and \(\mathbb D_{p,q}(A)\) be the set of degenerate Kasparov
\(KR_{p,q}\) cycles.

Two cycles \((F_j,\psi_j)\), \(j=0,1\), are called \emph{homotopy
equivalent} if there is a Kasparov cycle
\((F,\psi)\equiv(F_t,\psi_t)\in\mathbb E_{p,q}\Parens{\sh C([0,1],A)}\)
such that \[
  (F,\psi)|_{t=j}=(F_j,\psi_j)\quad(j=0,1).
\] They are called \emph{unitarily equivalent} if there is an even real
unitary \(u\in M^s(A)\) such that \[
  \psi_0=u^*\psi_1u,\quad F_0=u^*F_1u.
\] The quotient of \(\mathbb E_{p,q}(A)\) with respect to the
equivalence relation generated by homotopy and unitary equivalence is
denoted by \(\smash{\overline{\mathbb E}}_{p,q}(A)\). It is an Abelian
semigroup with respect to the direct sum of cycles. (This operation is
well defined, as \(M^s(A)\oplus M^s(A)\cong M^s(A)\) by Kasparov's
stabilisation theorem.) Let \(\smash{\overline{\mathbb D}}_{p,q}(A)\) be
the sub-semigroup generated by the image of \(\mathbb D_{p,q}(A)\) in
\(\smash{\overline{\mathbb E}}_{p,q}(A)\).

By definition, \emph{Kasparov's \(KR_{p,q}\)-group} is the Abelian
monoid \[
  KR_{p,q}(A)\defi\smash{\overline{\mathbb E}}_{p,q}(A) \bigm/ \smash{\overline{\mathbb D}}_{p,q}(A). 
\] It is in fact an Abelian group. By dropping the reality constraints,
we may define in the same way the Abelian group \(K_{p,q}(A)\), for any
trivially graded C\(^*\)-algebra \(A\). This group is called
\emph{Kasparov's \(K_{p,q}\)-group}.
\EndKnitrBlock{definition}

\BeginKnitrBlock{proposition}
\protect\hypertarget{prp:kr-kasparov}{}{\label{prp:kr-kasparov} }Let \(A\)
be an ungraded real C\(^*\)-algebra resp. an ungraded C\(^*\)-algebra.
Then \(KR_{p,q}(A)\cong KR_{q-p}(A)\) resp.
\(K_{p,q}(A)\cong K_{q-p}(A)\).
\EndKnitrBlock{proposition}

The proof of the proposition is somewhat technical (although not hard).
We defer it to the end of the section. One immediate consequence of the
proposition is the following standard simplification, well-known in
\(KK\)-theory.

\BeginKnitrBlock{lemma}
\protect\hypertarget{lem:unnamed-chunk-4}{}{\label{lem:unnamed-chunk-4} }Any
element of \(KR_{p,q}(A)\) or \(K_{p,q}(A)\) can be represented by a
cycle \((F,\psi)\) for which all of the following expressions vanish: \[
\psi(k_a)F+F\psi(k_a),\quad\psi(j_\alpha)F+F\psi(j_\alpha),\quad F^*+F\quad(a=1,\dotsc,p,\alpha=1,\dotsc,q-1).
\]
\EndKnitrBlock{lemma}

\BeginKnitrBlock{proof}
{}By Proposition \ref{prp:kr-kasparov} and
standard simplifications in \(KK\)-theory \autocite[Proposition
17.4.2]{blackadar-kthyopalg}, we may assume that \(F\) is
skew-Hermitian. If there is some \(a\) such that the anti-commutator
\(\psi(k_a)F+F\psi(k_a)\neq0\), we may take \(a\) is minimal for this
property. Form \[
  \tilde F\defi\frac12\Parens{1-\Ad(\psi(k_a))}(F).
\] One computes easily that \(\tilde F\) is skew-Hermitian and
\(\tilde F^2+1\in A\otimes\knums\). As for \(b<a\), \(\Ad(\psi(k_b))\)
commutes with \(\Ad(\psi(k_a))\), we see that
\(\Ad(k_b)(\tilde F)=-\tilde F\) for all \(b\sle a\). Similarly,
\(\psi(j_\alpha)F+F\psi(j_\alpha)\in A\otimes\knums\). As \(\tilde F\)
is a \enquote{compact perturbation} of \(F\), \((\tilde F,\psi)\)
represents the same class as \((F,\psi)\) \autocite[Proposition
17.2.5]{blackadar-kthyopalg}. Thus, we may assume that \(F\)
anti-commutes with all the \(\psi(k_a)\). Arguing similarly, we may
assume also that it anti-commutes with all the \(\psi(j_\alpha)\).
\EndKnitrBlock{proof}

In the following sections, we will identify all (real) \(K\)-groups with
their counterparts in Kasparov's picture whenever we see fit to do so.
We shall use the notation \(KR_{q-p}(A)\) interchangeably to denote
\(KR_{p,q}(A)\) or any other realisation of \(KR\) theory. We end the
subsection by a proof of the above proposition.

\BeginKnitrBlock{proof}[of Proposition \ref{prp:kr-kasparov}]
{}We will
sketch the construction leading to the proof of the statement that
\(KR_{p,q}(A)\) is isomorphic to the Kasparov group
\(\smash{KKR(C\ell_{q-1,p},A\otimes C\ell_{0,1})}\), which is well-known
to be isomorphic to \(\smash{KR_{q-p}(A)}\). The complex case is proved
in completely the same way.

Let \((F,\psi)\) be a Kasparov \(KR_{p,q}\) cycle, and set \[
  K_a\defi\psi(k_a),\quad J_\alpha\defi\psi(j_\alpha)\quad(a=1,\dotsc,p,\alpha=1,\dotsc,q-1).
\] These operators are ungraded. We introduce a further Clifford algebra
\(C\ell_{0,1}\) into the picture, equipped now with its natural grading
and its odd generator \(j_1\). This allows us to define an \emph{even}
real \(*\)-morphism \[
  \phi:C\ell_{q-1,p}\longrightarrow\mathrm M^s(A\otimes C\ell_{0,1})=\mathrm M^s(A)\otimes C\ell_{0,1}
\] of \emph{graded} real C\(^*\)-algebras, as follows: \[
  \begin{cases}
    k_\alpha\longmapsto J_\alpha\otimes j_1&\text{for }\alpha=1,\dots,q-1,\\
    j_a\longmapsto K_a\otimes j_1&\text{for }a=1,\dotsc,p.
  \end{cases}
\] (Note here that the roles of \(\alpha\) and \(a\) are exchanged on
the side of the Clifford generators, as is necessary to perform the
desired parity reversal.) Then
\(G\defi F\otimes j_1\in\mathrm M^s(A\otimes C\ell_{0,1})\) is real and
odd, and all of the expressions \[
  G\phi(k_\alpha)+\phi(k_\alpha)G,\quad G\phi(j_a)+\phi(j_a)G,\quad G^2-1,\quad G^*-G,
\] for \(a=1,\dotsc,p\) and \(\alpha=1,\dotsc,q-1\), are contained in
\(A\otimes C\ell_{0,1}\otimes\knums\).

Suppose now that \(\sh H\) is a separable real Hilbert space such that
\(\knums=\knums(\sh H)\). Then \(\mathrm M^s(A\otimes C\ell_{0,1})\) is
the graded real C\(^*\)-algebra of adjointable operators on the standard
graded real Hilbert-C\(^*\)-module
\(\sh E=\sh H_{A\otimes C\ell_{0,1}}\) over \(A\otimes C\ell_{0,1}\). It
follows that \((\sh E,\phi, G)\) represents a class in
\(KKR(C\ell_{q-1,p},A\otimes C\ell_{0,1})\). The remainder of the proof
is along the lines of that of \autocite[§6, Corollary 1 to Theorem
2]{kasparov-operatork}.
\EndKnitrBlock{proof}

\hypertarget{subs:bdy-class}{%
\subsection{From quasi-particle vacua to boundary
classes}\label{subs:bdy-class}}

In this section, we assign to any IQPV of any given symmetry index, an
element in a suitable \(KR\) group of the boundary algebra
\(\mathbb A_\partial\). An intermediate step is given by the following
definition.

\BeginKnitrBlock{definition}[Quasi-particle vacua with boundary]
\protect\hypertarget{def:iqpv-bdy}{}{\label{def:iqpv-bdy}
\iffalse (Quasi-particle vacua with boundary) \fi{} }A \emph{disordered
IQPV with boundary} is a real skew-Hermitian
\(\widehat J\in\widehat{\mathbb A}\), \emph{i.e.} \[
  \smash{\widehat J}^*=-\widehat J=\smash{\widehat J}^\intercal
\] such that \(J\defi\vrho(\widehat J)\in\mathbb A\) is a disordered
IQPV. By Theorem \ref{thm:bb-ses}, this means that \[
  \smash{\widehat J}^2\equiv-1\pmod{\mathbb A_\partial\otimes\knums(\ell^2(\nats))}.
\] A tuple
\((\widehat J;\phi)=(\widehat J; K_1,\dotsc, K_r,J_1,\dotsc, J_s)\) is
called a \emph{disordered IQPV with boundary of symmetry index
\((r,s)\)} if \(\widehat J\) is a disordered IQPV with boundary and
moreover, the tuple \((J;\phi)=(J;K_1,\dotsc, K_r,J_1,\dotsc, J_s)\) is
a disordered IQPV of index \((r,s)\), where \(J\defi\vrho(\widehat J)\).

Explicitly, this means the following:
\(\widehat J\in\widehat{\mathbb A}\) is a real skew-Hermitian operator;
\(K_a\), \(J_\alpha\), define a real \(*\)-morphism
\(\phi:C\ell_{r,s}\longrightarrow\endo W\); and the expressions \[
\widehat JK_a+K_a\widehat J,\quad\widehat JJ_\alpha+J_\alpha\widehat J,\quad\smash{\widehat J}^2+1\quad(a=1,\dotsc,r,\alpha=1,\dotsc,s)
\] all belong to \(\mathbb A_\partial\otimes\knums(\ell^2(\nats))\).
Similarly, in the complex case, we define \emph{charge-preserving
disordered IQPV with boundary} resp. \emph{disordered IQPV with boundary
of complex symmetry index \((r,s)\)} by replacing
\(\smash{\widehat{\mathbb A}}\) by \(\smash{\widehat{\mathrm A}}\) and
dropping the reality constraints.
\EndKnitrBlock{definition}

\BeginKnitrBlock{proposition}
\protect\hypertarget{prp:bdy-iqpv}{}{\label{prp:bdy-iqpv} }Any disordered
IQPV with boundary of (complex) symmetry index \((r,s)\) represents a
class in \(KR_{s-r+1}(\mathbb A_\partial)\) resp. in
\(K_{s-r+1}(\mathrm A_\partial)\).
\EndKnitrBlock{proposition}

\BeginKnitrBlock{proof}
{}By Proposition \ref{prp:kr-kasparov}, a
Kasparov \(KR_{r,s+1}\) cycle for \(\mathbb A_\partial\) represents a
class in the desired group \(KR_{s-r+1}(\mathbb A_\partial)\). Comparing
Definitions \ref{def:kasp-kr} and \ref{def:iqpv-bdy}, a disordered IQPV
with boundary \((\smash{\widehat{J}};\phi)\) of symmetry index \((r,s)\)
is distinguished from a Kasparov \(KR_{r,s+1}\) cycle only by lying in
\(\smash{\widehat{\mathbb A}}\) instead of
\(\mathrm M^s(\mathbb A_\partial)\).

Thus, it suffices to exhibit \(\smash{\widehat{\mathbb A}}\) inside of
\(\mathrm M^s(\mathbb A_\partial)=\mathrm M(\mathbb A_\partial\otimes\knums(\ell^2(\nats)))\).
To that end, recall from Theorem \ref{thm:bb-ses} and its proof that
\(\smash{\widehat{\mathbb A}}\) is isomorphic to the closed
\(*\)-subalgebra \(B\) of \(\mathbb A\otimes\blop(\ell^2(\nats))\)
generated by \[
  a\otimes1,\quad R_d\otimes\smash{\widehat S}^*\quad(a\in\mathbb A_\partial).
\] Moreover, \(\mathbb A_\partial\otimes\knums(\ell^2(\nats))\) is
isomorphic to the closed ideal \(I\subseteq B\) generated by
\(1\otimes\hat e\), where
\(\hat e=1-\smash{\widehat S}\smash{\widehat S}^*\). By the definition
of the multiplier algebra, there is a unique \(*\)-morphism
\(B\longrightarrow\mathrm M^s(\mathbb A_\partial)\) that is the identity
on \(I\) \autocite[Proposition 3.7 (i)]{busby-dc}. As \(B\) is unital,
it is surjective. To see that the \(*\)-morphism \(\vrho\) is injective,
it will be sufficient to prove that \(I\) is essential in \(B\)
\autocite[Proposition 3.7 (ii)]{busby-dc}. Let \(b\in B\) such that
\(bI=0\). Then \[
  b(\psi\otimes\ket n)=b(R_d^*\otimes\widehat S)^n(1\otimes\hat e)((R_d)^n\psi\otimes\ket0)=0\quad(\psi\in L^2(\Omega,P)\otimes\sh W,n\in\nats).
\] Since \(\psi\) and \(n\) were arbitrary and \(B\) is represented
faithfully on \(L^2(\Omega,P)\otimes\sh W\otimes\ell^2(\nats)\), we
conclude that \(b=0\), so that \(I\) is essential.
\EndKnitrBlock{proof}

By definition, a disordered IQPV with boundary
\((\widehat J; K_1,\dotsc, K_r,J_1,\dotsc, J_s)\) of symmetry index
\((r,s)\) determines a disordered bulk IQPV \[
      J=\vrho(\widehat J)\in\mathbb A
\] of the same symmetry index \((r,s)\). Since \(\vrho\) is surjective,
any bulk IQPV is thus determined. However, we can do better and always
give a canonical representative. Indeed, let
\((J;\phi)=(J; K_1,\dotsc, K_r,J_1,\dotsc, J_s)\) be a disordered IQPV
with of symmetry index \((r,s)\). We may thus define \begin{equation}
   \widehat J_\omega\defi qJ_\omega q\quad(\omega\in\Omega). 
\end{equation} Then
\((\widehat J;\phi)=(\widehat J; K_1,\dotsc, K_r,J_1,\dotsc, J_s)\) is a
disordered IQPV with boundary of symmetry index \((r,s)\), as follows
immediately from the following lemma. The same construction also works
in the case of a disordered IQPV of complex symmetry index \((r,s)\).

\BeginKnitrBlock{lemma}
\protect\hypertarget{lem:unnamed-chunk-8}{}{\label{lem:unnamed-chunk-8} }Let
\(a\in\mathbb A\). Then \(\vrho(qaq)=a\).
\EndKnitrBlock{lemma}

\BeginKnitrBlock{proof}
{}Recall the covariant pairs \((\pi,R)\) and
\((\widehat\pi,\widehat R)\) from Theorem \ref{thm:covcrossed} and
Proposition \ref{prp:halfsp-covcrossed}, together with their proofs. Let
\(f\in\sh C(\Omega)\otimes\endo W\) and \(x\in\hat\Lambda\). Then \[
  q\pi(f)R_xq=q\pi(f)qR_xq=\widehat\pi(f)\widehat R_x,
\] as \(\pi(f)\) commutes with \(q\). It follows from Theorem
\ref{thm:bb-ses} that \[
  \vrho(q\pi(f)R_xq)=\vrho(\widehat\pi(f)\widehat R_x)=\vrho(\widehat\pi(f))\vrho(\widehat R_x)=\pi(f)R_x.
\] Similarly, \[
  q\pi(f)R_{-x}q=\widehat\pi(f)\smash{\widehat R}_x^*, 
\] which implies that \[
  \vrho(q\pi(f)R_{-x}q)=\pi(f)R_x^*=\pi(f)R_{-x}.
\] The operators \(\pi(f)R_x\), for \(f\in\sh C(\Omega)\otimes\endo W\)
and \(x\in\Lambda\), span a dense subalgebra of \(\mathbb A\), in view
of Theorem \ref{thm:covcrossed}. Since
\(\hat\Lambda\cup(-\hat\Lambda)=\Lambda\), the assertion follows from
the linearity and continuity of \(\vrho\).
\EndKnitrBlock{proof}

\BeginKnitrBlock{definition}[Boundary class of an IQPV]
\protect\hypertarget{def:bound-IQPV}{}{\label{def:bound-IQPV}
\iffalse (Boundary class of an IQPV) \fi{} }Let
\(\phi:C\ell_{r,s}\longrightarrow\endo{W}\) be a real \(*\)-morphism
(resp. \(\phi:\mathbb C\ell_{r,s}\longrightarrow\endo{V}\) be a
\(*\)-morphism). Fix a reference disordered IQPV
\((J_{\mathrm{ref}};\phi)\) of (complex) symmetry index \((r,s)\). For
any disordered IQPV \((J;\phi)\) of symmetry index \((r,s)\), define its
\emph{boundary class} by \[
  [(J;\phi)]_\partial\defi[(\widehat J;\phi)]-[(\widehat J_{\mathrm{ref}};\phi)]\in KR_{s-r+1}(\mathbb A_\partial)\quad\text{(resp. }\in K_{s-r+1}(\mathrm A_\partial)\text{).}
\] Here, \([(\widehat J;\phi)]\) and
\([(\widehat J_{\mathrm{ref}};\phi)]\) are the classes represented by
the corresponding disordered IQPV with boundary of (complex) symmetry
index \((r,s)\).
\EndKnitrBlock{definition}

\BeginKnitrBlock{remark}
{} Readers inclined towards physics will wonder
how it is possible to assign a class in the \(K\)-theory of the boundary
algebra \(\mathbb A_\partial\) to a disordered IQPV in a given symmetry
class, as this would seem to contradict the boundary-anomalous nature of
a non-trivial topological phase. Observe, however, that the boundary
class is represented by an operator \(\smash{\widehat J}\) lying in (the
multiplier algebra of) the tensor product
\(\mathbb A_\partial\otimes\knums(\ell^2(\nats))\), rather than in
\(\mathbb A_\partial\) itself. That is, the boundary phase is not
exactly localised at the boundary, but instead spread out in its
vicinity.

For any disordered IQPV with boundary \((\widehat{J}';\phi)\) of the
same symmetry index, with \(\vrho(\widehat{J}')=J\), we have \[
    \vrho(\widehat{J}' - \widehat{J})=0.
  \] Thus,
\(K\defi \widehat{J'}-\widehat J\in \mathbb{A}_\partial \otimes \mathbb{K}\).
The homotopy \([0,1] \ni t \longmapsto \smash{\widehat{J}} + tK\)
induces a homotopy between \((\smash{\widehat{J}};\phi)\) and
\((\smash{\widehat{J}'};\phi)\). Thus \((\smash{\widehat{J}'};\phi)\)
and \((\smash{\widehat{J}};\phi)\) define the same class. In this sense,
the boundary class for a given bulk IQPV is unique.
\EndKnitrBlock{remark}

\BeginKnitrBlock{example}[Boundary classification for symmetry class $BD\mathrm I$]
\protect\hypertarget{exm:unnamed-chunk-11}{}{\label{exm:unnamed-chunk-11}
{} }For the symmetry class \(BD\mathrm{I}\), up to stable
equivalence, we may choose any symmetry index \((r,s)\) where
\(r-s\equiv 1\pmod 8\), for instance \((r,s)=(1,0)\).

Recalling the proof of Proposition \ref{prp:kr-kasparov}, the \(KKR\)
class corresponding to the \(KR\) class represented by the IQPV with
boundary \((\widehat J; K_1)\) is \[
  [( \mathcal{H}_{\mathbb{A}_\partial} \otimes C\ell_{0,1}, \psi, \widehat{J} \otimes j_1 )] \in KKR(C\ell_{0,1}, \mathbb{A}_\partial \otimes C\ell_{0,1}),
\] where \[
  \psi: C\ell_{0,1} \longrightarrow\mathrm M^s(\mathbb A_\partial \otimes C\ell_{0,1})
\] is determined by \(\psi(j_1)\defi K_1 \otimes j_1\). This class is
equivalent to \[
  [(\mathcal{H}_{\mathbb A_\partial}, 1_\cplxs, \widehat{J}K_1 )] \in KKR(\cplxs, \mathbb{A}_\partial),
\] where \(\mathcal{H}_{\mathbb{A}_\partial}\) is graded by the operator
\(K_1\). Using an index morphism \autocite[Theorem
2.2.8]{schroder1993k}, we can identify this \(KKR\) class with the class
\[
  [p^+]-[p^-]\in K_0(\mathbb A_\partial),
\] where \(p^\pm\) denote the projections onto the submodules
\(\ker(\frac{1}{2}(1\pm K_1)\widehat{J}K_1\frac{1}{2}(1\mp K_1))\) of
the graded real Hilbert \(\mathbb A_\partial\)-module
\(\mathcal{H}_{\mathbb{A}_\partial}\). For \(K_1=\sigma_z\), we obtain
\[
  \widehat{J}K_1 = \begin{pmatrix} 0 & \hat{u} \\ \hat{u}^* & 0 \end{pmatrix},\quad\big[p_{\ker(\hat{u})} \big] - \big[p_{\ker(\hat{u}^*)} \big] \in K_0(\mathbb{A}_\partial).
\]
\EndKnitrBlock{example}

\hypertarget{sec:bb}{%
\section{Bulk-boundary correspondence with disorder}\label{sec:bb}}

This section is devoted to the statement and proof of
\emph{bulk-boundary correspondence} for disordered free-fermion
topological phases: In the two previous sections, we have attached, to
every disordered IQPV with symmetries, real \(K\)-theory classes in the
bulk and at the boundary. Bulk and boundary are connected \emph{via} the
sequence \eqref{eq:bb-ses}. This sequence induces a long exact sequence in
real \(K\)-theory, \emph{viz.} \begin{equation}
  \label{eq:bb-les}
  \begin{tikzcd}[column sep=4.5ex]
    { }\arrow[r,dotted]
    &KR_{s-r+2}\parens{\mathbb A_\partial}\rar{\iota^*}\arrow[draw=none]{d}[name=Z, anchor=center]{}
    &KR_{s-r+2}\Parens{\widehat{\mathbb A}}\rar{\vrho^*}
    &KR_{s-r+2}(\mathbb A)
    \arrow[rounded corners,
      to path={ -- ([xshift=2ex]\tikztostart.east)
                |- (Z.center)\tikztonodes
                -| ([xshift=-2ex]\tikztotarget.west)
                -- (\tikztotarget)}]{dll}[pos=0.8,description]{\partial}
    &\\
    &KR_{s-r+1}(\mathbb A_\partial)\rar{\iota^*}
    &KR_{s-r+1}\Parens{\widehat{\mathbb A}}\rar{\varrho^*}
    &KR_{s-r+1}(\mathbb A) \arrow[r,dotted]
    &{ }
  \end{tikzcd}
\end{equation} In particular, we obtain a connecting map \[
  \partial:KR_{s-r+2}(\mathbb{A})\longrightarrow KR_{s-r+1}(\mathbb{A}_\partial),
\] which we call the \emph{bulk-boundary map}. In this section, we will
show that \(\partial\) maps the bulk class attached to a disordered IQPV
with symmetries onto the corresponding boundary class. (The same also
holds for the complex symmetry classes.)

\BeginKnitrBlock{theorem}[Bulk-boundary correspondence]
\protect\hypertarget{thm:ComplexBound}{}{\label{thm:ComplexBound}
\iffalse (Bulk-boundary correspondence) \fi{} }Fix a reference
disordered IQPV \((J_{\mathrm{ref}};\phi)\) of (complex) symmetry index
\((r,s)\). For any disordered IQPV \((J;\phi)\) of (complex) symmetry
index \((r,s)\), we have \begin{equation}
  \partial[(J;\phi)]=[(J;\phi)]_\partial.
\end{equation} That is, the bulk-boundary map \(\partial\) maps the bulk
class \([(J;\phi)]\) attached to \((J;\phi)\) onto the boundary class
\([(J;\phi)]_\partial\) attached to \((J;\phi)\).
\EndKnitrBlock{theorem}

\hypertarget{proof-of-the-main-theorem}{%
\subsection{Proof of the main theorem}\label{proof-of-the-main-theorem}}

This subsection is devoted to the proof of Theorem
\ref{thm:ComplexBound}. It relies on the use of \(KKR\) theory; once
again, let us refer to
\autocite{blackadar-kthyopalg,kasparov-operatork,schroder1993k} for
accounts thereof. Let us begin by fixing some conventions. Throughout
the proof, we will consider the Clifford algebras with their natural
grading. All other C\(^*\)-algebras are ungraded. For a real
C\(^*\)-algebra \(B\), the standard Hilbert \(B\)-module will be denoted
by \(\sh H_B\).

Recall Roe's map \(\alpha\) from the proof of Theorem \ref{thm:DKR-KR}.
Since \(\alpha\circ\partial=\partial\circ\alpha\), we can compute the
effect of \(\partial\) on the bulk class in \(DKR\)-theory and
subsequently apply \(\alpha^{-1}\) to obtain a class in \(KKR\)-theory.

The non-trivial part of the definition of \(\alpha\) is given by the
connecting map \(\partial_Q\) in \(DKR\)-theory. We relate it to the
connecting map for the bulk-boundary sequence \eqref{eq:bb-ses}. To that
end, recall that \(\mathbb A_\partial\otimes\knums\) (where
\(\knums=\knums(\ell^2(\nats))\)) is an essential ideal in
\(\smash{\widehat{\mathbb A}}\), as was shown in the proof of
Proposition \ref{prp:bdy-iqpv}. Therefore, the canonical map
\(\mathbb A_\partial\otimes\knums\longrightarrow\mathrm M^s(\mathbb A_\partial)\)
factors as \(\Phi\circ\iota\) where
\(\Phi:\smash{\widehat{\mathbb A}}\longrightarrow\mathrm M^s(\mathbb A_\partial)\)
is a unique injective real \(*\)-morphism.

We obtain a commutative diagram with exact rows: \[
\begin{tikzcd}
  0\rar{}&\mathbb A_\partial\otimes\mathbb K\dar[mathdouble]{}\rar{\iota}&\widehat{\mathbb A}\dar{\Phi}\rar{\varrho}&\mathbb A\rar{}\dar{\widetilde\Phi}&0\\
  0\rar{}&\mathbb A_\partial\otimes\mathbb K\rar{}&\mathrm M^s(\mathbb A_\partial)\rar{\pi}&Q^s(\mathbb A_\partial)\rar{}&0
\end{tikzcd}
\] From the naturality of connecting maps (which is a general fact for
\(\delta\)-functors, but can also be derived directly from the explicit
form of the connecting map given in \autocite[Proposition
3.4]{van_Daele2}), the following diagram commutes: \[
\begin{tikzcd}
  DKR_e\Parens{\mathbb A\otimes C\ell_{r,s+1}} \dar{\widetilde\Phi^*}\rar{\partial}&DKR\Parens{\mathbb A_\partial\otimes C\ell_{r+1,s+1} }\\
  DKR_{\tilde e}\Parens{Q^s(\mathbb A_\partial)\otimes C\ell_{r,s+1}}\arrow{ur}[swap]{\partial_Q},
\end{tikzcd}
\] where \(\tilde e\defi\widetilde\Phi(e)\). Therefore, \[
  (\alpha^{-1}\circ\partial)([x]-[e])=\Bracks{\Parens{\sh H_{\mathbb A_\partial}\otimes C\ell_{r,s+1} ,1_\cplxs,y}} - \Bracks{\Parens{\sh H_{\mathbb A_\partial}\otimes C\ell_{r,s+1} ,1_\cplxs,f}}
\] for \(y=\Phi(\hat x)\), \(f=\Phi(\hat e)\), \(\vrho(\hat x)=x\) and
\(\vrho(\hat e)=e\). Applying these considerations to the bulk class
with reference ORHU
\(e=J_{\mathrm{ref}}\otimes j_1\in\sh F(\mathbb A\otimes C\ell_{r,s+1})\),
we find by Equation \eqref{eq:bulk-explicit} that as an element of
\(KKR(\cplxs,\mathbb A_\partial\otimes C\ell_{r,s+1})\), the image
\(\partial[(J;\phi)]\) of the bulk class under \(\partial\) equals \[
  \Bracks{\Parens{\sh H_{\mathbb A_\partial}\otimes C\ell_{r,s+1},1_\cplxs,(\widehat J\otimes j_1)P+(\widehat J_{\mathrm{ref}}\otimes j_1)(1-P)}}-\Bracks{\Parens{\sh H_{\mathbb A_\partial}\otimes C\ell_{r,s+1},1_\cplxs,\widehat J_{\mathrm{ref}}\otimes j_1}}
\] where we write \(P\defi P^{r,s}\). This in turn equals \[
  \Bracks{\Parens{P(\sh H_{\mathbb A_\partial}\otimes C\ell_{r,s+1}),1_\cplxs,(\widehat J\otimes j_1)P}}-\Bracks{\Parens{P(\sh H_{\mathbb A_\partial}\otimes C\ell_{r,s+1}),1_\cplxs,(\widehat J_{\mathrm{ref}}\otimes j_1)P}}.
\] We denote the two parts of this difference by \(\partial_\mathrm I\)
and \(\partial_{\mathrm{I\!I}}\), respectively. To compare the
expression \(\partial_\mathrm I-\partial_{\mathrm{I\!I}}\) to the
boundary class, we may treat the parts independently. In what follows,
we shall focus on \(\partial_{\mathrm I}\), as the case of
\(\partial_{\mathrm{I\!I}}\) differs only in notation.

We set \(B\defi\mathbb A_\partial\otimes C\ell_{0,1}\) for brevity. Then
the boundary class is given in \(KKR(C\ell_{s,r},B)\), by the proof of
Proposition \ref{prp:kr-kasparov}. We will use some standard
isomorphisms in \(KK\)-theory to transfer the class computed above to
this \(KKR\) group. In preparation thereof, we apply the Clifford
algebra isomorphism
\(C\ell_{r,s+1}\cong C\ell_{0,1}\mathop{\widehat\otimes}C\ell_{r,s}\),
given by \[
j_1\longmapsto j_1\otimes 1,\quad j_{\alpha+1}\otimes 1\longmapsto 1\otimes j_{\alpha},\quad k_a\longmapsto -1\otimes k_a.
\] This isomorphism preserves Clifford orientation in the sense of
Kasparov. The image of the projection \(P\) now takes the form \[
  P'\defi\prod_{a=1}^r\tfrac12\Parens{1+(-1)^sK_a\otimes j_1\otimes k_a}\prod_{\alpha=1}^s \tfrac12\Parens{1+J_\alpha\otimes j_1\otimes j_\alpha}.
\]

The first non-trivial isomorphism in \(KK\)-theory that we apply is the
exterior tensor product with the unit class \[ 
  \tau\defi\Bracks{\Parens{C\ell_{s,r},\id_{C\ell_{s,r}},0}}\in KKR(C\ell_{s,r},C\ell_{s,r}).
\] The exterior product
\(\partial_\mathrm I\mathop{\widehat\otimes}\tau\in KKR\parens{C\ell_{s,r},B\mathop{\widehat{\otimes}}C\ell_{r,s}\mathop{\widehat{\otimes}}C\ell_{s,r}}\)
equals \[
  \Bracks{\Parens{(P'\otimes 1)(\sh H_B\mathop{\widehat{\otimes}}C\ell_{r,s}\mathop{\widehat{\otimes}}C\ell_{s,r}),\id_{C\ell_{s,r}},(J\otimes j_1\otimes 1\otimes 1)(P'\otimes 1)}}.
\] We now apply the orientation-preserving isomorphism
\(C\ell_{r,s}\mathop{\widehat{\otimes}}C\ell_{s,r}\cong C\ell_{s+r,s+r}\)
given by \[
  k_a\otimes 1\longmapsto (-1)^sk_{s+a},\quad j_\alpha\otimes1\longmapsto j_\alpha,\quad 1\otimes k_\alpha\longmapsto (-1)^sk_\alpha,\quad1\otimes j_a\longmapsto j_{s+a},
\] for \(a=1,\dotsc r\) and \(\alpha=1,\dotsc,s\). This isomorphism
converts \(P'\otimes1\) into the projection \[
  P''=\prod_{a=1}^r\tfrac12\Parens{1+K_a\otimes j_1\otimes k_{s+a}}\prod_{\alpha=1}^s\tfrac12\Parens{1+J_\alpha\otimes j_1\otimes j_\alpha},
\] so it identifies \(\partial_{\mathrm I}\mathop{\widehat\otimes}\tau\)
with the class \[
  \Bracks{\Parens{P''(\sh H_B\mathop{\widehat{\otimes}} C\ell_{s+r,s+r}),\vphi,(\widehat J\otimes j_1\otimes1)P'}}\in KKR\Parens{C\ell_{s,r},B\mathop{\widehat{\otimes}}C\ell_{s+r,s+r}},
\] where \(\vphi\) is defined by \[
  \vphi(k_\alpha)\defi(-1)^s\otimes1\otimes k_\alpha,\quad\vphi(j_a)=1 \otimes 1\otimes j_{s+a}\quad(a=1,\dotsc,r,\alpha=1,\dotsc,s).
\]

The next step is to apply a unitary equivalence. We define commuting
unitaries \[
U_\alpha\defi\tfrac1{\sqrt2}\Parens{1-J_\alpha\otimes j_1\otimes k_\alpha},\quad V_a\defi\tfrac1{\sqrt2}\Parens{1-K_a \otimes j_1 \otimes j_{s+a}}\quad(a=1,\dotsc r,\alpha=1,\dotsc,s).
\] Then we have \[
  \Ad(U_\alpha)(1\otimes 1\otimes k_\alpha)=-J_\alpha\otimes j_1\otimes 1,\quad\Ad(V_a)(1\otimes1\otimes j_{s+a})=K_a\otimes j_1\otimes1,
\] whereas \(1\otimes1\otimes c\) is fixed for any other generator \(c\)
of \(C\ell_{r+s,r+s}\). Similarly, \(\Ad(U_\alpha)\) fixes any
\(K_a\otimes j_1\otimes1\) and \(J_\beta\otimes j_1\otimes1\) for
\(\beta\neq\alpha\), and \(\Ad(V_a)\) fixes any
\(J_\alpha\otimes j_1\otimes1\) and \(K_b\otimes j_1\otimes1\) for
\(b\neq a\). Both actions fix \(\smash{\widehat J}\otimes j_1\otimes1\).

Hence, applying the unitary equivalence induced by
\(V_1\dotsm V_rU_1\dotsm U_s\) identifies
\(\partial_{\mathrm I}\mathop{\widehat\otimes}\tau\) with the class \[
  \Bracks{\Parens{\sh H_B\mathop{\widehat{\otimes}}Q(C\ell_{r+s,r+s}),\psi\otimes1,\widehat J\otimes j_1\otimes Q}}\in KKR\Parens{C\ell_{s,r},B\mathop{\widehat{\otimes}}C\ell_{r+s,r+s}}
\] where \(Q=\prod_{i=1}^{r+s}\frac12(1+k_ij_i)\) and \(\psi\) is
defined by \[
  \psi(k_\alpha)\defi (-1)^{s+1}J_\alpha\otimes j_1,\quad\psi(j_a)\defi K_a\otimes j_1\quad(a=1,\dotsc,r,\alpha=1,\dotsc,s).
\] The automorphism of \(C\ell_{s,r}\) given by
\(k_\alpha\longmapsto(-1)^{s+1}k_\alpha\) and \(j_a\longmapsto j_a\)
preserves orientation, as \(\smash{(-1)^{s(s+1)}=1}\), and thus acts
trivially on \(KKR\). We may therefore remove the signs in the
definition of \(\psi\) without changing the class.

The final step of the proof will be to apply stability. By
\autocite[Corollary 2.4.9]{schroder1993k}, the stability isomorphism
equals the intersection product with the class \[
  \alpha\defi\Bracks{\Parens{(\sh K\oplus\Pi\sh K)\mathop{\widehat\otimes}S,1\otimes c,T\otimes1}}\in KKR(C\ell_{r+s,r+s},\cplxs).
\] Here, \(\sh K\) is a separable Hilbert space, \(\Pi\sh K\) is an odd
copy thereof, \((c,S)\) is the unique irreducible real
\(*\)-representation of \(C\ell_{r+s,r+s}\) (see Propositions
\ref{prp:cliff11} and \ref{prp:cliffsimple}), and \[
  T = \begin{pmatrix} 0 & T_1^* \\ T_1^{\phantom*} & 0 \end{pmatrix},\quad T_1^{\phantom*}T_1^*=1,\quad T_1^*T_1^{\phantom*}=1-p,
\] where \(T_1\in\blop(\sh K)\) and \(p\) is a rank one projection.
Arguing as in the proof of \autocite[Theorem 2.4.7]{schroder1993k}, we
see that the intersection product with \(\alpha\) yields the class \[
  \Bracks{\Parens{\sh H_B\otimes c(Q)(S),\psi\otimes1,\widehat J\otimes j_1\otimes1}}\in KKR(C\ell_{s,r},B).
\] But \(c(Q)(S)\) has dimension one and contributes neither to the
action nor to the Fredholm operator, so we conclude \[
  \partial_{\mathrm I}\mathop{\widehat\otimes}\tau\mathop{\widehat\otimes}\nolimits_{C\ell_{r+s,r+s}}\alpha=\Bracks{\Parens{\sh H_B,\psi,\widehat J\otimes j_1}}\in KKR(C\ell_{s,r},B).
\] The same holds for \(\partial_\mathrm I\) replaced by
\(\partial_{\mathrm{I\!I}}\) and \(\widehat J\) replaced by
\(\widehat J_{\mathrm{ref}}\). Because \[
  [(J;\phi)]_\partial=\Bracks{\Parens{\sh H_B,\psi,\widehat J\otimes j_1}}-\Bracks{\Parens{\sh H_B,\psi,\widehat J_{\mathrm{ref}}\otimes j_1}},
\] this completes the proof of the theorem.

\hypertarget{corollaries-to-the-main-theorem}{%
\subsection{Corollaries to the main
theorem}\label{corollaries-to-the-main-theorem}}

\BeginKnitrBlock{corollary}
\protect\hypertarget{cor:bdy-local}{}{\label{cor:bdy-local} }Fix
\(r,s\sge0\) and let the reference IQPV in Theorem
\ref{thm:ComplexBound} be local, \emph{i.e.}
\(J_\mathrm{ref}\in\mathcal{C}(\Omega) \otimes \endo W\) (resp.
\(\endo{V}\) in the complex case). Then the boundary class simplifies to
\[
  \Bracks{\Parens{J;\phi}}_\partial=\Bracks{\Parens{\widehat J;\phi}},
\] so that bulk-boundary correspondence reads \[
  \partial[(J;\phi)]=\Bracks{\Parens{\widehat J;\phi}}.
\]
\EndKnitrBlock{corollary}

\BeginKnitrBlock{proof}
{}By assumption, \(J_{\mathrm{ref}}\) commutes
with \(q\), so that \(\widehat J_\mathrm{ref}\) is unitary in
\(\smash{\widehat{\mathbb A}}\). Hence, the class
\([(\hat{J}_\mathrm{ref};\phi)]\) is trivial.
\EndKnitrBlock{proof}

\BeginKnitrBlock{corollary}
\protect\hypertarget{cor:bdy-trivial}{}{\label{cor:bdy-trivial} }Fix
\(r,s\sge0\) and suppose that the reference IQPV in Theorem
\ref{thm:ComplexBound} is local (see the previous corollary). Let
\((J;\phi)\) be a disordered IQPV of (complex) symmetry index \((r,s)\)
such that its boundary class is trivial, \emph{i.e.}
\(\partial[(J;\phi)]=0\).

Then \(\widehat J\) is homotopic to a (real) skew-Hermitian unitary
anti-commuting with the \(K_a,J_\alpha\). More precisely, there is a
family \(\smash{\widehat J_t}\in\smash{\widehat{\mathbb A}}\),
\(t\in[0,1]\), continuous in the strong-\(*\) operator topology, such
that \((\smash{\widehat{J_t}};\phi)\) are disordered IQPV with boundary
of (complex) symmetry index \((r,s)\), and moreover,
\(\smash{\widehat J_0}=\smash{\widehat J}\) and
\((\smash{\widehat J}_1)^2=-1\).
\EndKnitrBlock{corollary}

\BeginKnitrBlock{proof}
{}This is immediate from Theorem
\ref{thm:ComplexBound}, Corollary \ref{cor:bdy-local}, and the Theorem
of Skandalis \autocite[Theorem 18.5.3]{blackadar-kthyopalg}.
\EndKnitrBlock{proof}

\BeginKnitrBlock{remark}
{}Expressed in physics terms, the statement of
Corollary \ref{cor:bdy-trivial} is the following: The boundary phase
attached to a disordered bulk phase of free fermions is topologically
trivial if and only if it exhibits no stable gapless localised boundary
states.
\EndKnitrBlock{remark}

\BeginKnitrBlock{remark}
{}By the exactness of the long exact sequence
in \(K\)-theory \eqref{eq:bb-les}, the image of the map \(\partial\) is
equal to the kernel of \[
  \iota^*:KR_{s-r+1}(\mathbb A_\partial)\longrightarrow KR_{s-r+1}(\widehat{\mathbb A}).
\] By \autocite[Theorem 1.5.5]{schroder1993k}, this coincides with
\(\ker({\id}-(\alpha_d)^*)\) where the automorphism
\(\alpha_d=\Ad(u_d)\) is defined by considering
\(\mathbb A_\partial\subseteq\mathbb A\) in a natural fashion. This
kernel is trivial if and only if \(\alpha_d\) is stably homotopic to the
identity. While this holds for the clean case without disorder, it may
fail in general. Thus, \(\partial\) may not be surjective.
\EndKnitrBlock{remark}

We can compare the boundary phases corresponding to different sides of a
boundary. To that end, observe that \(1-q\) is the projection
corresponding to the left half-space semilattice \(-\hat\Lambda\). In
the following proposition, let \(\widehat J_+\defi\widehat J\) and
\(\widehat J_-\defi(1-q)J(1-q)\).

\BeginKnitrBlock{proposition}
\protect\hypertarget{prp:unnamed-chunk-5}{}{\label{prp:unnamed-chunk-5} }Let
\((J_\mathrm{ref};\phi)\) be a reference IQPV of (complex) symmetry
index \((r,s)\). For any disordered IQPV \((J;\phi)\) of (complex)
symmetry index \((r,s)\), we have \[
  [(J;\phi)]^+_\partial=-[(J;\phi)]^-_\partial.
\] Here, we set \[
  [(J;\phi)]^\pm_\partial\defi\Bracks{\Parens{\widehat J_\pm;\phi}}-\Bracks{\Parens{\widehat J_{\mathrm{ref},\pm};\phi}}.
\]
\EndKnitrBlock{proposition}

\BeginKnitrBlock{proof}
{}It is sufficient to prove that the sum
\([(\widehat{J}_+;\phi)]+[(\widehat{J}_-;\phi)]\) is trivial for any
\(J\). The sum is represented by \[
  \Parens{\widehat J_+\oplus\widehat J_-,1_2\otimes\phi}
\] Here, by definition \[
  \widehat J_+\oplus\widehat J_-=qJq\oplus(1-q)J(1-q),
\] where \(q\) is the Szegő projection of Definition
\ref{def:half-covalg}. The right-hand side is the starting point \(F_0\)
of the path \[
  [0,1]\ni t\longmapsto F_t\defi 
  \begin{pmatrix}
    qJq & tqJ(1-q)\\ 
    t(1-q)Jq & (1-q)J(1-q) 
  \end{pmatrix}.
\] The operators \(F_t\) are skew-Hermitian and \[
  F_t^2=
  \begin{pmatrix}
    q(-1+(t^2-1)J(1-q)J)q&0\\ 
    0&(1-q)(-1+(t^2-1)JqJ)
  \end{pmatrix}.
\] The operator \(qJ(1-q)Jq\) lies in the kernel of the canonical
projection \(\vrho=\vrho_+\) of the right half-space algebra. Indeed,
the expression \(qO_1(1-q)O_2q\) vanishes if \(O_1\) and \(O_2\) are
among the generators \(f\in\sh C(\Omega)\otimes\endo{W}\) and
\(u_1,\dotsc,u_{d-1}\), together with their adjoints, as all of these
commute with \(q\). Furthermore, we compute \[
  (1-q)u_dq=0,\quad(1-q)u_d^*q=u_d^*e.
\] Because \(J\) is the limit of non-commutative polynomials in the
generators \(f\) and \(u_1,\dotsc,u_d\), together with their adjoints,
it follows that \(qJ(1-q)Jq\) lies in the closed two-sided ideal
generated by \(e\). This is the kernel of \(\vrho_+\) by Theorem
\ref{thm:bb-ses}. A similar statement holds for the lower right corner
of the matrix representing \(F_t^2\). Hence, \((F_t)\) is an operator
homotopy, so since \(F_1^2=-1\), it follows that the class represented
by \(F_0\) is trivial.
\EndKnitrBlock{proof}

\hypertarget{subs:disorder}{%
\subsection{Invariance under disorder}\label{subs:disorder}}

So far, the precise nature of the space \(\Omega\) of disorder
configurations has not been important. Bulk-boundary correspondence
holds for any choice of \(\Omega\). However, the \(K\)-theory of the
bulk and boundary algebras will in general depend on \(\Omega\). We now
address a specific type of disorder where this dependence vanishes.

In practice, disorder is introduced chiefly through impurities. In this
case, the electrons experience random perturbations to the on-site
atomic potential. This leads to the model \[
  \Omega=\Omega_0^\Lambda=\prod_{x\in\Lambda}\Omega_0
\] for some compact Hausdorff space \(\Omega_0\), where the action of
\(\Lambda\) is by translation in the parameter set \(\Lambda\). This
dynamical system, known in mathematics as a \emph{Bernoulli shift}, was
suggested by Bellissard as a model for the disorder space (see
\autocite{ps16} for an exposition).

If \(\Omega_0\) is the space of possible perturbations of the on-site
potential, then, in keeping with the literature, we assume that
\(\Omega_0\) is a compact convex subset of \(\reals^d\). (The physical
justification for this assumption is discussed at length in
\autocite{kuehne-prodan}.) Mathematically, one may just as well make the
following more general hypothesis.

\BeginKnitrBlock{proposition}
\protect\hypertarget{prp:CleanLimit}{}{\label{prp:CleanLimit} }Suppose that
\(\Omega\) is a Bernoulli shift with contractible \(\Omega_0\). Let
\(\mathbb A^0\), \(\widehat{\mathbb A}^0\), and \(\mathbb A_\partial^0\)
denote the algebras \(\mathbb A\), \(\smash{\widehat{\mathbb A}}\), and
\(\mathbb A_\partial\), respectively, for the case of no disorder
(\emph{i.e.}, where \(\Omega\) is a one-point space). There is a
commutative diagram \begin{equation}
\begin{tikzcd}
  0\rar{}&\mathbb A_\partial\otimes\knums(\ell^2(\nats))\rar{\iota}\dar{}&\widehat{\mathbb A}\rar{\vrho}\dar{}&\mathbb A\rar{}\dar{}&0\\
  0\rar{}&\mathbb A_\partial^0\otimes\knums(\ell^2(\nats))\rar{\iota^0}&\widehat{\mathbb A}^0\rar{\vrho^0}&\mathbb A^0\rar{}&0
\end{tikzcd}
\label{eq:homotopy-diag}
\end{equation} where the rows are the short exact sequence
\eqref{eq:bb-ses} for \(\mathbb A\) and \(\mathbb A^0\), respectively, and
the vertical maps are homotopy equivalences of real C\(^*\)-algebras.

In particular, the bulk-boundary map \(\partial\) is surjective and
\begin{align}
  KR_{s-r+2}(\mathbb{A}) &= KR_{s-r+2}(\mathbb A^0) = \bigoplus_{i=0}^d\binom di KR_{s-r+2-i}(\mathbb{C}), \label{eq:kr-torus}\\
  K_s(\mathrm{A})&=K_s(\mathrm A^0)=\mathbb{Z}^{2^{d-1}}, \label{eq:k-torus}
\end{align}
\EndKnitrBlock{proposition}

\BeginKnitrBlock{remark}
{}Equation \eqref{eq:kr-torus} identifies
\(KR_{s-r+2}(\mathbb A)\) with the real \(K\)-theory of the
\(d\)-dimensional (Brillouin) torus. The summands correspond to the
well-known \(KO\) groups of the point, see for example \autocite[Table
2]{abs}. The component of the bulk class of a disordered IQPV
\((J;\phi)\) of symmetry index \((r,s)\) corresponding to the summand
for \(i=d\) is what is often referred to in the literature as the
\emph{strong invariant}. The other components are summarily called
\emph{weak invariants}. As Theorem \ref{thm:ComplexBound} and
Proposition \ref{prp:CleanLimit} show, at least for the type of disorder
considered here, this distinction is not mathematically compelling.

In \autocite[Proposition 4.2.4]{ps16}, the \(K\)-theory of the bulk
algebra \(\mathrm A\) is computed under the assumption that \(\Omega\)
is a Bernoulli shift with \(\Omega_0\) a convex compact set. The proof
is based on the Pimsner--Voiculescu exact six-term sequence and relies
on the statement \autocite[Proposition 4.2.1]{ps16} that the action
\(\alpha_d\) of the translation in the direction transverse to the
boundary on \(\sh C(\Omega)\) is homotopic to the identity. It is
however not true in general that the homotopy \(\xi_t\) given in the
proof is actually an isotopy on \(\Omega\). For instance, let \(d=1\),
\(\Omega_0=[-1,1]\), and \(\omega_n\defi(-1)^n\). Then
\(\xi_{1/2}(0)=0\), but also \(\xi_{1/2}(\omega)=0\), so that
\(\xi_{1/2}\) is not injective. Hence, the proof as given in
\autocite{ps16} appears to be incomplete.
\EndKnitrBlock{remark}

\BeginKnitrBlock{proof}[of Proposition \ref{prp:CleanLimit}]
{}By
hypothesis, there is a point \(\omega_0\in\Omega\) and a homotopy
\(h=(h_t)\) of maps \(\Omega_0\longrightarrow\Omega_0\) where
\(h_0=\omega_0\) is the constant map and \(h_1={\id}_{\Omega_0}\).

Consider the inclusion
\(\eta:\endo{W}\longrightarrow\sh C(\Omega)\otimes\endo{W}\) and the
evaluation \[
  \eps:\sh C(\Omega)\otimes\endo{W}\longrightarrow\endo{W}:f\longmapsto f(x\longmapsto\omega_0)
\] at the constant sequence \(x\longmapsto\omega_0\). We have
\(\eps\circ\eta={\id}\). Moreover, for any \(t\in[0,1]\), \[
  \phi_t(f)(\omega)\defi f(h_t(\omega))\quad(f\in\sh C(\Omega)\otimes\endo{W},\omega\in\Omega)
\] defines a homotopy \((\phi_t)\) of real \(*\)-endomorphisms of
\(\sh C(\Omega)\otimes\endo{W}\) such that \[
  \phi_0=\eta\circ\eps,\quad\phi_1={\id}_{\sh C(\Omega)\otimes\endo{W}}.
\] By definition, the morphisms \(\eps\), \(\eta\), and \(\phi_t\) are
\(\Lambda\)-equivariant, and by restriction, also equivariant with
respect to \(\Lambda_\partial\) and \(\hat\Lambda\). By Theorem
\ref{thm:covcrossed}, Proposition \ref{prp:halfsp-covcrossed}, and
Theorem \ref{thm:bb-ses}, the existence of the diagram
\eqref{eq:homotopy-diag} follows.

The homotopy invariance of \(K\)-theory gives
\(KR_\bullet(\mathbb A)=KR_\bullet(\mathbb A^0)\). In view of
\autocite[Theorem 1.5.5]{schroder1993k}, \(\partial\) is surjective.
Since by Theorem \ref{thm:covcrossed}, \[
  \mathbb A^0=(\cplxs\rtimes\Lambda)\otimes\endo{W}=C^*(\Lambda)\otimes\endo{W},
\] where \(C^*(\Lambda)\) is the group C\(^*\)-algebra of \(\Lambda\),
we find that \[
  KR_\bullet(\mathbb A)=KR_\bullet(C^*(\Lambda))=KR_\bullet(\textstyle\bigotimes^d\sh C(\mathbb S^1)).
\] From this, Equation \eqref{eq:kr-torus} follows by \autocite[Theorem
1.5.4]{schroder1993k}. Equation \eqref{eq:k-torus} is a special case.
\EndKnitrBlock{proof}

\hypertarget{sec:discussion}{%
\section{Discussion}\label{sec:discussion}}

We have derived a framework for the bulk and boundary classification of
disordered free-fermion topological phases and proved that these are
naturally related. In this final section, we complement our previous
results by a discussion, and by drawing some physical conclusions.

\hypertarget{anomalous-nature-of-the-boundary-phase}{%
\subsection{Anomalous nature of the boundary
phase}\label{anomalous-nature-of-the-boundary-phase}}

The physicists among our readers may wonder how to interpret our main
result from the heuristics of anomaly inflow. Indeed, built on a
tight-binding model, the boundary algebra \(\mathbb{A}_\partial\) is
UV-complete and as such does not allow for the anomaly characteristic of
the spatial boundary of a system in a non-trivial topological phase.
Yet, the bulk-boundary correspondence established in Theorem
\ref{thm:ComplexBound} does assign to the anomalous system boundary a
class in the \(KR\)-theory of the algebra \(\mathbb A_\partial\).

To resolve the apparent conflict, one may observe that whereas the bulk
\(K\)-theory class of a free-fermion ground state in symmetry class
\(s\) resides in degree \(s+2\), the degree of the boundary class is
shifted by one to \(s+1\). The index shift may be traded for a
suspension, \emph{viz.} \[
    KR_{s+1}(\mathbb A_\partial)=KR_{s+2}(S\mathbb A_\partial),
\] where the suspension \(S\) is effected by the tensor product with the
algebra \(\sh C_0(i\reals)\). So, morally, the boundary phase is not
represented by a vector bundle (with the corresponding Clifford
symmetries) over the boundary; rather, it is represented by a vector
bundle over the \emph{suspension} of the boundary.

\hypertarget{the-role-of-the-trivial-phase}{%
\subsection{The role of the trivial
phase}\label{the-role-of-the-trivial-phase}}

Our construction of bulk classes depends on the choice of a trivial
phase. Indeed, we have constructed bulk classes as deviations from an
arbitrarily chosen reference IQPV. This circumstance is imposed by the
group structure of \(K\)-theory, which measures differences of certain
homotopy classes, rather than the classes themselves.

From a physical point of view, free-fermion topological phases do not
directly have a group structure, whereas differences of such phases do.
Therefore, it is advantageous for the physical interpretation of the
bulk classes as free-fermion topological phases that the choice of
reference IQPV is not fixed by the mathematical model, and instead
allows for adjustments according to the physical system of interest.

The situation is different for the boundary classes. Any reference
disordered IQPV \(J_\mathrm{ref}\) that has a \emph{unitary} lift
\(\smash{\widehat J_\mathrm{ref}}\) (in particular, any local reference
IQPV in the sense of Corollaries \ref{cor:bdy-local} and
\ref{cor:bdy-trivial}) will itself already represent a degenerate
Kasparov class, trivial in the associated \(K\)-group. This corresponds
to the understanding in physics that a boundary free-fermion topological
phase is trivial if there are no stable localised boundary states that
close the gap.

To state bulk-boundary correspondence cleanly, our definition of
boundary classes avoids the requirement that the reference IQPV be
trivial in this sense. It instead measures the deviation from an
arbitrary reference boundary class. The bulk-boundary map is sensitive
to the choice of reference phase and preserves this dependence.

\hypertarget{the-role-of-the-internal-space-w}{%
\subsection{\texorpdfstring{The role of the internal space
\(W\)}{The role of the internal space W}}\label{the-role-of-the-internal-space-w}}

Throughout this article, we have relied on a tight-binding approximation
in which the relevant degrees of freedom close to the chemical potential
split into a spatial contribution, corresponding to the translational
lattice of atomic sites \(\Lambda\), and a finite-dimensional internal
space \(V\) resp. \(W=V\oplus V^*\) independent of the spatial part. It
is important that the choice of this splitting is performed in such a
way that the only relevant spatial symmetries (in mean) of the
Hamiltonians of interest are the lattice translations.

This is shown by the following \emph{Gedankenexperiment} due to
Fu--Kane--Mele \autocite{Fu-Kane-Mele}. Assume that \(W\) is doubled to
\(W_2\defi W\oplus W\) while \(\Lambda\) is reduced to \(\Lambda_2\) by
merging atomic positions adjacent in one selected spatial direction. Let
\(\mathbb A_2\) be the bulk algebra defined in terms of \(W_2\) and
\(\Lambda_2\), and suppose we are in the setting of Proposition
\ref{prp:CleanLimit}. Although the \(K\)-groups of \(\mathbb A\) and
\(\mathbb A_2\) are isomorphic by the proposition, the natural
\enquote{doubling} map \(\phi_2:\mathbb A\longrightarrow\mathbb A_2\) is
not an isomorphism, not even on \(K\)-theory. Indeed, Fu--Kane--Mele
show for clean systems that, although the strong invariant is preserved,
some weak invariants are annihilated by doubling; by the proposition,
this also holds in the presence of disorder. Fu--Kane--Mele view the
annihilation of weak invariants as an indication that they are not
stable under disorder. However, as we have seen, all invariants are
preserved by the kind of disorder we consider. This is corroborated by
the work of Ringel--Kraus--Stern \autocite{rks}.

The problem lies in the fact that \(\phi_2\) maps an IQPV \(J\) in
\(\mathbb A\) to one in the algebra \(\mathbb A_2\) which forgets the
invariance (in mean) of \(J\) under the translations removed by the
doubling process. By incorporating these lost translations into the
definition of the algebra as symmetries of
\(\sh W_2=\ell^2(\Lambda_2)\otimes W_2\cong\sh W\), we obtain a new
algebra, isomorphic to \(\mathbb A\) under the doubling map, thus
removing the perceived inconsistencies.

\hypertarget{numerical-topological-invariants}{%
\subsection{Numerical topological
invariants}\label{numerical-topological-invariants}}

In this paper, we have defined bulk and boundary classes for disordered
IQPV in the symmetry classes of the Tenfold Way. For applications, it is
desirable to have numerical indices available which make this
topological classification amenable to computation. The derivation and
in-depth study of all such indices for our topological classes is beyond
the scope of this paper. Instead, we shall give an overview of the
literature~and then discuss the prominent special case of the strong
topological phase in some detail.

There are two major approaches to the construction of numerical
topological indices. One uses the pairing with cyclic cohomology
introduced by Connes, see \autocite[Chapter 3.3]{connes}. Attached to an
algebra \(A\), there are cyclic cohomology groups \(HC^n(A)\), dual to
\(K\)-theory in the sense that there is a canonical pairing \[
    K_n(A)\otimes HC^n(A)\longrightarrow\cplxs\quad(n=0,1).
\] This pairing has been used to write the quantised Hall conductance in
the IQHE, expressed in units of \(\smash{\frac{e^2}h}\), as the result
of pairing the \(K\)-theory class defined by the QHE Hamiltonian with a
canonical cyclic co-cycle \autocite{bellissard-etal,ps16}.

Although a pairing with cyclic cohomology can also be defined for the
\(KR\)-groups of real \(K\)-theory appearing in the context of the
present work, the result is unsatisfactory. Indeed, the \(KR\)-group of
the bulk algebra may have torsion, that is, it may contain non-zero
elements of finite order. For instance, the \(KR\)-theory of the torus
has \(\ints/2\ints\) summands in certain degrees. Any \(\cplxs\)- or
\(\ints\)-valued pairing will therefore be degenerate. Kellendonk
addresses this problem in recent work \autocite{kellendonk-index}.

A different approach to the construction of numerical topological
invariants is to replace cyclic cohomology by the \(K\)-homology groups
\(KR^n(\mathbb A)=KKR\parens{\mathbb A\otimes C\ell_{0,n},\cplxs}\) of
the bulk algebra \(\mathbb A\) and to use the Kasparov product \[
    KKR\Parens{C\ell_{s,r},\mathbb{A} \otimes C\ell_{0,1}}\otimes KKR\Parens{\mathbb A\otimes C\ell_{0,n},\cplxs}\longrightarrow KR_{s+1-r-n}(\cplxs).
\] Recall here that
\(KR_{s+1-r}(\mathbb A)=KKR\parens{C\ell_{s,r},\mathbb{A} \otimes C\ell_{0,1}}\)
is the group harbouring the bulk classes of free-fermion topological
phases that we have constructed. In principle, this approach avoids the
problems with \(\cplxs\)- or \(\ints\)-valued pairings, as the
right-hand side may be any summand appearing in the \(KR\)-theory of the
torus.

The difficulty lies in finding suitable classes in \(K\)-homology to
insert into the pairing. For \(n=d\), where \(d\) is the dimension of
lattice \(\Lambda\) of translations, such a \(K\)-homology class is
constructed in \autocite{bourne-carey-rennie}, leading to an index
formula for the strong topological invariant. A complete picture is as
yet unavailable for the case of \(n<d\) corresponding to the weak
invariants, although some contributions to this end have been made in
\autocite{bourne-sb-weak}.

Let us examine the strong invariant more closely. As we have seen above,
in the clean limit, it is defined as the top-dimensional \(K\)-theory
invariant of the \(d\)-dimensional torus. As long as the disorder space
\(\Omega\) meets the assumptions of Proposition \ref{prp:CleanLimit},
this definition still makes sense. In general, the definition depends on
the choice of a point \(\omega\in\Omega\) fixed by the action of the
translational lattice \(\Lambda\). (For the Bernoulli shifts of
Proposition \ref{prp:CleanLimit}, the fixed points are the constant
sequences.) Evaluation at \(\omega\) defines a map \[
    \lambda_{d-1}^\omega:KR_{s-r+1}(\mathbb{A}_\partial)\longrightarrow KR_{s-r+2-d}(\mathbb{C})
\] generalising the strong invariant
\autocite{bourne-kellendonk-rennie}.

As the notation suggests, \(\lambda_{d-1}^\omega\) might depend on
\(\omega\). In the situation of Proposition \ref{prp:CleanLimit}, it is
independent thereof. More generally, if the action of \(\Lambda\) on
\(\Omega\) is ergodic for an invariant probability \(\mathbb P\) on
\(\Omega\), then it is \(\mathbb P\)-a.s. constant as a function of
\(\omega\). (This is really a generalisation of the former statement, as
Bernoulli shifts admit ergodic measures.)

So far, we have assumed that the boundary is perpendicular the \(d\)th
coordinate direction. For a fixed choice of disordered IQPV of a given
symmetry index, one may observe topological features at different
boundaries. Although they may in general depend on the choice of the
boundary, we have for the strong topological invariant: \[
    (-1)^i\lambda_{d-1}^\omega \circ \partial_i= (-1)^j\lambda^\omega_{d-1} \circ \partial_j\quad(i,j=1,\dotsc,d),
\] as follows from \autocite[Corollary 3.6]{bourne-kellendonk-rennie}.
Here, \(\partial_i\) is the bulk-boundary map for a boundary plane
perpendicular to the \(i\)th coordinate axis. Thus, pull-backs to the
bulk of the strong topological invariants for different boundaries
differ only by the sign corresponding to the change of orientation. From
our explicit form of the bulk-boundary correspondence, we conclude that
all boundaries exhibit gapless boundary states, as long as the strong
topological bulk phase is non-trivial.

\begin{acknowledgements}
The authors wish to thank Matthias Lesch for his remarks on Lemma 5 in a
previous version of the manuscript, which have helped us improve a
statement which otherwise could easily have led to misconceptions. Our
thanks also extend to an anonymous referee, whose comments have aided us
in enhancing the article. Research supported by Deutsche
Forschungsgemeinschaft (DFG), Projektnummer 316511131, TRR 183, project
A03 (all authors), Projektnummer 282448306, AL 698/3-1 (AA), and
Projektnummer 49599460, GSC 260 (CM), as well as QM\(^2\)--Quantum
Matter and Materials as part of the Institutional Strategy of the
University of Cologne in the Excellence Initiative (all authors). We
(AA, CM) thank the Department of Mathematics of the University of
California at Berkeley and the Institute for Theoretical Physics at the
University of Cologne for their hospitality during the preparation of
this article. CM acknowledges the participation in the School on
``\(KK\)-theory, Gauge Theory, and Topological Phases'' (February
27--March 10, 2017) at the Lorentz Center, Leiden, and the Thematic
Programme on ``Bivariant \(K\)-theory in Geometry and Physics''
(November 6--30, 2018) at the Erwin Schrödinger Institute for
Mathematical Physics, Vienna.
\end{acknowledgements}
\printbibliography

\end{document}